\documentclass[useAMS,usenatbib]{mn2e}
\usepackage{graphicx}
\usepackage{url}
\usepackage{multirow}

\newcommand{\m}{^{\rm c}}
\newcommand{\AlII}{Al\,\textsc{ii}}
\newcommand{\AlIII}{Al\,\textsc{iii}}
\newcommand{\ArI}{Ar\,\textsc{i}}
\newcommand{\CI}{C\,\textsc{i}}
\newcommand{\CII}{C\,\textsc{ii}}

\newcommand{\CrII}{\textrm{Cr}\,\textsc{ii}}

\newcommand{\FeII}{Fe\,\textsc{ii}}

\newcommand{\HI}{\textrm{H}\,\textsc{i}}

\newcommand{\HII}{\textrm{H}\,\textsc{ii}}

\newcommand{\Lya}{Ly$\alpha$}
\newcommand{\Lyb}{Ly$\beta$}

\newcommand{\NHI}{$N(\textrm{H}\,\textsc{i})$}

\newcommand{\NI}{N\,\textsc{i}}
\newcommand{\NII}{N\,\textsc{ii}}

\newcommand{\NiII}{Ni\,\textsc{ii}}
\newcommand{\OI}{O\,\textsc{i}}

\newcommand{\SII}{S\,\textsc{ii}}
\newcommand{\SiII}{Si\,\textsc{ii}}

\newcommand{\ZnII}{\textrm{Zn}\,\textsc{ii}}

\def\rahr{^{\rm h}}
\def\ramin{^{\rm m}}
\def\rasec{\!\!^{\rm s}}
\def\decdeg{^{\circ}}
\def\decmin{'}
\def\decsec{\!\!''}

\def\ltsima{$\; \buildrel < \over \sim \;$}
\def\simlt{\lower.5ex\hbox{\ltsima}}
\def\gtsima{$\; \buildrel > \over \sim \;$}
\def\simgt{\lower.5ex\hbox{\gtsima}}

\title[The most metal-poor DLAs]
{The most metal-poor damped Ly$\alpha$ systems:
Insights into chemical evolution in the very metal-poor regime\thanks{
Based on observations collected at the European Organisation for Astronomical Research 
in the Southern Hemisphere, Chile [VLT programs 67.A-0078(A), 69.A-0613(A), 083.A-0042(A), 085.A-0109(A)], and
at the W.M. Keck Observatory, which is operated as a scientific partnership among the California Institute of Technology, the University of California and the National Aeronautics and Space Administration. The Observatory was made possible by the generous financial support of the W.M. Keck Foundation.
Keck telescope time was granted by NOAO, through the Telescope
System Instrumentation Program (TSIP). TSIP is funded by NSF.}}

\author[Cooke et al.]{Ryan Cooke$^{1,2}$\thanks{email: rcooke@ast.cam.ac.uk}, 
Max Pettini$^{1,2}$, Charles C. Steidel$^{3}$, Gwen C. Rudie$^{3}$, Poul E. Nissen$^{4}$\\
$^1$Institute of Astronomy, Madingley Road, Cambridge, CB3 0HA\\
$^2$Kavli Institute for Cosmology, Madingley Road, Cambridge, CB3 0HA\\
$^3$California Institute of Technology, MS 249-17, Pasadena, CA 91125, USA\\
$^4$Department of Physics and Astronomy, University of Aarhus, 8000 Aarhus C, Denmark}

\begin{document}

\date{Accepted . Received ; in original form }
\pagerange{\pageref{firstpage}--\pageref{lastpage}} 
\pubyear{2011}

\maketitle

\label{firstpage}

\begin{abstract}
We present a high spectral resolution survey of the most metal-poor damped 
Ly$\alpha$ absorption systems (DLAs) aimed at probing the nature and nucleosynthesis 
of the earliest generations of stars. Our survey comprises 22 systems with
iron abundance less than 1/100 solar; observations of seven of these 
are reported here for the first time.
Together with recent measures of the abundances of C and O in Galactic
metal-poor stars, we reinvestigate the 
trend of C/O in the very metal-poor regime and we compare, for the first time,
the O/Fe ratios in the most metal-poor DLAs and in halo stars. 
We confirm the near-solar values of C/O in DLAs at the lowest metallicities 
probed, and find that their distribution is in agreement with that seen in 
Galactic halo stars. 
We find that the O/Fe ratio in very metal-poor (VMP) 
DLAs is essentially constant, and shows very little 
dispersion, with a mean [$\langle$O/Fe$\rangle$]\,$=+0.39\pm0.12$ ,
in good agreement with the values measured in Galactic 
halo stars when the oxygen abundance 
is measured from the [\OI]\,$\lambda 6300$ line. 
We speculate that such good agreement in the observed abundance trends 
points to a universal origin for these metals. In view of this agreement, 
we construct the abundance pattern for a typical very metal-poor DLA and 
compare it to model calculations of Population II and Population III nucleosynthesis 
to determine the origin of the metals in VMP DLAs. 
Our results suggest that the most metal-poor DLAs may have been enriched by 
a generation of metal-free stars; however, given that abundance measurements
are currently available for only a few elements, we cannot 
yet rule out an additional contribution from Population II stars. 
\end{abstract}


\section{Introduction}

The initial conditions for cosmic chemical evolution
are of fundamental importance to our understanding
of galaxy formation and the process 
of galactic chemical evolution. These conditions, 
set by the yields of the first few generations of stars, 
depend on various (largely unknown) factors including
the form of the primordial stellar initial mass function and 
the uniformity of the enrichment of the intergalactic 
medium (IGM; \citealt{BroLar04,KarBroBla11}).
In order to pin down the initial conditions 
of cosmic chemical evolution, one should seek 
to understand the origin and relative 
abundances of the metals in the least chemically 
evolved systems.

The most metal-poor damped \Lya\ systems (DLAs), 
for example, are usually interpreted as 
distant protogalaxies at an early 
stage of chemical evolution \citep{Ern06,Coo11}. 
Whilst the origin of their metals is still 
largely unknown, recent hydrodynamical simulations 
suggest that such systems might have been enriched 
by just a few supernova events \citep{Bla11}. 
If this is indeed the case, 
the most metal-poor DLAs provide 
a simple route to study the first stages of 
chemical enrichment in our Universe. 

By definition, DLAs have a neutral hydrogen
column density in excess of 
$2\times10^{20}$ \HI\ atoms cm$^{-2}$ (\citealt{Wol86}; 
see also the review by \citealt{WolGawPro05}),
which acts to self-shield the gas from the ultraviolet
background radiation of quasars (QSOs) and 
galaxies \citep{HarMad01}. This results in the gas having
a simple ionization structure subject to
negligible corrections for unseen ion stages \citep{Vla01}, 
quite unlike the \Lya\ forest clouds 
that trace the low density regions 
of the IGM (e.g. \citealt{SimSarRau04}).
The main concerns that limit abundance studies
in DLAs are line saturation 
and the possibility that dust may hide some fraction 
of the metals \citep{Vla04}. 
These concerns are alleviated when the metallicity
of the DLA is below $\sim10^{-2}$ Z$_{\odot}$, which 
is also the regime where we expect to uncover the enrichment 
signature of the earliest generations of stars.

The recent interest in the most 
metal-poor DLAs \citep{Pet08,Pen10}
complements the ongoing local studies of 
metal-poor stars in the halo of the Milky Way
\citep{Cay04,BeeChr05,Sud08,Fre10}. These stars
are believed to have condensed out of near-pristine
gas (perhaps a metal-poor DLA itself?), 
that was enriched by only a few earlier generations of stars.
Thus, the first generation of stars can also 
be studied through the signature retained 
in the stellar atmospheres of the most 
metal-poor stars in the halo of our Galaxy.
However, unlike the relative ease with which one can
measure the abundances of metal-poor DLAs,
deriving element abundances from the stellar atmospheres
of metal-poor stars is not straightforward \citep{Asp05}.
Systematic uncertainties in the derived
abundances are introduced by assuming that the
spectral line being examined forms
in a region that is in local
thermodynamic equilibrium (LTE), as well as 
the need to account for three-dimensional (3D) 
effects in the 1D stellar atmosphere models. 

These effects are particularly acute
for oxygen, where several different
abundance indicators are known to produce
contradictory estimates in the low-metallicity
regime \citep{Gar06}. Despite the efforts
of many authors, our uncertainty in the
derived oxygen abundances has sparked
an ongoing debate as to the trend of
[O/Fe]\footnote{We adopt the standard notation:
[A/B] $\equiv \log (N_{\rm A}/N_{\rm B}) - \log (N_{\rm A}/N_{\rm B})_{\odot}$,
where $N_{\rm A,B}$ refers to the number of atoms in element A and B.}
in the Milky Way when [Fe/H] $\simlt-1.0$.
A history of the relevant discussion on 
[O/Fe] is provided by \citet{McW97}, with 
further details given in Section~\ref{sec:ofe}. 
In brief, at low metallicity, both O and Fe 
are produced exclusively by type-II supernovae (SNe II) 
and the winds from their progenitors. 
When [Fe/H] $\simgt-1.0$, there 
is a drop in [O/Fe] due to the 
\emph{delayed} contribution of Fe from 
type-Ia supernovae (SNe Ia). 
Thus, the [O/Fe] ratio is most commonly
used to measure the time
delay between SNe II and the onset
of SNe Ia. At the lowest metallicity, 
however, one can use the [O/Fe] ratio 
as a measure of the relative production 
of $\alpha$- to Fe-peak elements by 
the first few generations of massive stars. 

Another key diagnostic ratio at low 
metallicity that may shed light on the nature 
of the early generations of stars was uncovered 
by \citet{Ake04} who reported 
a rather surprising evolution of [C/O] 
with decreasing O abundance in their sample of 
34 halo stars (see also \citealt{Spi05}).
In disc and halo stars when the oxygen 
abundance is $\simgt-1.0$, [C/O] steadily rises 
from [C/O] $\sim-0.5$ to solar. 
When [O/H] $\simlt -1.0$, galactic chemical 
evolution models that \emph{only} consider 
the nucleosynthetic products of 
Population II stars predict [C/O] to 
decrease or plateau, contrary to the observed 
trend. The increase in [C/O] with decreasing 
metallicity has thus been interpreted as evidence for 
an increased carbon yield from either Population III 
stars \citep{ChiLim02,UmeNom03,HegWoo10}
or rapidly-rotating low-metallicity 
Population II stars \citep{Chi06}. At first, 
concerns were raised regarding the accuracy 
of the derived C and O abundances, since the 
lines used are subject to large non-LTE corrections. 
\citet{Fab09a}, however, performed a 
non-LTE analysis of the same lines, with 
further contraints from additional \CI\ lines,
to confirm the reality of the stellar [C/O] 
trend. These results depend somewhat on the 
adopted cross sections for collisions of \CI\ 
and \OI\ atoms with electrons and hydrogen
atoms, but for all probable values, [C/O] 
increases with decreasing metallicity 
when [O/H] $< -2.0$.

To summarize, at present there are still some 
remaining concerns that prevent us from accurately 
measuring C and O abundances in the 
atmospheres of metal-poor halo stars. 
These difficulties have prompted a few teams  
to focus on very metal-poor\footnote{Herein,
we adopt ``very metal-poor'' to be those DLAs with 
[Fe/H] $<-2.0$, in line with the classification 
scheme for stars proposed by \citet{BeeChr05}.}
(VMP) DLAs where the absorption lines of \CII\ and \OI\ 
may be unsaturated and the abundances of C and O 
can be measured with confidence. Unfortunately, these 
near-pristine DLAs are rare, falling in the tail of 
the metallicity distribution function of DLAs \citep{Pro07}. 
Thus, only a handful of confirmed VMP DLAs are 
known at present. The first  high 
spectral resolution survey 
($R \simeq 40000$, full width at half maximum, FWHM\,$\simeq 7$ km s$^{-1}$)
for VMP DLAs was conducted by \citet{Pet08},
whose specific goal was to study the relative
abundances of the CNO group of elements as 
a probe of early nucleosynthesis. Indeed, this 
was the first study to independently confirm 
the increased [C/O] abundance at low metallicity, 
suggesting that near-solar values of [C/O] are
commonplace in this metallicity regime. 

The [C/O] trend reported by \citet{Pet08} has 
also been independently noted by \citet{Pen10} 
in a medium spectral resolution  
($R\simeq 5000$, FWHM\,$\simeq 60$ km s$^{-1}$) 
survey of 35 DLAs (a preliminary report of 
this study can be found in \citealt{Pen08}). 
In many of their systems, the \CII\ and \OI\ lines were 
thought to be affected by line saturation, leaving only 
five DLAs to test the trend in C/O.
Interestingly, this sample of DLAs suggests that [C/O] 
continues to rise to \emph{supersolar} values 
when [O/H]\,$\simlt -3$.

\begin{table*}
\centering
\begin{minipage}[c]{1.0\textwidth}
    \caption{\textsc{Journal of Observations}}
    \begin{tabular}{@{}lcrrllcccl}
    \hline
    \hline
   \multicolumn{1}{c}{QSO}
& \multicolumn{1}{c}{$g^{\rm a}$} 
& \multicolumn{1}{c}{$z_{\rm em}$}
& \multicolumn{1}{c}{$z_{\rm abs}$}
& \multicolumn{1}{c}{Telescope/}
& \multicolumn{1}{c}{Wavelength}
& \multicolumn{1}{c}{Resolution}
& \multicolumn{1}{c}{Integration}
& \multicolumn{1}{c}{S/N$^{\rm b}$}
& \multicolumn{1}{c}{Program ID}\\
   \multicolumn{1}{c}{ }
& \multicolumn{1}{c}{(mag)}
& \multicolumn{1}{c}{ }
& \multicolumn{1}{c}{ }
& \multicolumn{1}{c}{Instrument }
& \multicolumn{1}{c}{ Range (\AA)}
& \multicolumn{1}{c}{(km~s$^{-1}$)}
& \multicolumn{1}{c}{Time (s)}
& \multicolumn{1}{c}{ }
& \multicolumn{1}{c}{}\\
    \hline
J0311$-$1722  & 17.7  & 4.039  & 3.73400  & VLT/UVES     & 4370--6410$^{\rm c}$  & 6.9   & 3\,600           & 15  & 69.A-0613(A)$^{\rm d}$     \\
J0831$+$3358  & 19.5  & 2.427  & 2.30364  & KECK/HIRESb  & 3130--5970$^{\rm c}$  & 7.3   & 18\,600          & 15  & A185Hb                  \\
J1001$+$0343  & 19.2  & 3.198  & 3.07841  & VLT/UVES     & 3740--6650$^{\rm c}$  & 7.3   & 33\,700          & 40  & 083.A-0042(A)           \\
J1037$+$0139  & 19.4  & 3.059  & 2.70487  & VLT/UVES     & 3640--6650$^{\rm c}$  & 7.3   & 26\,075          & 45  & 083.A-0042(A)           \\
J1340$+$1106  & 19.0  & 2.914  & 2.50792  & VLT/UVES \&  & 3483--6652$^{\rm c}$  & 7.3   & 29\,800 \&       & 50  & 085.A-0109(A) \&        \\
              &       &        & 2.79583  & VLT/UVES \&  & 3483--9396$^{\rm c}$  & 10.3  & 10\,800 \&       & 40  & 67.A-0078(A)$^{\rm d}$ \&  \\
              &       &        &          & KECK/HIRES   & 4648--7044$^{\rm c}$  & 8.1   & 8\,100           & 25  & U11H$^{\rm d}$             \\
J1419$+$0829  & 18.9  & 3.030  & 3.04973  & VLT/UVES     & 3710--6652$^{\rm c}$  & 7.3   & 29\,800          & 43  & 085.A-0109(A)           \\
    \hline
    \end{tabular}
    \smallskip

 $^{\rm a}$Magnitudes are SDSS $g$-band, except for J0311$-$1722 (not covered by the SDSS) which is $R$-band \citep{Per01}.\\
 $^{\rm b}$Indicative signal-to-noise ratio at 5000\,\AA\ (or 6000\,\AA\ in the case of J1340$+$1106).\\
 $^{\rm c}$With some wavelength gaps.\\
 $^{\rm d}$Spectra downloaded from either the UVES or HIRES data archives$^{4,5}$.\\
    \label{tab:obs}
\end{minipage}
\end{table*}

Such surveys for VMP DLAs are 
most useful for studying the 
general properties of \emph{entire clouds} 
of near-pristine gas before they form stars. 
In this contribution, we build on our ongoing 
survey for the most metal-poor DLAs as 
probes of early nucleosynthesis \citep{Pet08,Coo11}. 
With additional systems drawn from the literature, 
the total sample presented herein 
amounts to 22 VMP DLAs with abundance 
measurements derived from high spectral resolution data. 
From this sample, we confirm the elevated [C/O] values 
in these systems and, for the 
first time, present the trend of [O/Fe] in the 
most metal-poor DLAs. For both of these diagnostic ratios, 
we comment on the implications our findings 
have for local studies of Galactic metal-poor 
halo stars. Finally, we construct the abundance 
pattern of a typical VMP DLA for the elements
C, N, O, Al, Si and Fe,
and compare it to model calculations of Population II and 
Population III nucleosynthesis.  

This paper is arranged as follows. 
In Section~\ref{sec:obs} we detail the 
processing and preparation of the data. 
In Section~\ref{sec:pfaa} we explain the profile 
fitting procedure used for our new sample 
of VMP DLAs, which are discussed in 
Section \ref{sec:indobj}. The accuracy of our abundance 
analysis is discussed in Section~\ref{sec:abund_anal}, before 
we investigate the behaviour of [C/O] and 
[O/Fe] in VMP DLAs and compare with stellar data, 
in Section~\ref{sec:dvs}. Finally, we discuss the 
implications for our findings in Section~\ref{sec:disc}, 
before summarizing our main results and 
drawing our conclusions in Section~\ref{sec:conc}. 

\section{Observations and Data Reduction}
\label{sec:obs}

\subsection{Target Selection}

Even at the relatively low spectral resolution afforded by the Sloan Digital Sky 
Survey (SDSS), one can easily recognise DLAs in the spectra of quasars, 
owing to the characteristic damping wings of the \Lya\ 
absorption line profile. Subsequent identification of associated metal 
line absorption leads to a rough estimate of the gas-phase metallicity. 
Candidate metal-poor DLAs are then identified as those DLAs that appear to 
exhibit no metal line absorption; their absorption features are unresolved at
the spectral resolution of the SDSS. However, when these candidates are 
re-observed with echelle spectrographs of high resolution 
(R$\simgt 30000$, FWHM$\simlt 10\,\,{\rm km~s}^{-1}$), the 
metal absorption lines are resolved, and in many cases it is possible 
to measure elemental abundances with confidence \citep{Pet08}.

The most recent trawls through SDSS spectra of $\sim8000$
quasars with $z_{\rm em}\simgt2.2$ has 
yielded a sample of $\sim1000$
DLAs \citep{Not09,ProWol09}, of which 
$\sim400$ are classified as `metal-poor' \citep{Pen10}.\footnote{In this context,
a DLA is classed as `metal-poor'  if it has fewer than three significantly ($4\sigma$) 
detected metal absorption lines at the spectral resolution of the SDSS.}  
From compilations such as these, we selected 
a handful of metal-poor DLA candidates that 
exhibit \emph{no discernible metal-line absorption} at the 
spectral resolution of the SDSS, and 
re-observed these with echelle spectrographs, 
giving higher priority to candidates with: 
(i)   bright quasars, so as  to efficiently obtain spectra with signal-to-noise ratios $S/N \simgt 20$ in the continuum;
(ii)  DLAs where the difference between $z_{\rm abs}$ and $z_{\rm em}$ is minimized 
      so that the absorption lines of interest
      (e.g. \OI\,$\lambda1302$ and \CII\,$\lambda1334$)
      are not blended with unrelated \Lya\ forest lines;
(iii) quasars whose emission redshift is below $z_{\rm em}\simlt3.3$, so there is an
      improved chance that other lines of interest 
      (e.g. \OI\,$\lambda1039$ and \CII\,$\lambda1036$) 
      are not blended with \Lya\ forest lines;
(iv)  DLAs at the low end of the column density distribution 
      function -- that are still DLAs -- to ensure that even the
      strongest metal absorption lines are unsaturated, allowing 
      us to measure the metal ion column densities with confidence; and
(v)   quasars with more than one metal-poor DLA candidate in their spectra.

Our survey to date consists of 12 DLAs with [Fe/H]\,$\leq -2.0$.
Initial results for four of these
were published by \citet{Pet08}, while a fifth DLA, showing a pronounced
C enhancement relative to Fe, was the subject of a recent study by \citet{Coo11}.
The observations and analysis of the remaining seven DLAs, including one from
the European Southern Observatory's (ESO) Ultraviolet and
Visual Echelle Spectrograph (UVES) data archive, are presented here.
To our own data, we add a collection of published abundance measurements
in ten VMP DLAs, selected as described in Section~\ref{sec:lit_dlas}, 
to assemble an overall sample of abundance measurements in 22 VMP DLAs, 
all obtained from  high resolution spectra ($R \simgt 30\,000$).

\subsection{Echelle spectroscopic follow-up}

In order to achieve the high signal-to-noise ratio and spectral 
resolution required for accurate DLA abundance measurements,
we observed our prime candidates with echelle
spectrographs on $8-10$\,m class telescopes.
Most of our candidates were observed with the 
UVES spectrograph \citep{Dek00}, which is mounted 
on UT2 at the Very Large Telescope facility. 
An additional system was observed with 
the W.~M.~Keck Observatory's 
High Resolution Echelle Spectrometer
(HIRES, \citealt{Vog94}) on the Keck \textsc{i} telescope.
Table~\ref{tab:obs} lists details of the observations 
of seven VMP DLAs reported here for the first time.
For J1340$+$1106, we include details of some additional data,
of comparable spectral resolution to ours, retrieved
from the 
ESO\footnote{http://archive.eso.org/eso/eso\_archive\_main.html} 
and Keck Observatory\footnote{https://www2.keck.hawaii.edu/koa/public/koa.php} 
data archives (program IDs 67.A-0078(A) and U11H respectively).
We have also retrieved UVES spectra of the quasar J0311$-$1722 
from the ESO data archive 
(program ID 69.A-0613(A); see \citealt{Per05}), 
since the metal lines for the VMP DLA along this sightline 
have not been previously analysed. 

Our own UVES observations [program IDs 
083.A-0042(A) \&\ 085.A-0109(A)] employed 
a $1.2''$ wide slit, resulting in a spectral 
resolution $R\sim41\,000$ 
(velocity FWHM $\approx7.3$ km s$^{-1}$) 
sampled with $\sim3$ pixels. 
We used dichroic \#1 to split 
the quasar light into the blue and red 
spectroscopic arms containing the 
HER\_5 filter and SHP700 filter 
respectively. The resulting central 
wavelength for each arm was 
3900\,\AA\ (blue) and 5640\,\AA\ (red). 
Both the blue- and red-sensitive CCDs 
used $2\times2$ on-chip binning.
For our HIRES observations 
(program ID A185Hb) we used the 
C5 decker (a $7.0 \times 1.148$\,arcsec slit) which,
with sub-arcsec seeing, 
gave a spectral resolution of 
$R\sim41\,000$ (cf. \citealt{Coo11}), 
also sampled with $\sim3$ pixels.
We employed the ultraviolet cross-disperser
with no filters, and used $2\times2$ 
on-chip binning.

\subsection{Data Reduction}

We used the standard UVES data reduction pipeline\footnote{We used 
version 4.3.0, available from:\\
http://www.eso.org/sci/software/pipelines/}
provided by ESO to reduce the UVES data. 
The UVES reduction pipeline performs the 
usual steps relevant to echelle data 
reduction. The preliminary steps include bias 
subtraction, flat fielding, and background 
subtraction. The echelle orders are then traced 
using a flat field frame taken with a 
pinhole decker, and 1D 
spectra extracted. The data are 
wavelength calibrated with reference to 
a ThAr lamp.

The HIRES data were reduced with the \textsc{makee} data 
reduction pipeline developed by Tom Barlow. 
\textsc{makee} performs the same reduction 
steps as outlined above, but a trace frame 
is not always readily available. When
available, the orders were traced using a 
flat-field frame taken with a pinhole decker.
Otherwise, the science exposure of the 
quasar itself was used when a satisfactory trace 
could be made. Failing this, a trace frame was 
generated with a suite of purpose-built 
\textsc{python} programs, using the science 
frame as a guide to trace the echelle orders.

Following these initial reduction steps,
for each object we combined the science 
exposures using the software package 
\textsc{uves\_popler}\footnote{\textsc{uves\_popler} can be downloaded from\\
http://astronomy.swin.edu.au/$\sim$mmurphy/UVES\_popler},
maintained by Michael Murphy.
This software merges individual echelle orders,
and maps the data onto a vacuum heliocentric wavelength scale.
Finally, we normalized the data by dividing out
the quasar continuum and emission lines.
Using the approximate redshift derived from 
each DLA's SDSS discovery spectrum, we then 
prepared the final data for analysis by extracting
a $\pm 150$ km s$^{-1}$ window around the pixel with highest 
optical depth near all available absorption lines of interest. 
Finally, a further fine adjustment 
to the continuum was applied to these 
extracted portions of the spectra when necessary.

\section{Profile Fitting}
\label{sec:pfaa}

For DLAs with a metallicity 
below 1/100 $Z_{\odot}$, 
the metal line absorption is typically 
concentrated in only a few clouds of low 
velocity dispersion \citep{Led06,Mur07,Pro08}. 
By assuming that 
a Maxwellian distribution accurately 
describes the velocities of the dominant atoms 
within the neutral cloud, we can model 
a DLA's absorption lines by a Voigt profile.
To this end, we employed
the Voigt profile fitting software
\textsc{vpfit} to derive the cloud 
parameters for all DLAs
in our sample.\footnote{\textsc{vpfit} is available 
from http://www.ast.cam.ac.uk/${\sim}$rfc/vpfit.html}

\begin{table}
\centering
    \caption{\textsc{Absorption Components of Low Ion Transitions }}
    \begin{tabular}{@{}cccc}
    \hline
Component
& $z_{\rm abs}$
& $b$
& Fraction$^{\rm a}$\\
Number
& 
& (km~s$^{-1}$)
& $N$(\SiII) \\
\hline
\multicolumn{4}{c}{\textbf{J0311$-$1722}: DLA at $z_{\rm abs}=3.73400$, $\chi^{2}$/dof$=0.80$} \\
1 & 3.733862 $\pm$ 0.000017 &  5.6 $\pm$ 1.2  &  0.18 \\
2 & 3.733998 $\pm$ 0.000007 &  2.7 $\pm$ 0.8  &  0.23 \\
3 & 3.734035 $\pm$ 0.000028 & 14.4 $\pm$ 1.5  &  0.20 \\
4 & 3.734439 $\pm$ 0.000002 &  4.6 $\pm$ 0.2  &  0.39 \\
\hline
\multicolumn{4}{c}{\textbf{J0831$+$3358}: DLA at $z_{\rm abs}=2.30364$, $\chi^{2}$/dof$=1.20$} \\
1 & 2.303565 $\pm$ 0.000004 &  4.4 $\pm$ 0.2  &  0.49 \\
2 & 2.303720 $\pm$ 0.000004 &  6.1 $\pm$ 0.3  &  0.51 \\
\hline
\multicolumn{4}{c}{\textbf{J1001$+$0343}: DLA at $z_{\rm abs}=3.07841$, $\chi^{2}$/dof$=1.42$} \\
1 & 3.078413 $\pm$ 0.000002 & 7.0 $\pm$ 0.1 &  1.00 \\
\hline
\multicolumn{4}{c}{\textbf{J1037$+$0139}: DLA at $z_{\rm abs}=2.70487$, $\chi^{2}$/dof$=1.08$} \\
1 &  2.704870 $\pm$ 0.000002 & 5.9 $\pm$ 0.2 &  1.00 \\
\hline
\multicolumn{4}{c}{\textbf{J1340$+$1106}: DLA at $z_{\rm abs}=2.50792$, $\chi^{2}$/dof$=1.38$} \\
1 &  2.507649 $\pm$ 0.000003  & 2.0 $\pm$ 0.4 & 0.19  \\
2 &  2.507921 $\pm$ 0.000001  & 5.8 $\pm$ 0.1 & 0.81  \\
\hline
\multicolumn{4}{c}{\textbf{J1340$+$1106}: DLA at $z_{\rm abs}=2.79583$, $\chi^{2}$/dof$=1.76$} \\
1 &  2.7955454 $\pm$ 0.0000018  & 9.2 $\pm$ 0.1 & 0.21  \\
2 &  2.7958272 $\pm$ 0.0000007  & 6.55 $\pm$ 0.05 & 0.79  \\
\hline
\multicolumn{4}{c}{\textbf{J1419$+$0829}: DLA at $z_{\rm abs}=3.04973$, $\chi^{2}$/dof$=1.45$} \\
1 &  3.049649 $\pm$ 0.000002 & 3.5 $\pm$ 0.1 & 0.43  \\
2 &  3.049835 $\pm$ 0.000002 & 6.4 $\pm$ 0.1 & 0.57  \\
\hline
    \end{tabular}
\begin{flushleft}
$^{\rm a}${Fraction of the total column density of \SiII.}\\
\end{flushleft}
\label{tab:cloud_models}
\end{table}

\begin{table*}
\caption{\textsc{Adopted metal line laboratory wavelengths and oscillator strengths}}
\centering
    \begin{tabular}{lr@{.}lr@{.}lp{0.7cm}lr@{.}lr@{.}lp{0.7cm}lr@{.}lr@{.}l}
    \hline
   \multicolumn{1}{l}{Ion}
& \multicolumn{2}{l}{Wavelength}
& \multicolumn{2}{c}{$f$}
&& \multicolumn{1}{l}{Ion}
& \multicolumn{2}{l}{Wavelength}
& \multicolumn{2}{c}{$f$}
&& \multicolumn{1}{l}{Ion}
& \multicolumn{2}{l}{Wavelength}
& \multicolumn{2}{c}{$f$}\\
   \multicolumn{1}{c}{}
& \multicolumn{2}{c}{(\AA)}
& \multicolumn{2}{c}{}
&& \multicolumn{1}{c}{}
& \multicolumn{2}{c}{(\AA)}
& \multicolumn{2}{c}{}
&& \multicolumn{1}{c}{}
& \multicolumn{2}{c}{(\AA)}
& \multicolumn{2}{c}{}\\
    \hline
\CII    & 1036&3367      & 0&118     &&   \AlII   & 1670&7886      & 1&740     &&   \FeII   & 1063&1764      & 0&0547    \\
\CII    & 1334&5323      & 0&1278    &&   \AlIII  & 1854&71829     & 0&559     &&   \FeII   & 1081&8748      & 0&0126    \\
\CII*   & 1335&6627      & 0&01277   &&   \AlIII  & 1862&79113     & 0&278     &&   \FeII   & 1096&8769      & 0&0327    \\
\CII*   & 1335&7077      & 0&115     &&   \SiII   &  989&8731      & 0&171     &&   \FeII   & 1125&4477      & 0&0156    \\
\NI     & 1134&4149      & 0&0278    &&   \SiII   & 1020&6989      & 0&0168    &&   \FeII   & 1143&2260      & 0&0192    \\
\NI     & 1134&9803      & 0&0416    &&   \SiII   & 1190&4158      & 0&292     &&   \FeII   & 1144&9379      & 0&0830    \\
\NI     & 1199&5496      & 0&1320    &&   \SiII   & 1193&2897      & 0&582     &&   \FeII   & 1260&533       & 0&0240    \\
\NI     & 1200&2233      & 0&0869    &&   \SiII   & 1260&4221      & 1&18      &&   \FeII   & 1608&4509      & 0&0577    \\
\NI     & 1200&7098      & 0&0432    &&   \SiII   & 1304&3702      & 0&0863    &&   \FeII   & 1611&20034     & 0&00138   \\
\NII    & 1083&9937      & 0&111     &&   \SiII   & 1526&7070      & 0&133     &&   \FeII   & 2344&21296     & 0&1142    \\
\OI     &  925&446       & 0&000354  &&   \SiII   & 1808&01288     & 0&00208   &&   \FeII   & 2374&46033     & 0&0313    \\
\OI     &  936&6295      & 0&00365   &&   \SII    & 1250&578       & 0&00543   &&   \FeII   & 2382&76418     & 0&320     \\
\OI     &  948&6855      & 0&00631   &&   \SII    & 1253&805       & 0&0109    &&   \NiII   & 1317&217       & 0&057     \\
\OI     &  976&4481      & 0&00331   &&   \SII    & 1259&5180      & 0&0166    &&   \NiII   & 1370&132       & 0&056     \\
\OI     &  988&5778      & 0&000553  &&   \ArI    & 1048&2199      & 0&263     &&   \NiII   & 1454&842       & 0&0323    \\
\OI     &  988&6549      & 0&0083    &&   \ArI    & 1066&6598      & 0&0675    &&   \NiII   & 1709&6042      & 0&0324    \\
\OI     &  988&7734      & 0&0465    &&   \CrII   & 2056&25693     & 0&1030    &&   \NiII   & 1741&5531      & 0&0427    \\
\OI     & 1039&2304      & 0&00907   &&   \CrII   & 2062&23610     & 0&0759    &&   \NiII   & 1751&9157      & 0&0277    \\
\OI     & 1302&1685      & 0&048     &&   \multicolumn{7}{c}{}\\
\hline
\end{tabular}
\label{tab:atomic}
\end{table*}

\textsc{vpfit} uses a chi-squared minimization 
algorithm to simultaneously fit multiple Voigt 
profiles to a set of absorption lines characterized 
by three free parameters: 
(1) the cloud's absorption redshift ($z_{\rm abs}$);
(2) the Doppler parameter of the absorbing gas ($b$ in km~s$^{-1}$); and
(3) the column density of the ion that gives rise to the absorption line.
When it was evident that the DLA metal 
absorption arises from more than one 
cloud component, we introduced additional 
components to reduce the $\chi^{2}$ 
(such that the $\chi^{2}$ divided by 
the number of degrees of freedom [dof] 
was close to $1.0$), whilst 
maintaining realistic errors on the derived 
parameters (i.e. $\simlt10\%$ uncertainty on $b$ 
and a redshift uncertainty less than 
the sampling size of $\sim2.5$ km s$^{-1}$). 
Throughout the fitting procedure we assumed 
that the dominant ions in \HI\ regions 
(e.g. \CII, \NI, \OI, \SiII, \FeII)
are kinematically associated with the same gas. 
We therefore fixed the redshift and 
Doppler parameter of each absorption 
component to be the same for each of these ions. 
The resulting cloud model parameters, including 
the reduced $\chi^{2}$, are provided in 
Table~\ref{tab:cloud_models} where, as
a guide, the last column lists the 
fraction of the total column density of \SiII\
in each component. 
The total column densities of 
available ions in each DLA are collected 
in Table~A1.
Table~\ref{tab:atomic} lists laboratory 
wavelengths and oscillator strengths of relevant
atomic transitions 
from the compilation by \citet{Mor03},
with subsequent updates by \citet{JenTri06}.


\section{Individual Objects}
\label{sec:indobj}

In this section, we briefly comment on the 
properties of each new DLA analysed 
in this paper.

\subsection{J0311$-$1722: DLA at $z_{\rm abs}=3.73400$}\label{sec:J0311m1722}

The VMP DLA along the line-of-sight to J0311$-$1722 
(J2000.0: $03\rahr11\ramin15.\rasec20$, $-17\decdeg22\decmin47.\decsec4$) 
was first identified by \citet{Per01}. 
Follow-up UVES spectroscopy  
by \citet{Per05} revealed a \Lya\ absorber 
at $z_{\rm abs}=3.734$ with 
$\log$\, [\NHI/cm$^{-2}]=19.48\pm0.10$,
which is lower than the conventional 
limit for DLAs set by \citet{Wol86}, 
$\log$\,[\NHI/cm$^{-2}] \geq 20.3$. 
Such systems are often referred to as sub-DLAs 
(\citealt{Per03a}) or super Lyman-limit systems, 
and are defined to have 
$19.0\leq \log$\,[\NHI/cm$^{-2}]\leq20.3$. 

\begin{figure*}
  \centering
  \includegraphics[angle=0]{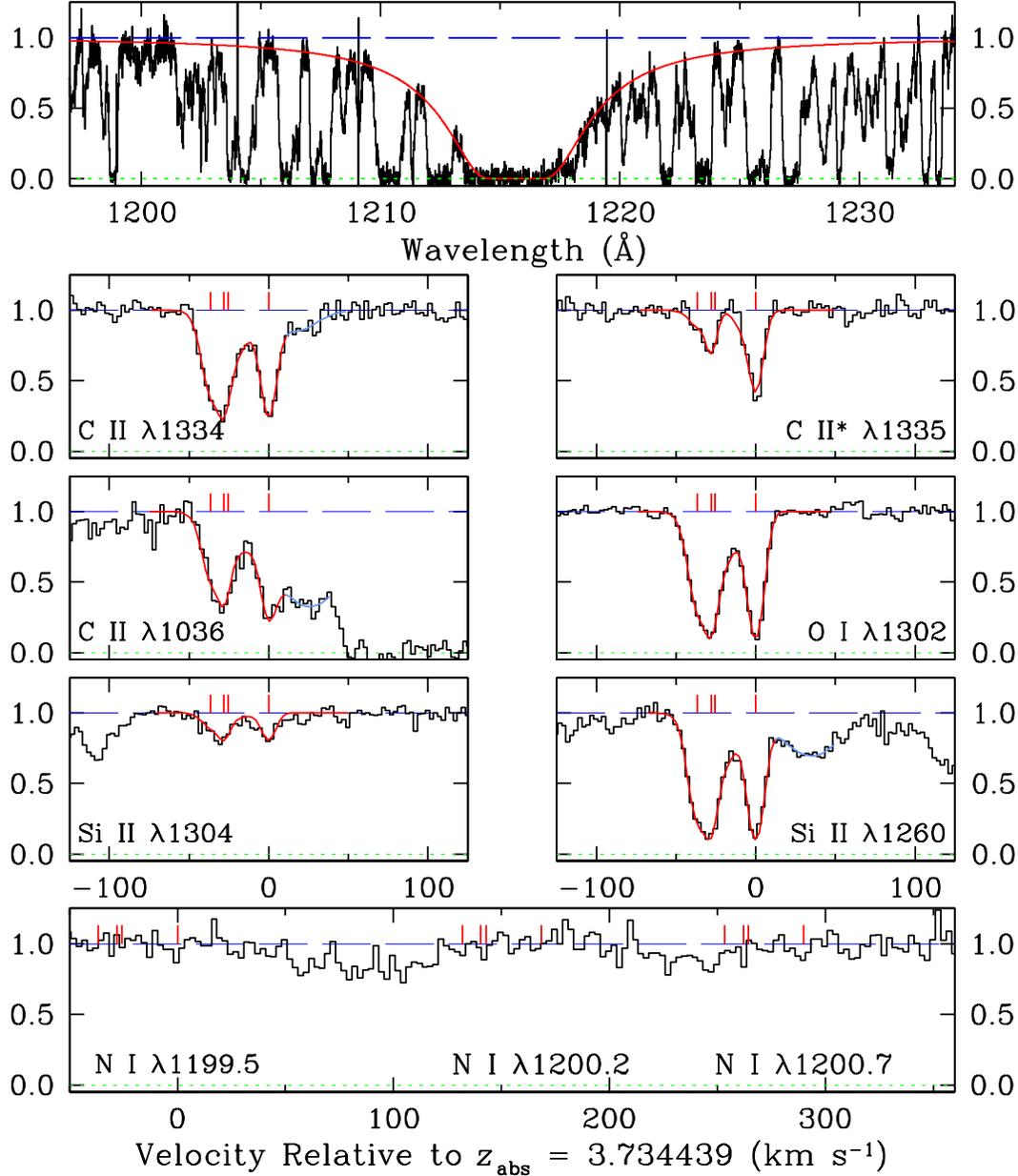}
  \caption{ 
The top panel shows the DLA towards J0311$-$1722 (black histogram) which exhibits a 
damped \Lya\ line at $z_{\rm abs}=3.73400$.
The red continuous line shows the theoretical Voigt profile for a neutral
hydrogen column density $\log[N({\rm H}\,\textsc{i})/{\rm cm}^{-2}]=20.30$. 
The remaining panels display a selection of metal lines, overlaid with the Voigt
profiles for the derived cloud model (in red) and blends (in light blue).
The red tick marks above the normalized continuum indicate the locations of the absorption components.
For all panels, the $y$-axis scale is residual intensity.
The normalized quasar continuum and zero-level are shown by the blue dashed and green dotted lines respectively.
  }
  \label{fig:J0311m1722}
\end{figure*}

\citet{Per05} derived 
their estimate of \NHI\ for this absorber 
from a consistent model fit to the first $6$ 
Lyman lines (from \Lya\ to Ly6), where the 
\Lya\ line produces the poorest fit. 
Having re-analysed these data, giving higher 
priority to the \Lya\ and \Lyb\ lines, 
we find that a better fit 
results if $\log$\,[\NHI/cm$^{-2}]=20.30\pm0.06$ 
(see top panel of Figure~\ref{fig:J0311m1722}), 
provided that the 
Doppler parameter is $b\simlt30$ km s$^{-1}$. 
Possibly,  the value of \NHI\ 
derived by \citet{Per05} was 
driven by the higher order Lyman lines, 
which may be blended with other Lyman forest
lines,
and resulted in a Doppler parameter 
much larger than 30\,km s$^{-1}$. 

\begin{table}
\centering
\caption{\textsc{Ion column densities of the DLA in J0311$-$1722 at $z_{\rm abs}=3.73400$}}
    \begin{tabular}{@{}lp{1.8in}c}
    \hline
  \multicolumn{1}{l}{Ion}
& \multicolumn{1}{c}{Transitions used}
& \multicolumn{1}{c}{log $N$(X)/${\rm cm}^{-2}$}\\
    \hline
\HI    &  1025, 1215  &  20.30 $\pm$ 0.06 \\
\CII   &  1036, 1334  &  14.02 $\pm$ 0.08 \\
\CII*  &  1335        &  13.55 $\pm$ 0.06 \\
\NI    &  1200.2      &  $\le13.07^{\rm a}$ \\
\OI    &  1302        &  14.70 $\pm$ 0.08 \\
\SiII  &  1260, 1304  &  13.31 $\pm$ 0.07 \\
\FeII  &  1125        &  $\le13.76^{\rm a}$ \\
    \hline
    \end{tabular}
\begin{flushleft}
{$^{\rm a}\,3\sigma$ limiting rest frame equivalent width.}\\
\end{flushleft}
\label{tab:J0311m1722_cd}
\end{table}

In the lower panels of Figure~\ref{fig:J0311m1722}
we present a selection of the metal lines 
associated with this VMP DLA. 
All of the available metal absorption 
lines  are unsaturated, 
leading to reliable estimates of the 
metal column densities. A four component 
model was found to accurately 
reproduce the metal-line profiles, 
whilst maintaining reasonable estimates 
of the parameter errors (see Section~\ref{sec:pfaa}). 
The derived cloud 
model parameters are listed in 
Table~\ref{tab:cloud_models}.
Although both \CII\ lines ($\lambda 1334$ and $\lambda 1036$)
are blended on 
the red wing of component 4 
(centred near $v = 0$\,km~s$^{-1}$ in Figure~\ref{fig:J0311m1722}),
this does not greatly affect our final estimate 
of $N$(\CII), since component 4 has a 
well-determined Doppler parameter 
from the relative strengths of 
\SiII\,$\lambda1260$ and \SiII\,$\lambda1304$, 
and from other unblended transitions in 
regions of high signal-to-noise. In any case, 
the majority of the absorbing column is 
contributed by the first three 
components ($\sim60\%$). We list the total 
column density returned by \textsc{vpfit} 
for each available 
ion in Table~\ref{tab:J0311m1722_cd}.

In this table, we also provide upper 
limits for the column densities of 
several key ions that are undetected 
at the signal-to-noise of the data. 
Specifically, we calculate the 
$3\sigma$ limiting rest-frame equivalent 
width, $W_{0}$, over the velocity interval of absorption 
exhibited by the weakest transition, which 
in this case is  
\SiII\,$\lambda1304$. A $3\sigma$ 
upper limit to the undetected feature 
is then derived using the optically 
thin limit approximation,
$N = 1.13\times10^{20}\cdot W_{0}/\lambda^{2}f\,\,{\rm cm^{-2}}$.
For \NI\ we use the undetected $\lambda1200.2$ 
line to derive $W_{0}$(\NI) $\leq13\,{\rm m\AA}$, 
which implies log $N$(\NI)/${\rm cm}^{-2} \leq 13.07$.
Similarly for \FeII\,$\lambda1125$,
$W_{0}$(\FeII) $\leq10\,{\rm m\AA}$ 
implies log $N$(\FeII)/${\rm cm}^{-2} \le 13.76$.

\subsection{J0831$+$3358: DLA at $z_{\rm abs}=2.30364$}\label{sec:J0831p3358}

\begin{figure*}
  \centering
  \includegraphics[angle=0,width=0.75\textwidth]{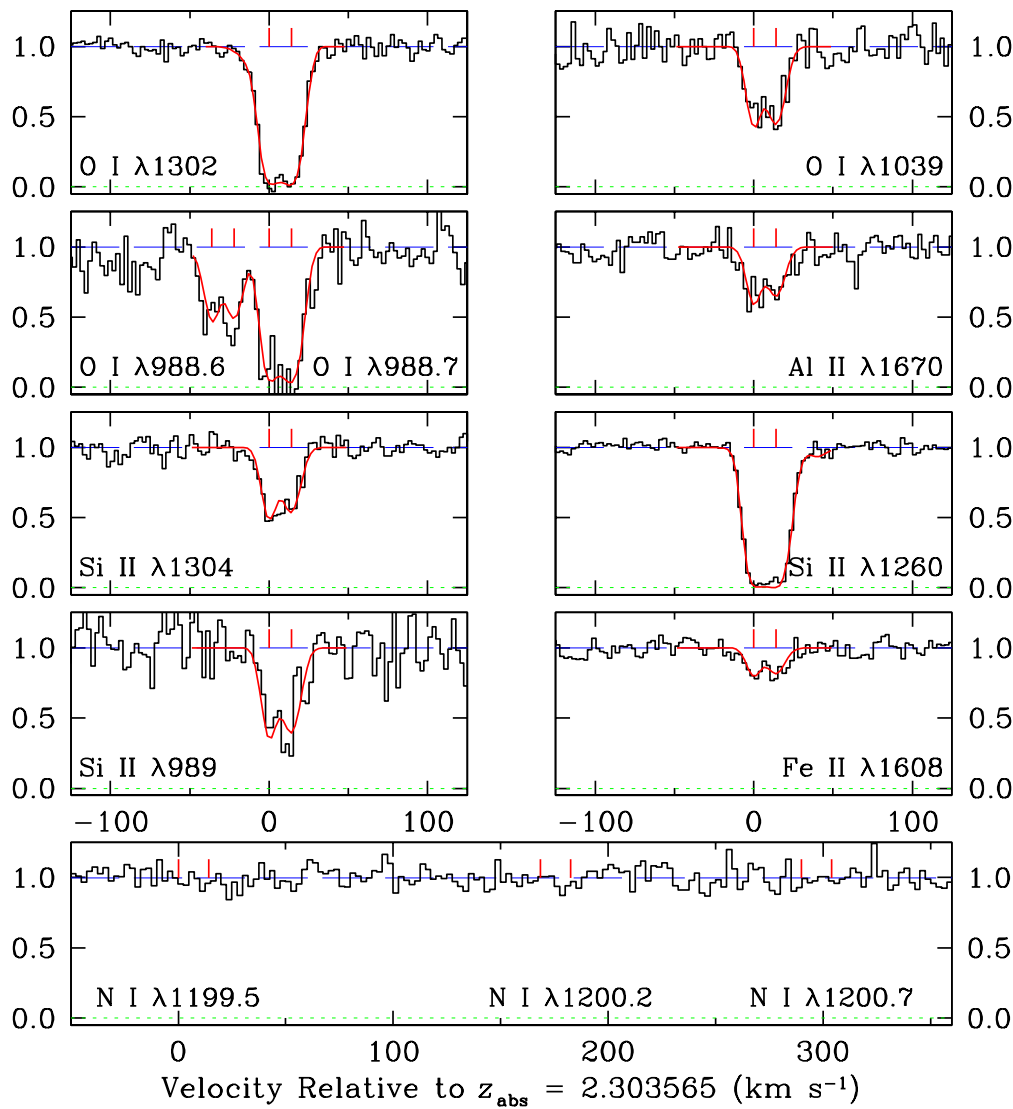}
  \caption{
Same as Figure~\ref{fig:J0311m1722} for a selection of metal lines associated with the $z_{\rm abs}=2.30364$ DLA towards J0831$+$3358.
  }
  \label{fig:J0831p3358}
\end{figure*}

\begin{table}
\centering
    \caption{\textsc{Ion column densities of the DLA in J0831$+$3358 at $z_{\rm abs}=2.30364$}}
    \begin{tabular}{@{}lp{1.8in}c}
    \hline
  \multicolumn{1}{l}{Ion}
& \multicolumn{1}{c}{Transitions used}
& \multicolumn{1}{c}{log $N$(X)/${\rm cm}^{-2}$}\\
    \hline
\HI    &  1215                   &  20.25$\pm$ 0.15$^{\rm a}$ \\
\NI    &  1199.5                 &  $\le12.78^{\rm b}$ \\
\OI    &  988.5, 988.6, 988.7, 1039, 1302        &  14.93 $\pm$ 0.05   \\
\AlII  &  1670                   &  12.19 $\pm$ 0.06   \\
\SiII  &  989, 1193, 1260, 1304  &  13.75 $\pm$ 0.04   \\
\SII   &  1250                   &  $\le13.75^{\rm b}$ \\
\FeII  &  1608                   &  13.33 $\pm$ 0.06   \\
    \hline
    \end{tabular}
\begin{flushleft}
$^{\rm a}${Penprase et al. (2010)}\\
$^{\rm b}${$3\sigma$ limiting rest frame equivalent width.}\\
\end{flushleft}
\label{tab:J0831p3358_cd}
\end{table}

We observed J0831$+$3358 with HIRES on 
2009 December 9 under good conditions 
with subarcsecond seeing. We used a 
$1.148''$ wide slit which, as measured 
by \citet{Coo11}, delivered a spectral 
resolution of 
$7.3$ km s$^{-1}$ FWHM. 
Unfortunately, the \Lya\ line at 
$\lambda_{\rm obs}\simeq4015$\AA\ falls 
in a gap between two of the CCDs on the 
HIRES detector mosaic. Thus, for this DLA, we 
adopt the \HI\ column density 
$\log[N({\rm H}\,\textsc{i})/{\rm cm}^{-2}]=20.25\pm0.15$ 
derived by \citet{Pen10} from their observations 
of this QSO at $R\simeq5000$, which is sufficient 
to resolve the broad damped profile of the \Lya\ line. 

\begin{figure*}
  \centering
  \includegraphics[angle=0]{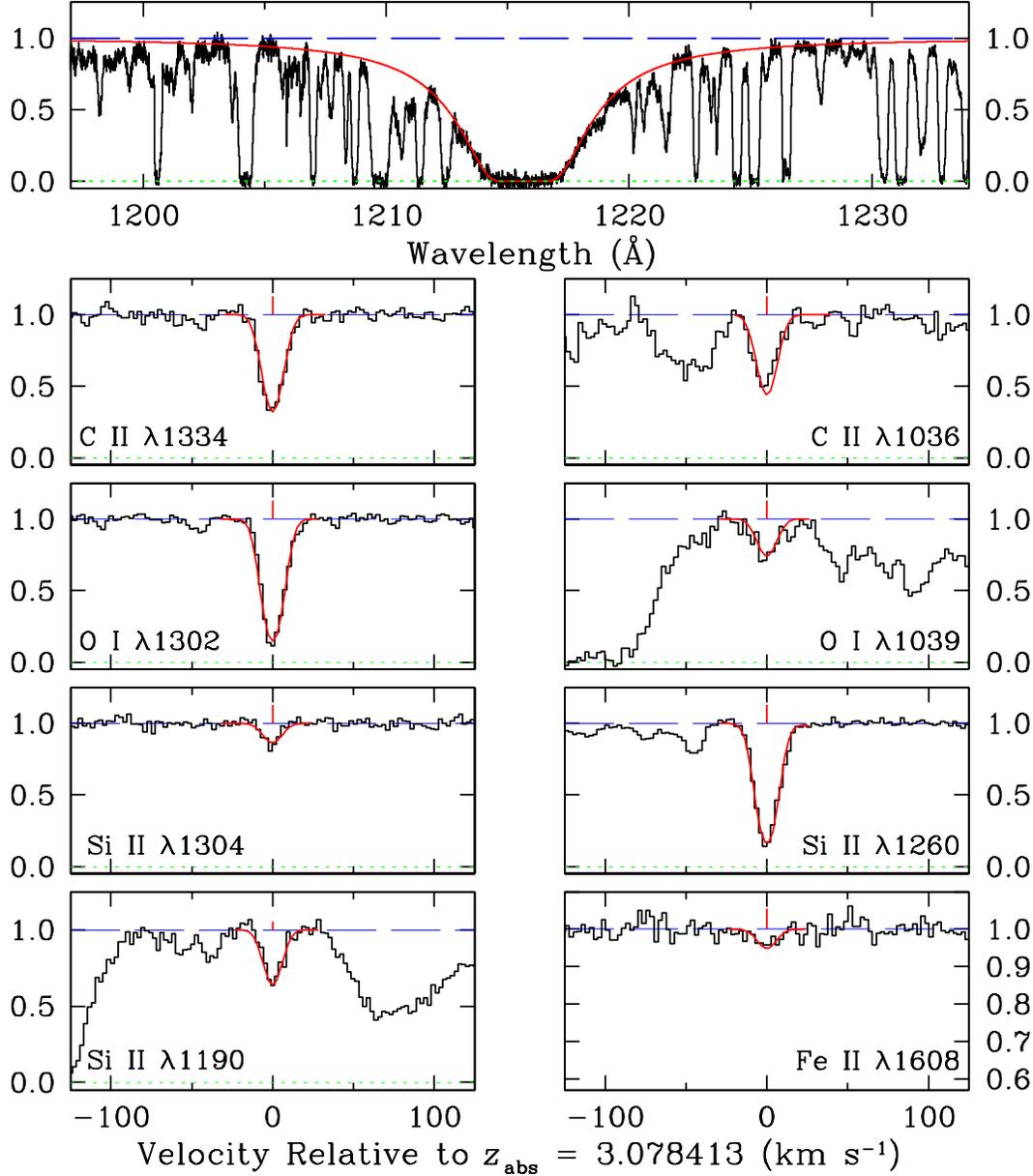}
  \caption{ 
Same as Figure~\ref{fig:J0311m1722}, for the DLA towards J1001$+$0343 (black histogram) which exhibits a 
damped \Lya\ line at $z_{\rm abs}=3.07841$ (top panel). 
Here, the red continuous line shows the theoretical Voigt profile for an \HI\ 
column density $\log[N({\rm H}\,\textsc{i})/{\rm cm}^{-2}]=20.21$. 
The remaining panels display a selection of metal lines. 
Note the different y-axis scale that is used for the 
weak \FeII\,$\lambda1608$ line (bottom-right panel).
  }
  \label{fig:J1001p0343}
\end{figure*}

The metal-lines in our data are well 
fit by a model with two components 
of roughly equal strength separated 
by $14$ km s$^{-1}$ 
(see Figure~\ref{fig:J0831p3358}). 
Details of the 
derived cloud model are presented in 
Table~\ref{tab:cloud_models}, with 
the associated column densities 
given in Table~\ref{tab:J0831p3358_cd}. 
The \CII\ lines are saturated in this 
DLA, however, we have a clean measurement 
of $N$(\OI) from a number of unsaturated \OI\ 
lines. In Table~\ref{tab:J0831p3358_cd} 
we also provide $3\sigma$ upper limits 
to the \NI\ and \SII\ column densities 
which are undetected at the signal-to-noise 
of our data.

\subsection{J1001$+$0343: DLA at $z_{\rm abs}=3.07841$}\label{sec:J1001p0343}

\begin{table}
\centering
    \caption{\textsc{Ion column densities of the DLA in J1001+0343 at $z_{\rm abs}=3.07841$}}
    \begin{tabular}{@{}lp{1.8in}c}
    \hline
  \multicolumn{1}{l}{Ion}
& \multicolumn{1}{c}{Transitions used}
& \multicolumn{1}{c}{log $N$(X)/${\rm cm}^{-2}$}\\
    \hline
\HI    & 1215                          & 20.21 $\pm$ 0.05 \\
\CII   & 1036, 1334                    & 13.58 $\pm$ 0.02 \\
\NI    & 1200.2                        & $\le 12.50^{\rm a}$ \\
\OI    & 1039, 1302                    & 14.25 $\pm$ 0.02 \\
\SiII  & 1190, 1193, 1260, 1304, 1526  & 12.86 $\pm$ 0.01 \\
\SII   & 1253                          & $\le 12.91^{\rm a}$ \\
\FeII  & 1608                          & 12.50 $\pm$ 0.14 \\
    \hline
    \end{tabular}
\begin{flushleft}
$^{\rm a}${$3\sigma$ limiting rest frame equivalent width.}\\
\end{flushleft}
  \label{tab:J1001p0343_cd}
\end{table}

This QSO was observed with UVES 
in service mode on the nights of 
2009 April 19 \&\ 29, 
2010 January 11 \&\ 27 and
2010 February 7. 
Our total integration time was 
$33700$\,s, yielding a S/N per
pixel of $\sim40$ at $5000$\,\AA.
The DLA inline to J1001$+$0343 was 
also investigated by \citet{Pen10}, 
being amongst the most metal-poor 
in their sample. From our observations 
we derive an \HI\ column density of 
$\log[N({\rm H}\,\textsc{i})/{\rm cm}^{-2}]=20.21\pm0.05$
from the wings of the \Lya\ absorption 
line. This compares well with the 
estimate by \citet{Pen10}, 
$\log[N({\rm H}\,\textsc{i})/{\rm cm}^{-2}]=20.15\pm0.10$.
We present our Voigt profile fit to the 
\Lya\ line in the top panel of Figure~\ref{fig:J1001p0343}.

\begin{figure*}
  \centering
  \includegraphics[angle=0]{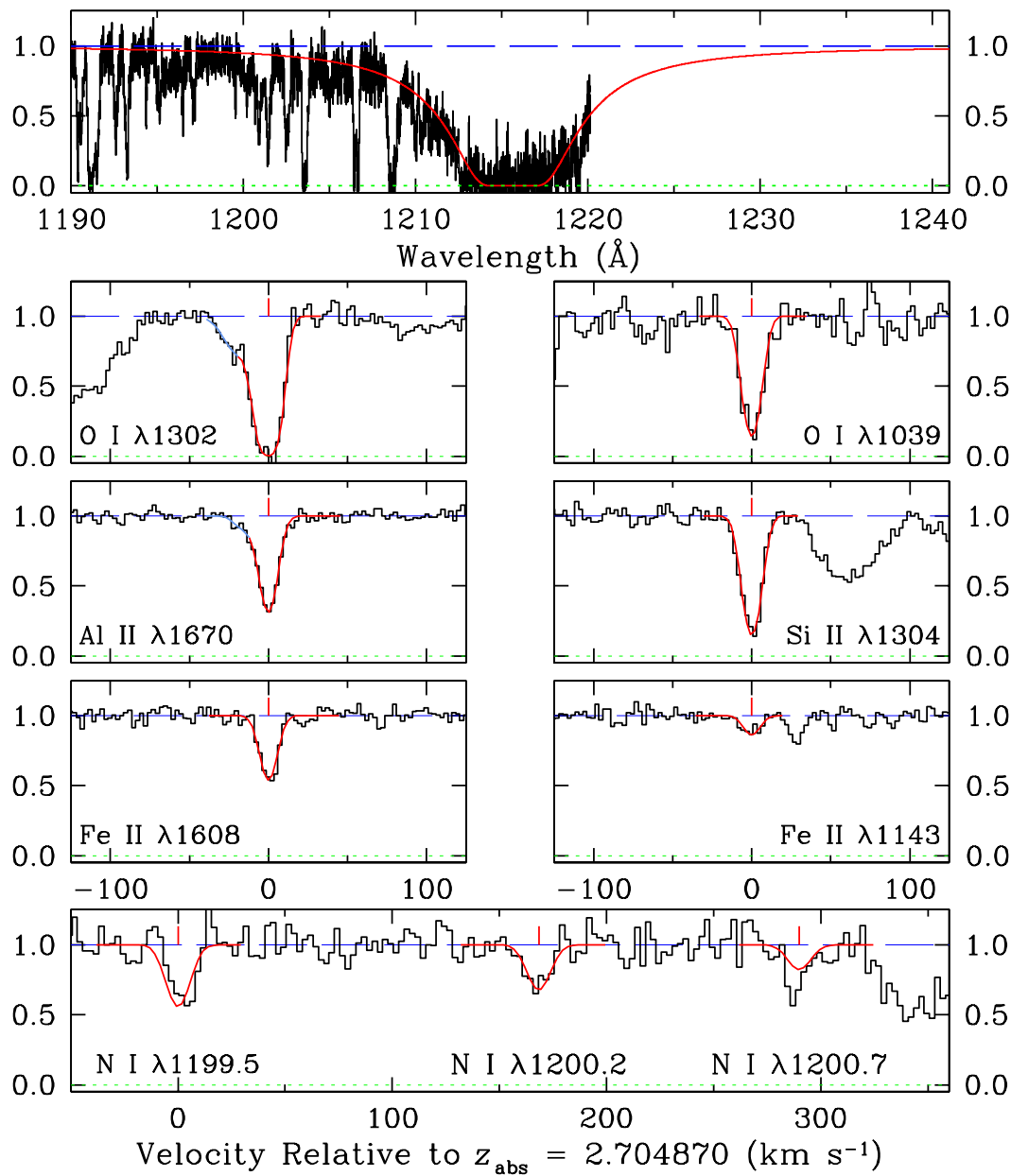}
  \caption{ 
Same as Figure~\ref{fig:J0311m1722}, for the DLA towards J1037$+$0139 (black histogram) which exhibits a 
damped \Lya\ line at $z_{\rm abs}=2.70487$ (top panel).
The red continuous line shows the theoretical Voigt profile for an \HI\ 
column density $\log[N({\rm H}\,\textsc{i})/{\rm cm}^{-2}]=20.50$. 
The remaining panels display a selection of metal lines.
}
  \label{fig:J1037p0139}
\end{figure*}

\begin{table}
\centering
    \caption{\textsc{Ion column densities of the DLA in J1037+0139 at $z_{\rm abs}=2.70487$}}
    \begin{tabular}{@{}lp{1.8in}c}
    \hline
  \multicolumn{1}{l}{Ion}
& \multicolumn{1}{c}{Transitions used}
& \multicolumn{1}{c}{log $N$(X)/${\rm cm}^{-2}$}\\
    \hline
\HI    & 1025, 1215              & 20.50 $\pm$ 0.08 \\
\NI    & 1199.5, 1200.2, 1200.7  & 13.27 $\pm$ 0.04 \\
\OI    & 1039, 1302              & 15.06 $\pm$ 0.04 \\
\AlII  & 1670                    & 12.32 $\pm$ 0.03 \\
\SiII  & 1260, 1304              & 13.97 $\pm$ 0.03 \\
\FeII  & 1143, 1144, 1608        & 13.53 $\pm$ 0.02 \\
    \hline
    \end{tabular}
    \smallskip
  \label{tab:J1037p0139_cd}
\end{table}

The remaining panels of Figure~\ref{fig:J1001p0343} 
showcase a number of the available absorption lines 
that were used in deriving the cloud model. In fact, 
all of the available metal absorption lines are
unsaturated in this DLA, thus providing reliable 
measurements of the elemental abundances. 
Several metal-line transitions of varying strength 
are well fit by a single component 
cloud model with a Doppler parameter of 
$b=7.0\pm0.1$ km s$^{-1}$ 
(see Table~\ref{tab:cloud_models}). The curve of growth 
analysis by \citet{Pen10} yielded a Doppler parameter 
of $7.5$ km s$^{-1}$, which is in good agreement 
with that found here. The derived column densities for all 
available ions are presented in Table~\ref{tab:J1001p0343_cd}.
We also provide $3\sigma$ upper limits 
to the \NI\ and \SII\ column densities 
which are undetected at the S/N 
of our data. \FeII\,$\lambda1608$ is detected 
at the $3.6\sigma$ level. The corresponding 
fit is presented in the bottom-right panel of 
Figure~\ref{fig:J1001p0343} (note the different
$y$-axis scale).

\begin{figure*}
  \centering
  \includegraphics[angle=0]{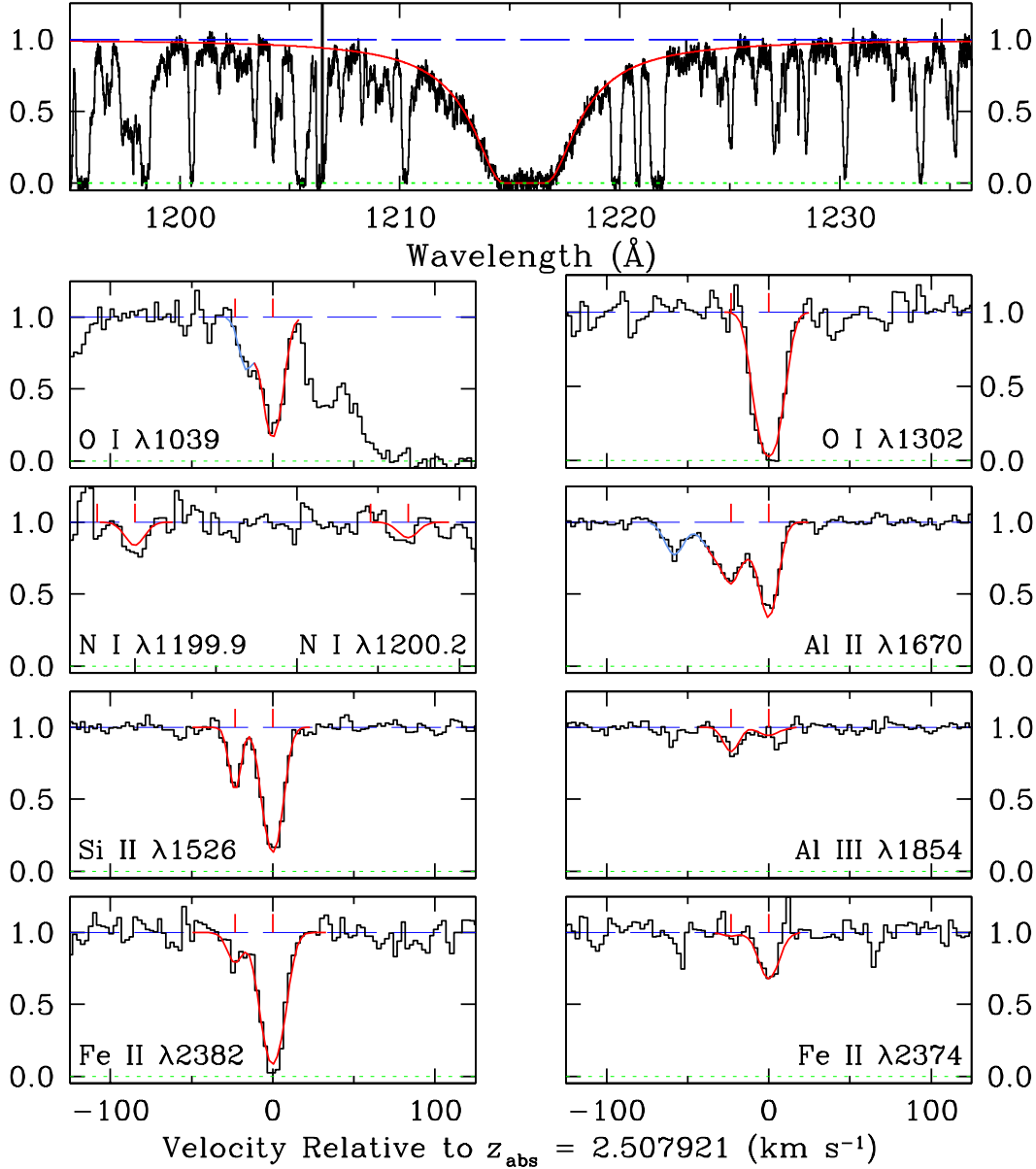}
  \caption{ 
Same as Figure~\ref{fig:J0311m1722}, for the DLA towards J1340$+$1106 (black histogram) 
which exhibits a 
damped \Lya\ line at $z_{\rm abs}=2.50792$ (top panel).
The red continuous line shows the theoretical Voigt profile for an \HI\ 
column density $\log[N({\rm H}\,\textsc{i})/{\rm cm}^{-2}]=20.09$. 
The remaining panels display a selection of metal lines.
  }
  \label{fig:J1340p1106a}
\end{figure*}

\subsection{J1037$+$0139: DLA at $z_{\rm abs}=2.70487$}\label{sec:J1037p0139}

This QSO was also observed in service mode with UVES 
on 2010 February 12--14 and again on 2010 March 5. 
J1037$+$0139 was also one of the QSOs observed independently 
by \citet{Pen10}. It is one of the faintest QSOs in 
our sample, requiring a total of $26075$\,s of integration 
to achieve a S/N$\sim 40$ at $5000$\,\AA. The \Lya\ line 
falls on the edge of the blue detector, but this does not 
affect the accuracy of the \HI\ column density since 
the blue wing of the damped \Lya\ line is still intact, 
and can be fit using the redshift derived from the 
well-defined narrow metal absorption lines. Moreover, we have 
access to \Lyb, which also exhibits damping wings 
(although not as strong as that of \Lya). We derive an
\HI\ column density of 
$\log[N({\rm H}\,\textsc{i})/{\rm cm}^{-2}]=20.50\pm0.08$,
which is consistent with estimates derived 
from the SDSS spectrum 
($\log[N({\rm H}\,\textsc{i})/{\rm cm}^{-2}]=20.45$; \citealt{ProWol09}), 
as well as that derived by \citet{Pen10} 
($\log[N({\rm H}\,\textsc{i})/{\rm cm}^{-2}]=20.40\pm0.25$). 
We present our Voigt profile fit to the \Lya\ 
line in the top panel of Figure~\ref{fig:J1037p0139}.

\begin{table}
\centering
    \caption{\textsc{Ion column densities of the DLA in J1340+1106 at $z_{\rm abs}=2.50792$}}
    \begin{tabular}{@{}lp{1.8in}c}
    \hline
  \multicolumn{1}{l}{Ion}
& \multicolumn{1}{c}{Transitions used}
& \multicolumn{1}{c}{log $N$(X)/${\rm cm}^{-2}$}\\
    \hline
\HI    & 1215                          &  20.09 $\pm$ 0.05  \\
\NI    & 1199.5, 1200.2                &  12.80 $\pm$ 0.04  \\
\OI    & 1039, 1302                    &  15.02 $\pm$ 0.03  \\
\AlII  & 1670                          &  12.27 $\pm$ 0.02  \\
\AlIII & 1854, 1862                    &  11.51 $\pm$ 0.10  \\
\SiII  & 1190, 1193, 1304, 1526        &  13.75 $\pm$ 0.02  \\
\FeII  & 1096, 1143, 2344, 2374, 2382  &  13.49 $\pm$ 0.02  \\
    \hline
    \end{tabular}
    \smallskip
  \label{tab:J1340p1106a_cd}
\end{table}

Again, the metal absorption is concentrated in a 
single component, in this case with a Doppler parameter of 
5.9 km s$^{-1}$. 
The corresponding column densities for all 
of the available ions are listed in Table~\ref{tab:J1037p0139_cd}.
For this system, both \CII\,$\lambda1036$ 
and $\lambda1334$ are saturated and blended;
however, we have a robust measure of the \OI\ and 
\FeII\ column densities from several unsaturated 
transitions. We also have a 
clear detection of the \NI\ triplet near 
$\lambda_{0}=1200$\,\AA. Selected metal 
absorption lines are reproduced in 
Figure~\ref{fig:J1037p0139}.

\subsection{J1340$+$1106: DLA at $z_{\rm abs}=2.50792$}\label{sec:J1340p1106a}

This QSO has previously been observed with both 
UVES (at a spectral resolution of $10.3$ km s$^{-1}$ FWHM; 
\citealt{LedPetSri03}) and 
HIRES (at a spectral resolution of $8.1$ km s$^{-1}$ FWHM; 
\citealt{Pro03}). However, given that this QSO 
intersects two VMP DLAs, one of which had potentially 
unsaturated \CII\ lines, we decided to 
reobserve it with UVES for $29800$\,s at a slightly 
higher spectral resolution of $7.3$ km s$^{-1}$, and
obtained complete spectral coverage from 
3500\,\AA\ to almost 1$\mu$m.

Since the broad damped profile of the \Lya\ line 
is independent  of the spectral resolution, 
we combined the three datasets to obtain a high 
S/N ratio near the damped \Lya\ line, from which we 
derived $\log[N({\rm H}\,\textsc{i})/{\rm cm}^{-2}]=20.09 \pm 0.05$. 
This model fit, along with the combined data, 
is shown in the top panel of Figure~\ref{fig:J1340p1106a}.
The profiles of the metal absorption lines, on the other hand, 
are narrow (FWHM $\simlt10$ km s$^{-1}$); therefore,
their observed profiles are not independent of the spectral resolution. 
Thus, we separately combined the data of equal spectral resolution 
and individually read these three reduced spectra into \textsc{vpfit}, 
which convolved the fitted model with the 
spectral resolution appropriate to the data. 
The upshot of proceeding in this way is that 
the cloud model is then largely driven by 
the dataset of highest S/N for each absorption 
line that is input. However, since we cannot 
combine all three datasets, in 
Figure~\ref{fig:J1340p1106a} we only 
present the dataset with the highest S/N 
near each absorption line.

\begin{figure*}
  \centering
  \includegraphics[angle=0]{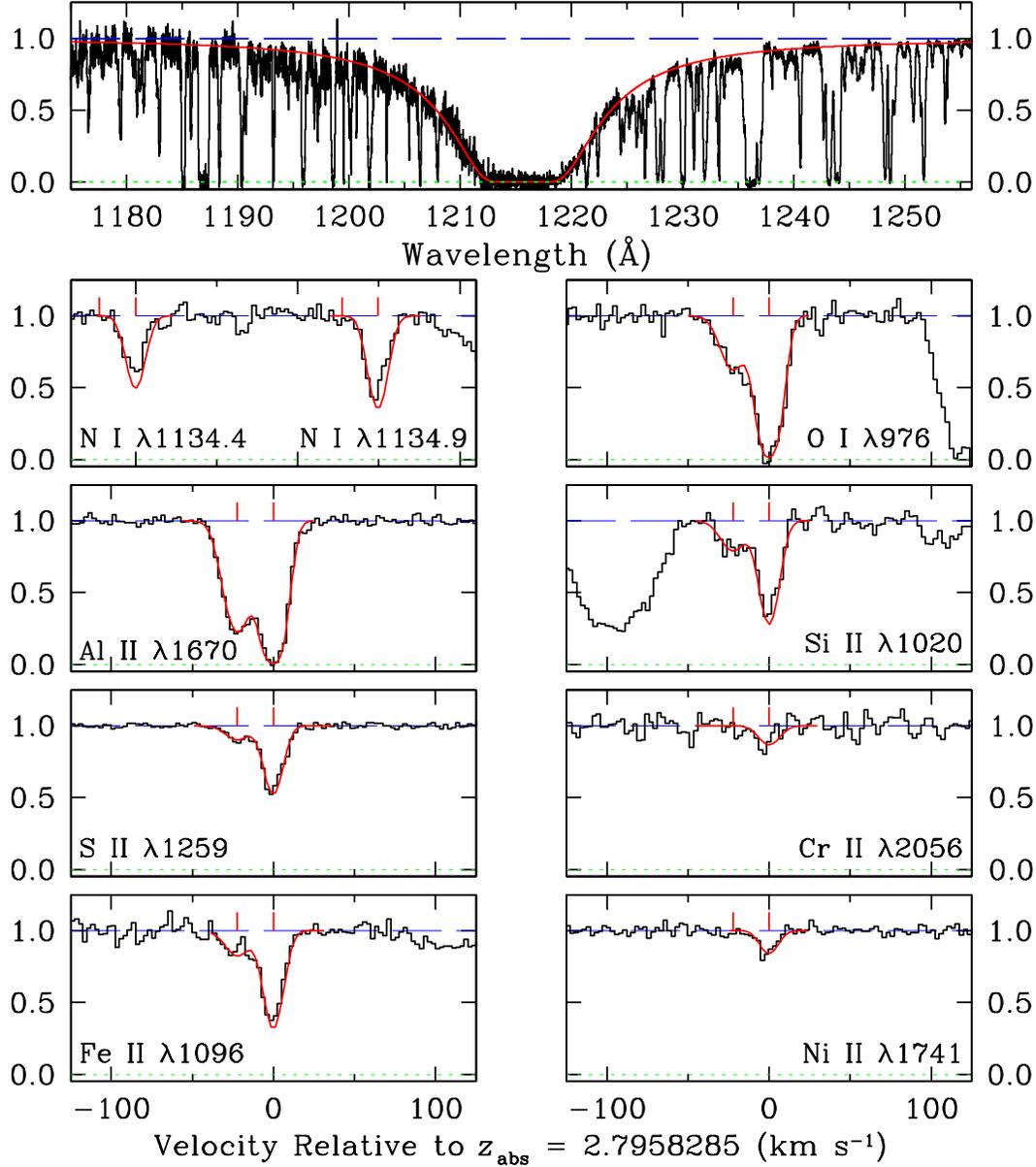}
  \caption{ 
Same as Figure~\ref{fig:J0311m1722}, for the DLA towards J1340$+$1106 (black histogram) which exhibits a 
damped \Lya\ line at $z_{\rm abs}=2.79583$ (top panel).
The red continuous line shows the theoretical Voigt profile for an \HI\ 
column density $\log[N({\rm H}\,\textsc{i})/{\rm cm}^{-2}]=21.00$. 
The remaining panels display a selection of metal lines.
  }
  \label{fig:J1340p1106b}
\end{figure*}

\begin{figure*}
  \centering
  \includegraphics[angle=0]{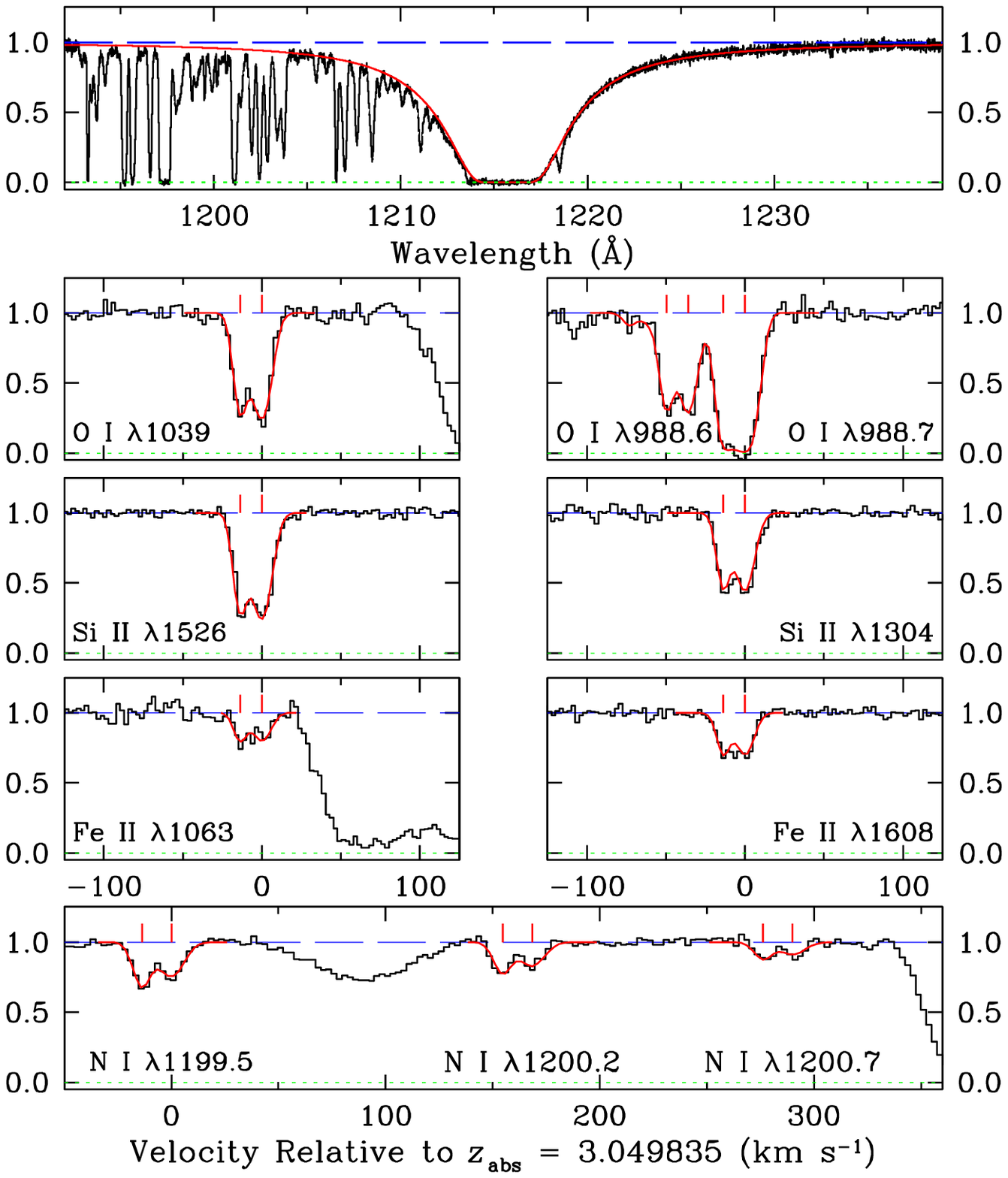}
  \caption{ 
Same as Figure~\ref{fig:J0311m1722}, for the DLA towards J1419$+$0829 (black histogram) which exhibits a 
damped \Lya\ line at $z_{\rm abs}=3.04973$ (top panel).
The red continuous line shows the theoretical Voigt profile for an \HI\ 
column density $\log[N({\rm H}\,\textsc{i})/{\rm cm}^{-2}]=20.40$. 
The remaining panels display a selection of metal lines.
  }
  \label{fig:J1419p0829}
\end{figure*}

Unfortunately, both of the
\CII\ absorption lines are 
saturated, and partially arise from nearby 
ionized gas, exhibiting a similar profile 
shape to the \AlII\,$\lambda1670$ line. 
In fact, most of the metal absorption lines 
exhibit a second component on the blue wing 
(at $v = -23$\,km s$^{-1}$ relative to the main component 
at $z_{\rm abs} = 2.507921$), which appears 
to arise from ionized gas. This is confirmed 
by the absence of this blue component in \OI\ 
absorption (see \OI\,$\lambda1302$, and 
note that the absorption on the blue wing 
of the \OI\,$\lambda1039$ line is due to 
unrelated absorption), which has long 
been known to accurately trace neutral gas 
\citep{FieSte71}. The presence 
of ionized gas is also confirmed by the 
higher \AlIII/\AlII\ ratio exhibited by 
the blue component.

We derived the cloud model for both of 
these components from the host of 
available \SiII\ and \FeII\ lines, with 
additional constraints coming from the two 
\OI\ lines. In Table~\ref{tab:cloud_models} we 
present the fitting results from both the 
blue component (component 1; which arises from 
nearby mildly ionized gas), and the main component 
(component 2) which we attribute to the DLA. 
Fixing the parameters of this cloud 
model, we then derived the column density for 
all ions in both components. 
In Table~\ref{tab:J1340p1106a_cd},
however, we only provide the column density 
for the single, dominant component that we 
attribute to the DLA. The model fits to the 
data are presented in the lower panels of 
Figure~\ref{fig:J1340p1106a} where, 
as stated above, we only present 
the model and data that correspond to 
the highest S/N for each absorption line. 

The detection of ions that arise in ionized gas, 
such as \AlIII, will later provide a useful 
means to test the accuracy of one of our 
underlying assumptions; that we can use 
the single dominant ion for each element to 
measure the elemental abundances in DLAs
(see Section~\ref{sec:ics}).

\subsection{J1340$+$1106: DLA at $z_{\rm abs}=2.79583$}\label{sec:J1340p1106b}

We now report on the second VMP DLA that is 
intersected by this QSO. Again, we treat the 
\Lya\ line and the metal lines of this DLA as 
detailed in Section~\ref{sec:J1340p1106a}. 
From the combined data we derive an \HI\ 
column density of 
$\log[N({\rm H}\,\textsc{i})/{\rm cm}^{-2}]=21.00\pm0.06$.
The combined spectrum in the region of the 
damped \Lya\ line, together with the 
profile fit, is reproduced in the 
top panel of Figure~\ref{fig:J1340p1106b}.

Turning now to the metal absorption lines, 
we have found that a two component cloud 
model (with Doppler parameters of 
$9.2\pm0.1$ and $6.55\pm0.05$ km s$^{-1}$ 
separated by $22$ km s$^{-1}$) provides a 
good fit to the data. 
This cloud model is perhaps the best determined 
in our dataset, given the high S/N of the data,
and the numerous atomic transitions available. 
Our Voigt profile fits to the metal lines are 
shown in the lower panels of Figure~\ref{fig:J1340p1106b}.
However, as discussed in Section~\ref{sec:J1340p1106a},
we only present the model and data that correspond to 
the highest S/N for each absorption line.
The column densities for all available ions 
are provided in Table~\ref{tab:J1340p1106b_cd}.
In this table, we have also provided the column 
densities for \AlIII\ and \NII, which are 
coincident with the DLA, but are typically 
associated with \HII\ regions.

\begin{table}
\centering
    \caption{\textsc{Ion column densities of the DLA in J1340+1106 at $z_{\rm abs}=2.79583$}}
    \begin{tabular}{@{}lp{1.8in}cc}
    \hline
  \multicolumn{1}{l}{Ion}
& \multicolumn{1}{c}{Transitions used}
& \multicolumn{1}{c}{log $N$(X)/${\rm cm}^{-2}$}\\
    \hline
\HI    & 1215                                                     &  21.00 $\pm$ 0.06  \\
\CII*  & 1335                                                     &  12.99 $\pm$ 0.04  \\
\NI    & 1134.4, 1134.9, 1199.5, 1200.7                           &  14.04 $\pm$ 0.02  \\
\NII   & 1083                                                     &  12.81 $\pm$ 0.09  \\
\OI    & 925, 976, 1302                                           &  16.04 $\pm$ 0.04  \\
\AlII  & 1670                                                     &  13.24 $\pm$ 0.03  \\
\AlIII & 1854, 1862                                               &  12.19 $\pm$ 0.08  \\
\SiII  & 1020, 1193, 1260, 1304, 1526, 1808                       &  14.68 $\pm$ 0.02  \\
\SII   & 1253, 1259                                               &  14.30 $\pm$ 0.02  \\
\ArI   & 1048, 1066                                               &  13.18 $\pm$ 0.02  \\
\CrII  & 2056, 2062                                               &  12.62 $\pm$ 0.11  \\
\FeII  & 1063.1, 1081, 1096, 1125, 1608, 1611, 2260, 2344, 2382   &  14.32 $\pm$ 0.01  \\
\NiII  & 1317, 1370, 1454, 1709, 1741, 1751                       &  13.08 $\pm$ 0.03  \\
    \hline
    \end{tabular}
    \smallskip
  \label{tab:J1340p1106b_cd}
\end{table}

\subsection{J1419$+$0829: DLA at $z_{\rm abs}=3.04973$}\label{sec:J1419p0829}

We recorded the spectrum of J1419$+$0829 ($z_{\rm em}=3.034$) 
for $29800$\,s with UVES in service mode, resulting in a S/N 
near 5000\AA\ (near the red wing of the damped \Lya\ line) 
of $\simeq40$. This high S/N in combination with a virtually 
uninterrupted red wing to the \Lya\ absorption allows 
a very accurate measurement of the \HI\ column density, 
$\log[N({\rm H}\,\textsc{i})/{\rm cm}^{-2}]=20.40\pm0.03$. 
The Voigt profile fit to the \Lya\ line is shown in the 
top panel of Figure~\ref{fig:J1419p0829}, with a selection of 
the associated metal absorption lines in the remaining panels. 

A cloud model with two components 
separated by $\sim 14$ km s$^{-1}$ 
(Table~\ref{tab:cloud_models})
provides a good fit to the data. 
This cloud model is well determined by a host
of \OI\ and \SiII\ lines (see Table~\ref{tab:J1419p0829_cd}).
Both \CII\ $\lambda 1334$ and $\lambda 1036$ 
are saturated; column densities for \NI, \OI\, \SiII,
and \FeII\ are listed in Table~\ref{tab:J1419p0829_cd}.

\subsection{The final VMP DLA sample}\label{sec:lit_dlas}

\begin{figure}
  \centering
  \includegraphics[angle=0,width=80mm]{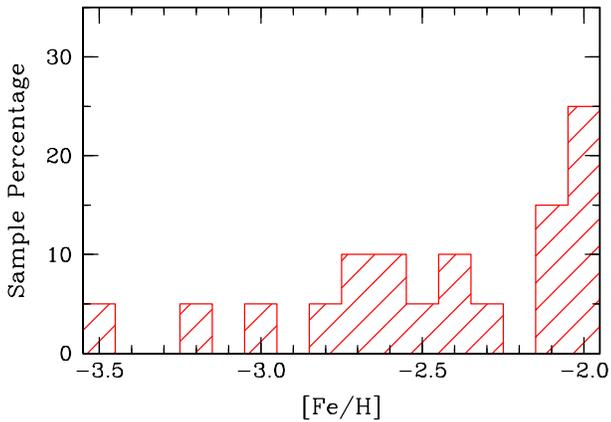}
  \caption{The metallicity distribution function for 
  the sample of VMP DLAs listed in Table~\ref{tab:mp_dlas_abund} 
  is shown as the red histogram.
  }
  \label{fig:mdf}
\end{figure}

In order to augment our survey of VMP DLAs, we have searched the
literature for known examples satisfying the following conditions:
(i) the QSO spectra were observed at high spectral resolution 
($R > 30\,000$) -- in practice this meant that the data were recorded
with either UVES or HIRES; (ii) [Fe/H]\,$\leq -2.0$; and 
(iii) at least one unsaturated \OI\ absorption 
from which [O/H] could be measured. 
These conditions were imposed to select a sample of measurements 
from the literature which is highly compatible 
to our own data and whose metal abundances 
could be adopted without reanalysing the spectra 
(although we referred all measurements 
to the same solar abundance scale -- see Section~\ref{sec:abund_anal}).
The literature trawl yielded an additional 
ten DLAs satisfying the above conditions;
together with our own observations they form a sample
of 22 VMP DLAs. The metallicity distribution 
function for this sample, which appears to 
tail off towards the lowest metallicities, 
is shown in Figure~\ref{fig:mdf}.

Relevant details of the full sample are collected in Table~\ref{tab:mp_dlas_abund},
where we list absorption redshifts, neutral hydrogen column densities
and element abundances for a selection of the metals most commonly
observed in VMP DLAs, including:
(1) C, N, O -- the first elements synthesised 
    in the chain of stellar nucleosynthesis;
(2) Al -- an odd atomic number element;
(3) Si -- an even atomic number element; and
(4) Fe -- an iron-peak element.
The uncertainty in each abundance includes the error in \HI. 
In Appendix~\ref{app:1}, we provide a similar table 
listing the column densities of each ion from which 
these abundances were derived.

\begin{table}
\centering
    \caption{\textsc{Ion column densities of the DLA in J1419+0829 at $z_{\rm abs}=3.04973$}}
    \begin{tabular}{@{}lp{1.8in}c}
    \hline
  \multicolumn{1}{l}{Ion}
& \multicolumn{1}{c}{Transitions used}
& \multicolumn{1}{c}{log $N$(X)/${\rm cm}^{-2}$}\\
    \hline
\HI    &  1025, 1215                                         &  20.40 $\pm$ 0.03  \\
\NI    &  1199.5, 1200.2, 1200.7                             &  13.28 $\pm$ 0.02  \\
\OI    &  936, 948, 976, 988.5, 988.6, 988.7, 1039, 1302     &  15.17 $\pm$ 0.02  \\
\SiII  &  989, 1190, 1193, 1260, 1304, 1526                  &  13.83 $\pm$ 0.01  \\
\FeII  &  1063.1, 1608                                       &  13.54 $\pm$ 0.03  \\
    \hline
    \end{tabular}
  \label{tab:J1419p0829_cd}
\end{table}


\begin{table*}
\centering
\begin{minipage}[c]{0.99\textwidth}
    \caption{\textsc{C, N, O, Al, Si, and Fe Abundance Measurements in VMP DLAs}}
    \begin{tabular}{@{}lrcccccccr}
    \hline
    \hline
   \multicolumn{1}{c}{\multirow{2}{*}{QSO}}
& \multicolumn{1}{c}{\multirow{2}{*}{$z_{\rm abs}$}} 
& \multicolumn{1}{c}{$\log N$\/(H\,{\sc i})}
& \multicolumn{1}{c}{\multirow{2}{*}{[C/H]}}
& \multicolumn{1}{c}{\multirow{2}{*}{[N/H]}}
& \multicolumn{1}{c}{\multirow{2}{*}{[O/H]}}
& \multicolumn{1}{c}{\multirow{2}{*}{[Al/H]}}
& \multicolumn{1}{c}{\multirow{2}{*}{[Si/H]}}
& \multicolumn{1}{c}{\multirow{2}{*}{[Fe/H]}}
& \multicolumn{1}{c}{\multirow{2}{*}{Ref.$^a$}}\\
  \multicolumn{1}{c}{}
& \multicolumn{1}{c}{}
& \multicolumn{1}{c}{(cm$^{-2}$)}
& \multicolumn{1}{c}{}
& \multicolumn{1}{c}{}
& \multicolumn{1}{c}{}
& \multicolumn{1}{c}{}
& \multicolumn{1}{c}{}
& \multicolumn{1}{c}{}
& \multicolumn{1}{c}{}\\    
  \hline
\multicolumn{10}{l}{\textbf{Our VMP DLA Sample}}\\
J0035$-$0918    & 2.34010  & $20.55\pm0.10$ & $-1.51\pm0.18$ & $-2.87\pm0.12$  & $-2.28\pm0.13$  & $-3.26\pm0.11$  & $-2.65\pm0.11$  & $-3.04\pm0.12$  &  2     \\
J0311$-$1722    & 3.73400  & $20.30\pm0.06$ & $-2.71\pm0.10$ & $\le-3.06$      & $-2.29\pm0.10$  & \ldots          & $-2.50\pm0.09$  & $\le-2.01$      &  1     \\
J0831$+$3358    & 2.30364  & $20.25\pm0.15$ & \ldots         & $\le-3.30$      & $-2.01\pm0.16$  & $-2.50\pm0.16$  & $-2.01\pm0.16$  & $-2.39\pm0.16$  &  1,4   \\
Q0913$+$072     & 2.61843  & $20.34\pm0.04$ & $-2.79\pm0.06$ & $-3.88\pm0.13$  & $-2.40\pm0.04$  & $-3.00\pm0.05$  & $-2.55\pm0.04$  & $-2.82\pm0.04$  &  3     \\
J1001$+$0343    & 3.07841  & $20.21\pm0.05$ & $-3.06\pm0.05$ & $\le-3.54$      & $-2.65\pm0.05$  & \ldots          & $-2.86\pm0.05$  & $-3.18\pm0.15$  &  1     \\
J1016$+$4040    & 2.81633  & $19.90\pm0.11$ & $-2.67\pm0.12$ & $\le-2.97$      & $-2.46\pm0.11$  & \ldots          & $-2.51\pm0.12$  & \ldots          &  3     \\
J1037$+$0139    & 2.70487  & $20.50\pm0.08$ & \ldots         & $-3.06\pm0.09$  & $-2.13\pm0.09$  & $-2.62\pm0.09$  & $-2.04\pm0.09$  & $-2.44\pm0.08$  &  1     \\
J1340$+$1106    & 2.50792  & $20.09\pm0.05$ & \ldots         & $-3.12\pm0.06$  & $-1.76\pm0.06$  & $-2.26\pm0.05$  & $-1.85\pm0.05$  & $-2.07\pm0.05$  &  1     \\
J1340$+$1106    & 2.79583  & $21.00\pm0.06$ & \ldots         & $-2.79\pm0.06$  & $-1.65\pm0.07$  & $-2.20\pm0.07$  & $-1.83\pm0.06$  & $-2.15\pm0.06$  &  1     \\
J1419$+$0829    & 3.04973  & $20.40\pm0.03$ & \ldots         & $-2.95\pm0.04$  & $-1.92\pm0.04$  & \ldots          & $-2.08\pm0.03$  & $-2.33\pm0.04$  &  1     \\
J1558$+$4053    & 2.55332  & $20.30\pm0.04$ & $-2.51\pm0.07$ & $-3.47\pm0.08$  & $-2.45\pm0.06$  & $-2.82\pm0.07$  & $-2.49\pm0.04$  & $-2.70\pm0.07$  &  3     \\
Q2206$-$199     & 2.07624  & $20.43\pm0.04$ & $-2.45\pm0.05$ & $-3.47\pm0.06$  & $-2.07\pm0.05$  & $-2.69\pm0.04$  & $-2.29\pm0.04$  & $-2.57\pm0.04$  &  3     \\
    &          &       &          &                 &          &          &             &        &        \\
\multicolumn{10}{l}{\textbf{Literature VMP DLAs}}\\
Q0000$-$2620    & 3.39012  & $21.41\pm0.08$ & \ldots         & $-2.54\pm0.08$  & $-1.68\pm0.13$  & \ldots          & $-1.86\pm0.08$  & $-2.01\pm0.09$  &  5     \\
Q0112$-$306     & 2.41844  & $20.50\pm0.08$ & \ldots         & $-3.17\pm0.09$  & $-2.24\pm0.11$  & \ldots          & $-2.39\pm0.08$  & $-2.64\pm0.09$  &  6     \\
J0140$-$0839    & 3.69660  & $20.75\pm0.15$ & $-3.05\pm0.17$ & $\le-4.20$      & $-2.75\pm0.15$  & $-3.37\pm0.16$  & $-2.75\pm0.17$  & $-3.45\pm0.24^{\rm b}$ &  7     \\
J0307$-$4945    & 4.46658  & $20.67\pm0.09$ & \ldots         & $-2.93\pm0.15$  & $-1.45\pm0.19$  & $-1.75\pm0.11$  & $-1.50\pm0.11$  & $-1.93\pm0.19$  &  8     \\
Q1108$-$077     & 3.60767  & $20.37\pm0.07$ & \ldots         & $\le-3.36$      & $-1.69\pm0.08$  & \ldots          & $-1.54\pm0.07$  & $-1.96\pm0.07$  &  6     \\
J1337$+$3153    & 3.16768  & $20.41\pm0.15$ & $-2.86\pm0.16$ & $\le-3.44$      & $-2.67\pm0.17$  & $-2.85\pm0.16$  & $-2.68\pm0.16$  & $-2.74\pm0.30$  &  9     \\
J1558$-$0031    & 2.70262  & $20.67\pm0.05$ & \ldots         & $-2.04\pm0.05\m$& $-1.50\pm0.05\m$& \ldots          & $-1.94\pm0.05\m$& $-2.03\pm0.05\m$&  10    \\
Q1946$+$7658    & 2.84430  & $20.27\pm0.06$ & \ldots         & $-3.51\pm0.07$  & $-2.14\pm0.06$  & \ldots          & $-2.18\pm0.06$  & $-2.50\pm0.06$  &  11    \\
Q2059$-$360     & 3.08293  & $20.98\pm0.08$ & \ldots         & $-2.86\pm0.08$  & $-1.58\pm0.09$  & \ldots          & $-1.63\pm0.09$  & $-1.97\pm0.08$  &  6     \\ 
J2155$+$1358    & 4.21244  & $19.61\pm0.10$ & $-2.09\pm0.12$ & \ldots          & $-1.80\pm0.11$  & $-2.13\pm0.20$  & $-1.87\pm0.11$  & $-2.15\pm0.25$  &  12    \\
    \hline
    \end{tabular}
    \smallskip

$^{\rm a}${References---1: This work;
2:  \citet{Coo11};
3:  \citet{Pet08};
4:  \citet{Pen10};
5:  \citet{Mol00}; 
6:  \citet{PetLedSri08};
7:  \citet{Ell10};
8:  \citet{Des01};
9: \citet{Sri10};
10: \citet{OMe06};
11: \citet{Pro02};
12: \citet{Des03}.
}\\
$^{\rm b}${\citet{Ell10} quote a $3\sigma$ upper limit to the \FeII\ column density of 
$\log N$\/(Fe\,{\sc ii})/cm$^{-2} < 12.73$. We have since rereduced these data 
(as described in Section~\ref{sec:obs}), and detected the \FeII\,$\lambda1608$ 
line at the $4\sigma$ level. The Fe abundance quoted here is derived 
using the optically thin limit approximation to measure $N$(\FeII).}
\\
$\m${The metal ion uncertainty for this measurement was not provided by the authors. Thus, we only quote the uncertainty in $N$(\HI).}
\\
\label{tab:mp_dlas_abund}
\end{minipage}
\end{table*}


\section{Abundance Analysis}
\label{sec:abund_anal}

As in previous DLA work, we assume  that each 
element resides in a single 
dominant ionization stage in the 
neutral gas. Thus, the abundance of 
a given element is found by taking 
the ratio of the dominant 
ions column density to that of \HI, 
and referring it to a solar scale 
(i.e. [X/H] = $\log (N$(X)/$N$(\HI)) $- \log$ (X/H)$_{\odot}$).
Throughout this article, 
we adopt the \citet{Asp09} 
solar scale, taking the photospheric, 
meteoritic, or the average of the two,
based on the suggestion by \citet{LodPlaGai09}.
The adopted abundances relevant to this 
work are collected in Table~\ref{tab:solar}.

\begin{table}
\caption{\textsc{Adopted solar abundances$^{\rm a}$}}
\centering
    \begin{tabular}{lcp{0.5cm}lc}
    \hline
   \multicolumn{1}{l}{X}
& \multicolumn{1}{c}{log(X/H)$_{\odot}$}
&& \multicolumn{1}{l}{X}
& \multicolumn{1}{c}{log(X/H)$_{\odot}$}\\
    \hline
C   & $-3.57$ & &   S   & $-4.86$ \\
N   & $-4.17$ & &   Ar  & $-5.60$ \\
O   & $-3.31$ & &   Cr  & $-6.36$ \\
Al  & $-5.56$ & &   Fe  & $-4.53$ \\
Si  & $-4.49$ & &   Ni  & $-5.79$ \\
\hline
\end{tabular}
    \smallskip

\begin{flushleft}
$^{\rm a}${From Asplund et al. (2009).}
\end{flushleft}
\label{tab:solar}
\end{table}

\begin{figure*}
  \centering
 {\hspace{-0.25cm} \includegraphics[angle=0,width=80mm]{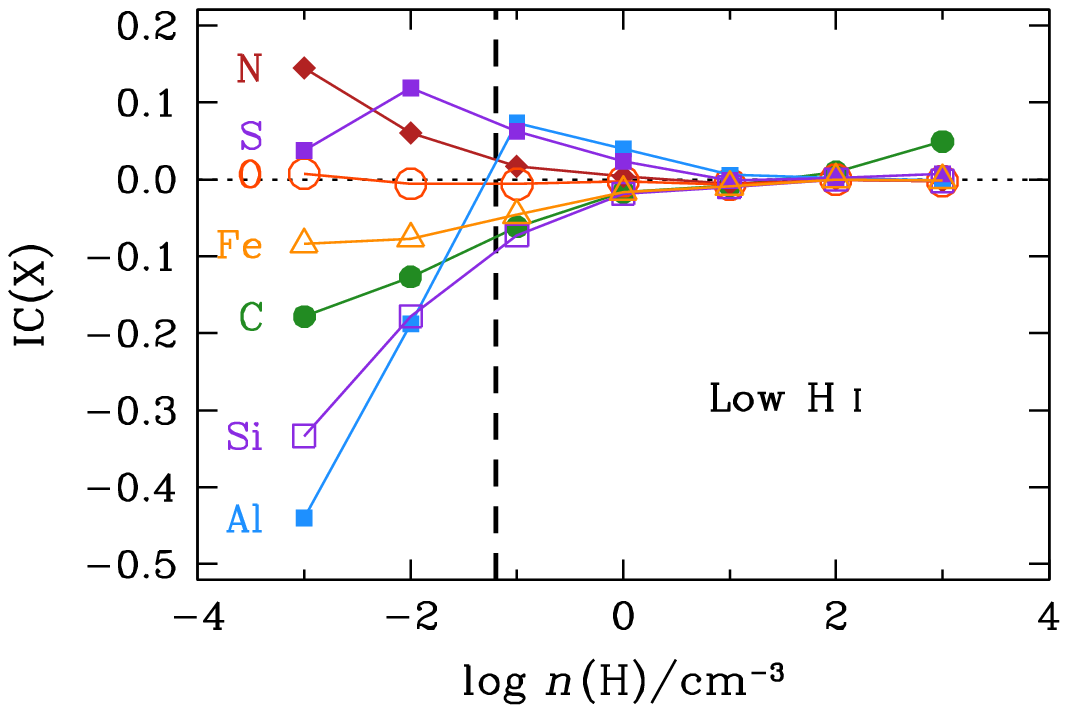}}
 {\hspace{0.25cm} \includegraphics[angle=0,width=80mm]{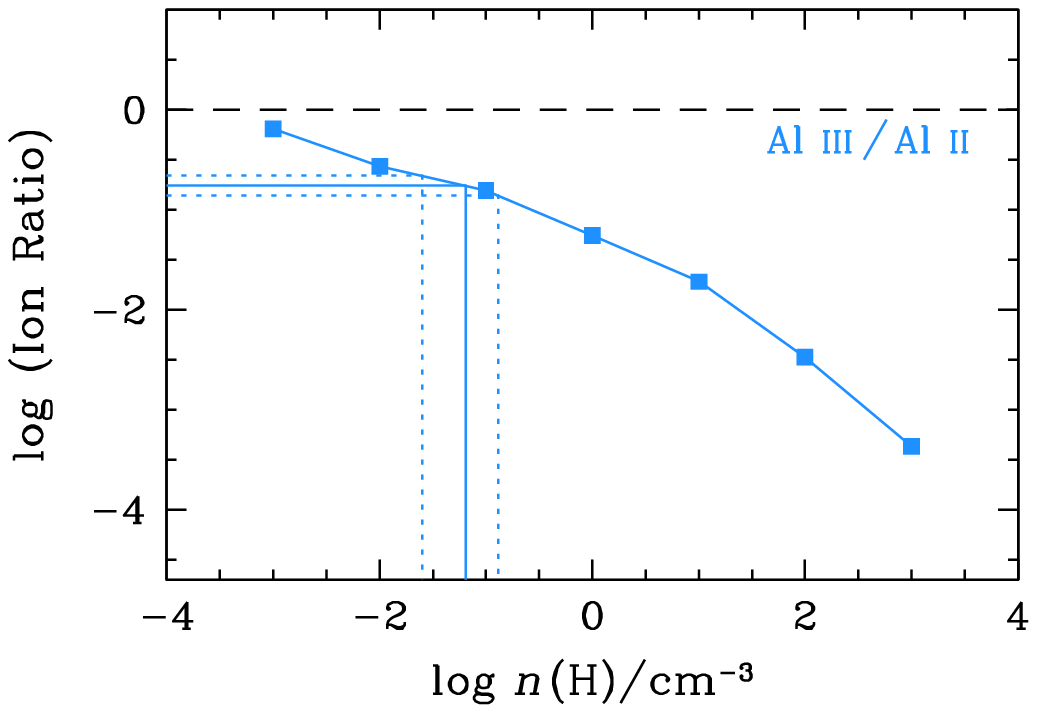}}\\
 {\vspace{0.4cm}}
 {\hspace{-0.25cm} \includegraphics[angle=0,width=80mm]{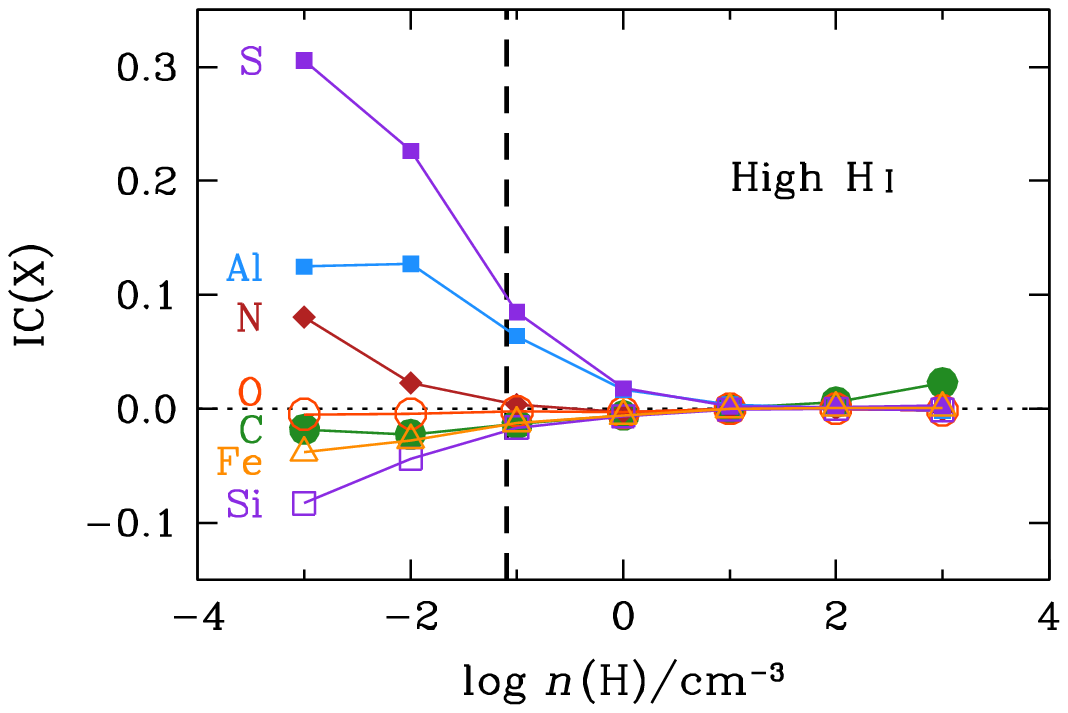}}
 {\hspace{0.25cm} \includegraphics[angle=0,width=80mm]{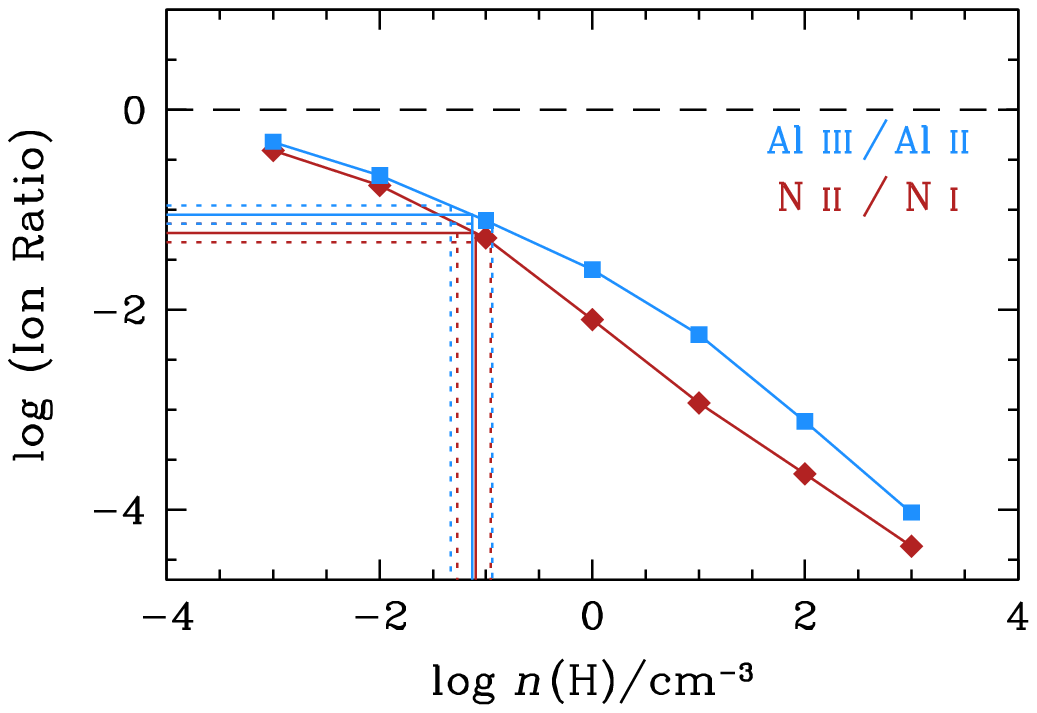}}\\
  \caption{Ionization corrections for two 
typical VMP DLAs, respectively with low and high \HI\ column density.
\emph{Left Panels:} The ionization corrections, 
as defined by Equation~\ref{eqn:ics}, for the most 
commonly observed elements in VMP DLAs are plotted 
as a function of gas volume density. 
The vertical dashed lines correspond to the estimated 
gas densities of the two DLAs 
(shown by the solid vertical line in the right panels).
\emph{Right Panels:} Column density ratios 
for successive ion stages of 
Al (blue squares connected by a solid line) and 
N (red diamonds connected by a solid line) vs. gas density. 
We also plot the observed values 
of the  \AlIII/\AlII\ and \NII/\NI\ ratios (solid horizontal lines) 
along with their uncertainties (dotted lines) 
for the two VMP DLAs we use as examples in this test
(see text for further details).
  }
  \label{fig:ics_gd}
\end{figure*}

Thus, measuring elemental abundances in 
VMP DLAs is a relatively straightforward 
process. Some uncertainty arises, however, 
if the observed metal-line absorption 
from ions that are dominant in \HI\ regions, 
does not perfectly trace the \HI\ gas. 
For example, if the dominant ion for a 
given element is also present in nearby 
\HII\ gas (often the case for singly ionized species; e.g. \CII, \SiII, \FeII), 
we will over-estimate the abundance of this element. 
On the other hand, if the dominant ion for 
a given element is mildly ionized in the \HI\ gas 
itself (usually the case for neutral species; e.g. \NI), 
we will under-estimate the element's abundance.
In addition to these ionization corrections, 
further uncertainties may be introduced into 
the abundance analysis if some refractory elements 
have condensed to form dust grains. Whilst both 
of these concerns are expected to be 
negligible in the VMP regime \citep{Vla01,Vla04}, 
we reassess their importance in the 
following sub-sections.

\subsection{Ionization Corrections}
\label{sec:ics}

Ionization corrections 
are known to be small when the 
neutral hydrogen column density is in 
excess of $\sim10^{20}$ atoms cm$^{-2}$.
Nevertheless, for each element X with the 
dominant ionization stage \textsc{n}, one 
needs to apply a small correction, IC(X), 
to recover the \emph{true} elemental 
abundance,
\begin{equation}\label{eqn:ics}
[{\rm X/H}] = [{\rm X}\,\textsc{n}/\HI] + \mathrm{IC(X)}.
\end{equation}

To estimate the magnitude of such corrections, 
we used the \textsc{cloudy} photoionization 
software developed by \citet{Fer98} to model 
two of the VMP DLAs from our sample, chosen to 
represent the range of \NHI\ values that we report: 
one DLA with a low \HI\ column density 
(the DLA with $\log[N({\rm H}\,\textsc{i})/{\rm cm}^{-2}]=20.09$ towards J1340$+$1106 at $z_{\rm abs}=2.50792$) 
and the other with a high $N$(\HI)
(the DLA with $\log[N({\rm H}\,\textsc{i})/{\rm cm}^{-2}]=21.00$ towards J1340$+$1106 at $z_{\rm abs}=2.79583$).
In both cases, we modelled the VMP DLA 
as a plane-parallel slab of constant 
volume density gas in the range
$-3 < \log[n({\rm H})/{\rm cm}^{-3}] < 3$, 
irradiated by the cosmic microwave background and UV 
background \citep{HarMad01} at the appropriate redshift. 
Using the solar abundance scale in Table~\ref{tab:solar}
we globally scaled the metal abundances of the VMP DLA 
to be $10^{-2}$ Z$_{\odot}$. 
The simulations were stopped once the 
\HI\ column density of the DLA was reached, 
at which point we output the simulated 
ion column densities of the slab. 

Once the above value of the background 
radiation field is assumed, 
the ionization correction for each element 
depends on the volume density of the gas 
(left panels of Figure~\ref{fig:ics_gd}), 
which may be estimated by considering 
the ratio of successive ion stages 
(right panels of Figure~\ref{fig:ics_gd}). 
For the low \NHI\ DLA being considered as an 
example here, we measure 
$N$(\AlIII)/$N$(\AlII)$ = -0.76 \pm 0.10$,
implying a gas density of 
$\log [n{\rm (H)/cm}^{-3}] \simeq -1.2 \pm 0.4$.
For the high \NHI\ DLA, we measure
$N$(\AlIII)/$N$(\AlII)$ = -1.05 \pm 0.09$ and 
$N$(\NII)/$N$(\NI)$ = -1.23 \pm 0.09$,
which are both consistent with a gas density of 
$\log [n{\rm (H)/cm}^{-3}] \simeq -1.1 \pm 0.2$. 
At these values of the gas density, 
it can be seen that the ionization corrections 
for the VMP DLAs in our sample are $\simlt0.1$ dex
for the main elements of interest.

We also performed the above calculations under 
the assumption of a softer background radiation 
field (i.e. an O star) rather than the \citet{HarMad01} 
background. Our results are quantitatively similar 
to those of \citet{Vla01}: corrections for all of the 
elements we are interested in are $\simlt0.1$ dex, 
except for \AlII\ which can require corrections of 
the order $\sim0.3$ dex for the lowest column density 
systems. At present, it is not yet clear whether
these very metal-poor systems harbour (or are nearby 
to) massive Population II or Population III stars. 
Based on the results described in this subsection, 
we have therefore not corrected any of the measured
abundances for ionization effects. 

\subsection{Dust Depletion}
\label{sec:dd}

To measure accurately element abundances 
in DLAs, the fraction of a given element that 
is not observed in the gas phase, but is 
instead locked up in dust grains, 
must also be considered. 
To account for this effect, 
one ideally considers the 
relative abundances of a 
refractory and a volatile element
(e.g. [Cr/Zn]), and compares this to 
the expected intrinsic nucleosynthetic 
ratio (typically the ratio seen in stars 
of comparable metallicity). 
Previous studies based on such a comparison
have shown that DLAs exhibit 
minimal dust depletion when [Fe/H] $\simlt-2.0$ 
\citep{Pet97,Ake05}. Unfortunately, the \CrII\ 
and \ZnII\ lines are too weak in the VMP 
regime to be measured, so one needs to resort 
to more abundant elements, such as Si and Fe, 
which are known to be depleted to different degrees 
(Fe is more readily incorporated into dust grains than Si).

The most metal-poor stars in the halo of our Galaxy 
suggest that the intrinsic nucleosynthetic 
ratio of Si/Fe is virtually independent of 
metallicity, corresponding roughly to a constant 
of [Si/Fe] $= +0.37 \pm 0.15$ \citep{Cay04}. 
Such a plateau was first seen in DLAs by \citet{ProWol02}, 
who found [Si/Fe] $\simeq+0.3$ when [Fe/H] $< -2$ 
(see updated version in \citealt{WolGawPro05}). 
A similar study was also conducted by \citet{Vla02}, 
who suggested there may still exist some mild 
depletion onto dust, resulting in a plateau of 
[Si/Fe] $\simeq+0.25$. 
From the 19 VMP DLAs in our sample, we find that 
[$\langle$Si/Fe$\rangle$] $= +0.32 \pm 0.09$, 
which is certainly consistent 
with minimal dust depletion. 
We therefore proceed under the assumption 
that dust has a negligible effect on our 
derived elemental abundances. 


\section{Comparing VMP DLAs and stars}
\label{sec:dvs}

DLAs likely experience very different chemical 
histories to that of the stars in the halo of our 
Galaxy. However, in the limit of decreasing 
metallicity, both stars and DLAs are polluted by 
very few previous generations of stars. 
In fact, since the physical conditions of the gas 
that gives rise to DLAs are conducive to forming 
stars \citep{Not08,Jor09}, it certainly seems 
plausible that some of the VMP stars in the 
halo of our Galaxy originally condensed out of 
a VMP DLA. If this is indeed true, we would 
therefore expect both VMP stars and DLAs to share 
similar chemical signatures at the lowest 
metallicities. In the following subsections, 
we test the validity of this expectation, 
by directly comparing the abundances of the 
VMP DLAs in this survey to the most recent 
abundance measurements of VMP stars. 

\subsection{Revisiting C/O at low metallicity}
\label{sec:co}

\begin{figure*}
  \centering
  \includegraphics[angle=0,width=140mm]{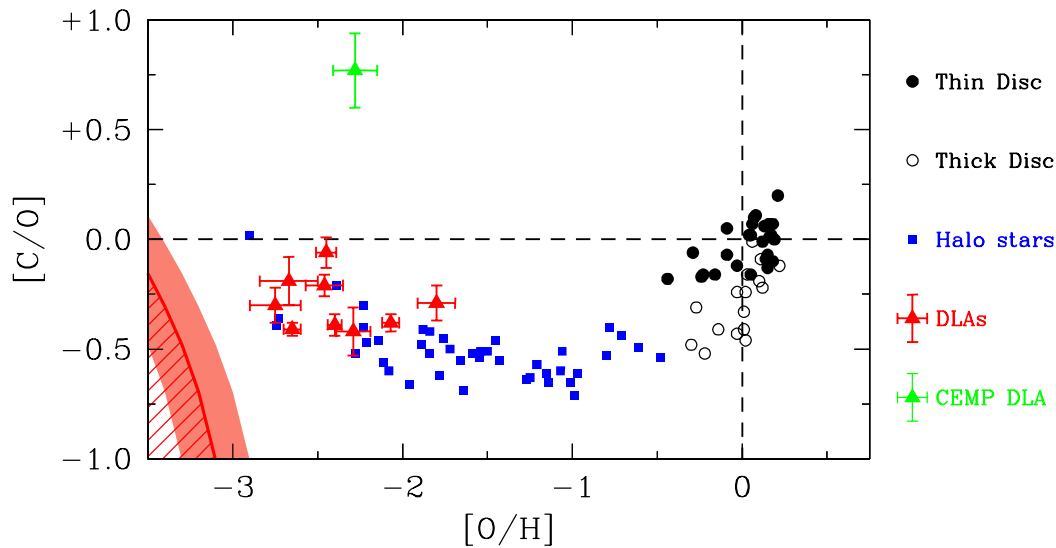}
  \caption{
C and O abundances in VMP DLAs (red triangles) observed at high spectral resolution. 
The green triangle represents the carbon-enhanced DLA reported by \citet{Coo11}. 
For comparison, we also plot the sample of metal-poor halo stars 
analysed by \citet{Fab09a}, where we adopt the values for 
C and O that are calculated assuming the \citet{Dra69} formula for 
collisions with hydrogen ($S_H = 1$; blue squares) 
We also show a sample of thin- and thick-disc stars 
(filled and open circles respectively) from \citet{BenFel06}.
The red hatched region corresponds to the transition discriminant 
outlined by \citet{FreJohBro07}, where the uncertainty in this 
relation is shown by the light red shaded region. All of the 
plotted data have been referred to the \citet{Asp09} solar 
abundance scale (see Section~\ref{sec:abund_anal}).
  }
  \label{fig:co}
\end{figure*}

We first consider the trend of C/O at low metallicity, 
which has received a great deal of attention 
in recent years. The evolution of the C/O ratio when 
[O/H] $\simgt -1.0$ has been known for a while; 
[C/O] increases linearly from $\sim-0.6$ 
to solar with increasing [O/H] (see Figure~\ref{fig:co}). 
This trend is thought to be due to the 
increased, metallicity-dependent, carbon yields of 
massive rotating stars, combined with the delayed 
release of carbon from low and intermediate mass stars
\citep[][]{Ake04}.
Based on current models of Population II nucleosynthesis,
below [O/H] $\sim-1.0$, [C/O] is predicted to 
decrease (or perhaps plateau) with 
decreasing metallicity. Indeed, such a plateau 
was first reported for a sample of 
halo stars by \citet{Tom92}, who measured 
[C/O] abundances down to [O/H] $\sim -1.7$. 

This work was extended to even lower 
oxygen abundances by \citet{Ake04} and \citet{Spi05}. 
Contrary to the supposed decrease in 
[C/O], these authors uncovered quite the 
opposite trend when [O/H] $\simlt -2.0$; 
an extrapolation of this trend suggests that 
[C/O] could reach near-solar values 
when [O/H] $\sim-3.0$. Three possibilities have 
been suggested to explain this behaviour: 
(1) the leftover signature of a high-carbon 
producing generation of Population III stars
\citep{ChiLim02,UmeNom03,HegWoo10}; 
(2) pollution from a previous generation 
of rapidly-rotating low-metallicity 
Population II stars \citep{Chi06}; or 
(3) systematic uncertainties in the adopted 
1D LTE analysis. 
The third of these possibilities has recently
been ruled out by \citet{Fab09a}, 
who conducted a more detailed analysis of the lines
used by \citet{Ake04}, thus confirming 
the stellar C/O trend at low metallicity. 

Such an `unexpected' trend is perhaps not 
so surprising in hindsight, since several other 
lines of evidence support a high-carbon producing 
generation of early stars, including: 
(1) the fact that a high carbon abundance is 
\emph{required} at early times to efficiently 
cool the gas, and drive the transition from 
Population III to Population II star formation 
\citep{FreJohBro07}; 
(2) observations of the three most iron-poor 
stars known to date have revealed that they 
all exhibit extreme carbon enhancements 
\citep{Chr02,Fre05,Nor07};
(3) in addition, the fraction of all 
carbon-enhanced metal-poor (CEMP) stars 
is thought to increase with decreasing metallicity 
\citep{BeeChr05}; 
(4) for at least a subset of these CEMP stars, 
it has been suggested that their carbon-enhancement 
reflects the composition of the cloud of gas 
from which the CEMP star first condensed 
\citep{Rya05,Aok07}; and 
(5) the observed fraction of CEMP stars in the 
outer halo component of our Galaxy is roughly twice that 
of the inner halo component, which favours the existence 
of a high carbon-producing source other than 
asymptotic giant branch stars \citep{Car11}. 

Additional evidence for a high carbon-producing 
early generation of stars has recently been provided 
by studies of the most metal-poor DLAs \citep{Pet08,Pen10}, 
which probe entire clouds of near-pristine gas. 
Such surveys find that DLAs and stars tell the same story 
in the metal-poor regime -- both show elevated, 
near-solar values of [C/O] when [O/H] $\simlt-2.0$.
In fact, the recent medium spectral resolution study 
by \citet{Pen10} 
suggests that [C/O] might further increase 
to \emph{super-solar} values at even lower 
metallicity, although this still remains 
uncertain due to saturation effects. 
Further evidence for an increased carbon yield by an 
early generation of stars has recently come to light
with the discovery by  
\citet{Coo11} of a metal-poor 
DLA which exhibits a C/Fe ratio $35$ times greater than solar. 

In Figure~\ref{fig:co}, we show the updated 
plot of [C/O] versus [O/H] for our full DLA survey, 
together with values 
for the metal-poor halo stars (blue squares) 
analysed by \citet{Fab09a}\footnote{The data shown 
in Figure~\ref{fig:co} refer to cross sections 
for collisions with hydrogen atoms based on the 
classical recipe of \citet{Dra69}, i.e. using a 
scaling factor $S_H = 1$. If hydrogen collisions 
are completely neglected ($S_H = 0$), this trend 
is `stretched' to higher [C/O] (by about 0.2 dex) 
and lower [O/H] (by about 0.3 dex at the 
lowest values of [O/H].}, and 
a sample of thin- and thick-disc stars with C and O 
abundances determined from forbidden lines (filled and open 
black circles respectively; \citealt{BenFel06}). 
All data have been corrected for the updated 
\citet{Asp09} solar abundance scale 
(see Section~\ref{sec:abund_anal}).
We first note that there is generally a 
good agreement -- both in the trend and 
the dispersion of [C/O] -- between the 
most metal-poor stars and DLAs. The new 
DLA measurements reported here confirm the 
initial indications from the more limited 
samples considered by \citet{Pet08} and \citet{Pen10}. 
The main departure from this trend is the 
CEMP DLA reported by \citet{Coo11}, 
represented by the green triangle in 
Figure~\ref{fig:co}. Aside from this 
system, no other DLA exhibits 
super-solar [C/O]. This statement is 
also true for the sample of seven 
\OI\ absorbers at $z_{\rm abs}\sim6$
recently reported by \citet[][see also \citealt{Bec06}]{Bec11}.

It is thus somewhat surprising that 
\citet{Pen10} found [C/O]\,$\ge 0.0$
for four out of the five VMP DLAs
in which they could measure this ratio.
The difference may be due to different
sample criteria between our survey and theirs:
while we have excluded from our analysis
absorption systems where the \CII\ lines are
saturated, such cases may be more difficult 
to recognise at the lower resolution
of the ESI spectra analysed by \citet{Pen10}.
VMP DLAs with super-solar C/O ratios may
well exist
(and indeed the CEMP DLA discovered by \citet{Coo11} is
one such example), but their C abundance is generally more difficult
to measure with confidence due to line saturation.
On the other hand, our survey is not biased against uncovering 
systems with \emph{lower} [C/O] values than those reported here;
thus, the VMP DLAs in our current 
sample define a \emph{lower-envelope} in the 
[C/O] versus [O/H] plane. 
This envelope appears to be in
good agreement with the envelope defined by VMP stars
(see Figure~\ref{fig:co}).

In Figure~\ref{fig:co} we also show the `Frebel criterion' 
(red hatched region; \citealt{FreJohBro07}; 
see also \citealt{BroLoe03}) which states that, 
if the fine-structure lines of \OI\ and \CII\ dominate 
the cooling in a near-pristine cloud of gas 
that has been enriched to some critical metallicity, 
then the first low mass Population II stars will form. 
Thus, given these conditions, no Population II star 
should be observed in this red hatched region 
(where the uncertainty in this region is given by 
the light red shaded band). DLAs, on the other hand, are 
not restricted by this criterion. Indeed, if a cloud 
of gas was to be observed in this Population II star 
`forbidden zone', it may very well form a collection of 
massive stars below or near the critical metallicity! 
Such systems, if found, would provide 
a unique window to study the transition 
from Population III to Population II star formation. 

\begin{figure*}
  \centering
  \centering
 {\hspace{0.25cm} \includegraphics[angle=0,width=140mm]{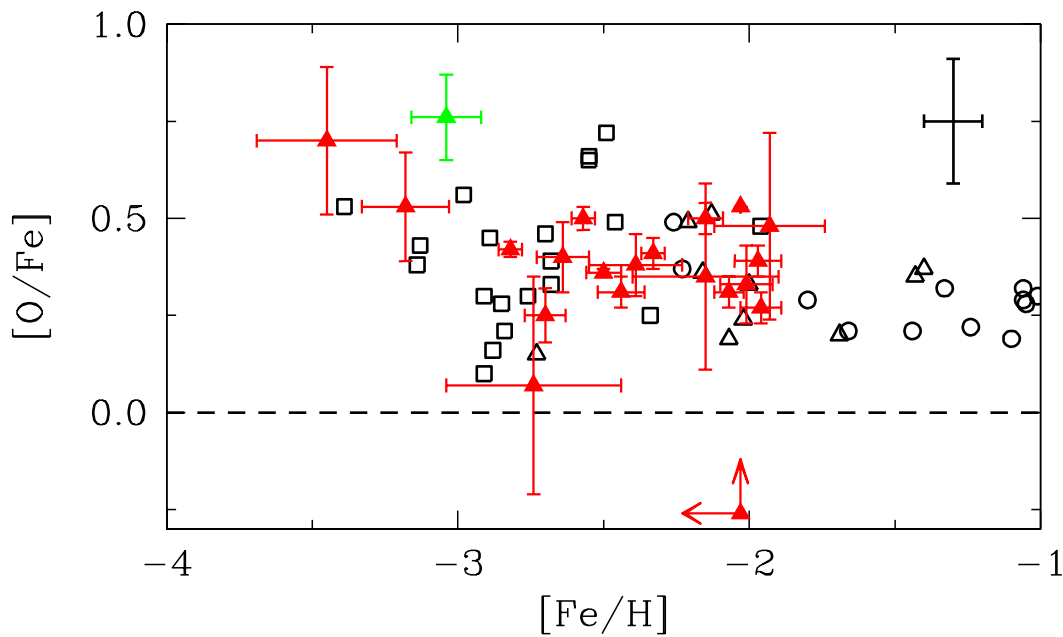}}\\
  \caption{
The [O/Fe] ratio in VMP DLAs (filled red triangles) 
where [O/H] has been measured from high resolution spectra. 
The green triangle refers to the carbon-enhanced DLA reported by \citet{Coo11};
the upper limit on [Fe/H] and corresponding
lower limit on [O/Fe] (red triangle with arrows) is for
the DLA towards J0311$-$1722 where \FeII\ absorption is not
detected. 
The VMP DLA sample is compared with the stellar abundance measurements 
by \citet{Nis02} (circles), \citet{Cay04} (squares) and 
\citet{Gar06} (triangles), all based  on
the [\OI]\,$\lambda6300$ line and  
corrected for 3D effects (see text for further details). 
The error bars at the top right corner of the plot indicate the 
typical errors in the stellar abundance measurements.
All the measurements in this plot are tabulated, together with their
individual errors, in Appendix~B.
  }
  \label{fig:ofe}
\end{figure*}

With these considerations in mind, 
we note that all DLAs in our sample will, 
perhaps unsurprisingly, form 
Population II stars. One might, therefore, 
be tempted to conclude that the stars in the halo 
of our Milky Way represent the same distribution 
that defines VMP DLAs. 
To test this possibility, we performed a linear fit 
to [C/O] versus [O/H] for the halo stars 
that have [O/H] $\le-2.0$, and calculated the 
deviations about this line 
for both stars and DLAs. 
A Kolmogorov-Smirnov (K-S) test 
between the calculated deviations reveals a 
$75\%$ chance that both VMP stars and DLAs 
are drawn from the same population, 
which is inconclusive given the 
present statistics. 
For this test 
we have used the C/O values in halo stars 
that were derived assuming efficient 
hydrogen collisions (with a scaling factor $S_H=1$; see footnote $9$). 
If instead we adopt the stellar values derived for 
inefficient collisions ($S_H=0$), we 
find a $90\%$ chance that both samples are 
drawn from the same population.
Such good agreement between halo stars 
and DLAs does not support the 
recent claim by \citet{TsuBek11}, 
who suggest that the 
initial mass function (IMF) 
of the stars that enriched 
metal-poor DLAs is different 
from the IMF of the stars that 
enriched Galactic halo stars with their metals. 
The good general agreement we find 
between stars and DLAs -- both exhibiting 
an elevated C/O ratio at the lowest 
metallicities probed -- points to a 
universal origin for their 
C/O `excess' in this regime. 

\subsection{The O/Fe debate in the metal-poor regime}
\label{sec:ofe}

We now turn to the relative abundances of 
oxygen and iron at low metallicity. 
Whilst we cannot do justice to 
the extensive literature on this topic, 
we outline below the basic facts 
that are relevant to our discussion, 
and direct the interested reader to the 
comprehensive review by \citet{McW97}. 

The largest oxygen yield comes from the 
most massive stars that explode as SNe II. 
The budget for iron, on the other hand, 
is largely contributed by SNe Ia which 
typically explode $\sim1$ Gyr later 
(see e.g. \citealt{Gre10}).\footnote{Iron 
may also be contributed by SNe Ia that 
`promptly' explode at early times ($\sim0.1$ Gyrs). 
For the relevant details, we direct the 
interested reader to the 
discussion by \citet{ManDelPan06}.}
Therefore, at early times (when the 
metallicity is low), one expects O to be 
enhanced relative to Fe. At later times, 
when the delayed contribution of Fe 
from SNe Ia kicks in, there is 
a break in the O/Fe trend which is 
then expected to decrease. 
Thus, the relative abundance of 
O and Fe allows us to measure the 
relative contribution of SNe Ia and SNe II
(see e.g. the qualitative discussion 
by \citealt{WheSneTru89}).

In the Milky Way, the break in
 [O/Fe] occurs roughly 
at [Fe/H]\,$\simeq -1.0$. 
Whilst there is sound agreement regarding 
the nature of the trend in [O/Fe] when [Fe/H] $\simgt-1.0$,
the behaviour 
of [O/Fe] when [Fe/H] $\simlt-1.0$
is less certain. 
This disagreement stems 
from the uncertainty of the oxygen abundances 
measured in metal-poor stars: there 
are four different indicators of the oxygen 
abundance, and to some extent they all disagree
with one another 
in the metal-poor regime \citep{Gar06}. 

Perhaps the most reliable [O/H] indicator 
at low metallicity is the forbidden 
[\OI]\,$\lambda 6300$ line which, despite being 
subject to 3D corrections of $\sim-0.2$ dex when 
[Fe/H] $\sim-2.0$ \citep{Nis02,ColAspTra07}, 
is known to form in LTE \citep{Asp05}. 
Unfortunately, this line becomes very weak 
when [Fe/H] $<-2.0$ and its detection
requires data of high S/N.
After accounting for 3D corrections to the O abundance,
most authors conclude that the O/Fe ratio
is approximately constant at 
[O/Fe] $\simeq+0.4$ for 
[Fe/H]\,$\simlt -1$
\citep{Nis02,Cay04,Gar06}, 
with perhaps a slight increase 
towards the lowest metallicities.

The most commonly used diagnostic for 
measuring [O/H] in stars is the \OI\ triplet 
near 777\,nm ($\lambda=7771.9$, $7774.2$, $7775.4$\,\AA), 
despite the fact that it suffers 
from large non-LTE corrections \citep{Fab09b}.
However, contrary to the nearly constant value of [O/Fe] below
[Fe/H]\,$\simlt -1$ deduced from the 
[\OI]\,$\lambda 6300$ line, 
an LTE analysis of the \OI\ triplet leads 
to a quasi-linear 
increase in [O/Fe] with decreasing metallicity 
(see e.g. \citealt{FulJoh03}). This discrepancy 
is often blamed on the uncertain (negative) 
non-LTE corrections to the \OI\ triplet.
In order to make headway with the 
[O/Fe] conflict, \citet{Fab09b} performed a detailed 
non-LTE analysis of the \OI\ triplet, 
and found corrections amounting to 
$\simgt 0.5$ dex when [Fe/H] $=-3.0$, 
increasing rapidly at lower metallicities 
(see also \citealt{Fab09a}). After 
accounting for such corrections, 
\citet{Fab09b} concluded that almost 
all diagnostics are now conceivably 
consistent, and that [O/Fe] exhibits 
a roughly flat plateau with values between 
$+0.4$ and $+0.6$ when [Fe/H]\,$\simlt -1$.

In contrast to the profusion of stellar 
studies of [O/Fe] in the VMP regime, this ratio 
has received relatively little attention in DLAs so far. 
The reason is that the most readily 
available \OI\ absorption lines are 
almost always saturated, 
and the weakest lines are often blended with 
unrelated absorption in 
the \Lya\ forest \citep{ProWol02}. 
For these reasons, several authors have 
investigated the [O/Fe] trend in sub-DLAs, 
where the \OI\ absorption lines are weaker 
\citep{Per03b,OMe05}. However, uncertain 
negative ionization corrections to the 
\FeII\ lines might become important for such
systems,
complicating the interpretation. Other 
authors have instead used [S/Zn] as a proxy 
for [O/Fe] \citep{Nis07}, but these lines 
become too weak in the VMP regime. 

In fact, there have only been two studies 
in the literature that consider [O/Fe] in 
DLAs. The first was conducted by 
\citet{PetLedSri08}, who reported an [O/Fe] 
plateau of $+0.32\pm0.10$ from their 
sample of 13 DLAs with [Fe/H]\,$< -1.0$, 
(three of which have [Fe/H]\,$< -2.0$). 
The second, more 
recent, study was conducted by \citet{Pen10} 
whose sample includes  five DLAs with [Fe/H]\,$< -2.0$. 
Their measurements, however, have 
large uncertainty ($\sim\pm0.4$ dex), 
so it is difficult to discern the underlying trend. 

Our survey constitutes the largest sample of 
high resolution measures of \OI\ and \FeII\ absorption 
in DLAs. The [O/Fe] values are 
plotted in Figure~\ref{fig:ofe} where, for
comparison, we also show a selection of 
[O/Fe] measurements in Galactic stars based on 
the forbidden [\OI]\,$\lambda6300$ line 
\citep{Nis02,Cay04,Gar06}
with 3D corrections applied as we now describe. 

The [\OI]\,$\lambda6300$ line corresponds to a 
forbidden transition between two levels of the ground 
configuration of the \OI\ atom, 
which are closely coupled via collisions. 
Because nearly all oxygen atoms are in the ground state 
in the atmospheres of late-type stars, 
one expects LTE to prevail; this is 
confirmed by detailed statistical equilibrium calculations 
\citep{Kis93}. Non-LTE effects on the derived iron 
abundances are also negligible, when lines from the dominating 
ionization stage (\FeII) are considered \citep{Mas11}. 
The 3D -- 1D corrections are, however, significant 
for metal-poor stars;  [O/H] derived from the [\OI]\,$\lambda6300$ 
line decreases and [Fe/H] from \FeII\ lines increases slightly. 
As calculated by \citet{Nis02}, the net effect on [O/Fe] 
for metal-poor main-sequence stars can be approximated by the 
expression [O/Fe]$_{\rm 3D} -$ [O/Fe]$_{\rm 1D} = 0.11$\,[Fe/H], 
whereas [Fe/H]$_{\rm 3D} -$ [Fe/H]$_{\rm 1D} = -0.04$\,[Fe/H]. 
\citet{Cay04} assumed that the same corrections are also valid for 
metal-poor red giants. We have verified that this is approximately 
correct by applying 3D corrections for giant stars as calculated by 
\citet{ColAspTra07} for the [\OI]\,$\lambda6300$ line and 
the Fe lines used by \citet{Cay04}. The same 3D corrections 
are then also expected for the cool subgiants studied 
by \citet{Gar06}, because they have atmospheric parameters 
intermediate between those of the main-sequence and red giant stars. 
For reference, we list in Appendix~\ref{app:2} values
of [O/Fe] and [Fe/H] corrected for 3D effects, as
well as the  [O/Fe] and [Fe/H] 
values measured in our sample of VMP DLAs. 

\begin{figure}
  \centering
  \includegraphics[angle=0,width=80mm]{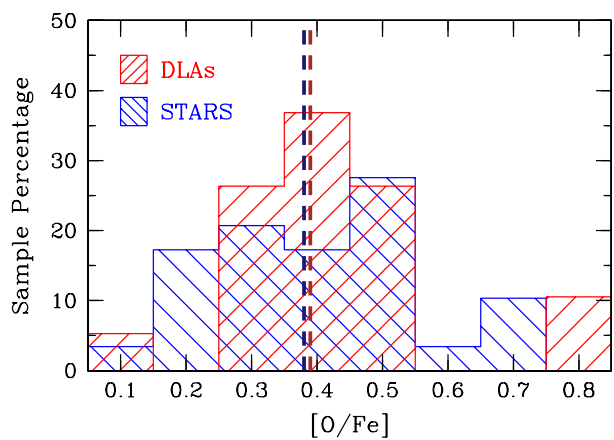}
  \caption{
  The distribution of [O/Fe] values (where [Fe/H] $\le-2.0$) in DLAs for our sample (red histogram) 
  as compared to the compilation of [O/Fe] in stars measured from the 
  [\OI]\,$\lambda6300$ line (blue histogram). 
  The dashed vertical red and blue lines indicate the median values for DLAs and stars 
  respectively, corresponding to a difference of only $\sim0.01$ dex.
  }
  \label{fig:ofe_hist}
\end{figure}

Once the above-mentioned 3D corrections are applied, 
it can be seen that stars and DLAs share a similar 
trend of [O/Fe] with decreasing metallicity. 
To illustrate this point, we show the two 
samples (where [Fe/H] $\le-2.0$) as histograms 
in Figure~\ref{fig:ofe_hist}. A K-S test\footnote{We have 
not included the star CS\,22949$-$037 in this test 
(from \citealt{Cay04}), since it exhibits a peculiar 
abundance pattern, with [O/Fe] $=+1.54$.} reveals that the 
probability the two data sets are drawn from the same 
parent population is $71\%$. 

The DLA values of [O/Fe] exhibit relatively little scatter
given the errors. In the range $-3\le$ [Fe/H] $\le-2$, 
[O/Fe] is consistent with a constant value:
[$\langle$O/Fe$\rangle$] $= +0.35\pm0.09$. 
This is in good agreement with the mean value reported by 
\citet{PetLedSri08}, [$\langle$O/Fe$\rangle$] $= +0.32\pm0.10$, 
for $-2.0\simlt$ [Fe/H] $\simlt-1.0$. Interestingly, 
there may be a hint in our data that [O/Fe] 
increases further when [Fe/H]\,$\simlt -3$
(see Figure~\ref{fig:ofe}), which would presumably 
be indicative of a contribution from more massive stars. 
However, more data are required to confirm this `trend' 
which at present is suggested by the two most metal-poor 
DLAs in our sample. 
For the moment, we simply conclude that the 
`cosmic' trend of [O/Fe] in the VMP regime 
($-3\le$ [Fe/H] $\le-2$) reaches a plateau of $\sim+0.35$, 
and is remarkably tight, especially given the 
observational errors. 

In closing, the DLA measurements of [O/Fe] 
help resolve the controversy
regarding the relative abundances of O and Fe 
metal-poor Galactic stars.
It is plausible that VMP DLAs harbour the reservoir 
of neutral gas that will later condense 
to form a population of VMP stars.
It is thus expected that towards the 
lowest metallicities both stars and DLAs 
should exhibit a similar trend. We conclude that,
unless there are marked differences
between the chemical evolution histories
of DLAs and the early Galaxy, our results
and those of \citet{PetLedSri08}
favour an approximately constant plateau
of stellar [O/Fe] values when [Fe/H]\,$\simlt -1$, 
with perhaps a mild increase 
with decreasing [Fe/H]. 
In any case, given the on-going improvement
in the accuracy of the stellar models and the
increasing samples of
VMP DLAs, we anticipate that this issue 
may well be settled in the near future.

\section{Discussion}
\label{sec:disc}

\subsection{The typical VMP DLA}
\label{sec:typ_VMPDLA}

\begin{table}
\centering
\caption{\textsc{The mean value and dispersion in X/O for each element of the VMP DLAs in our sample.}}
    \begin{tabular}{lccc}
    \hline
    \multicolumn{1}{l}{X}
&    \multicolumn{1}{c}{[$\langle$X/O$\rangle$]}
&    \multicolumn{1}{c}{$\sigma_{\rm [\langle X/O\rangle]}$}
&    \multicolumn{1}{c}{$n_{\rm [\langle X/O\rangle]}$}\\
    \hline
C   &  $-0.28$   &  $0.12$  &  $9$   \\
N   &  $-1.05$   &  $0.19$  &  $13$  \\
Al  &  $-0.44$   &  $0.13$  &  $11$  \\
Si  &  $-0.08$   &  $0.10$  &  $21$  \\
Fe  &  $-0.39$   &  $0.12$  &  $20$  \\
    \hline
    \end{tabular}
\label{tab:typ_dla}
\end{table}

With the large sample of measurements we have 
assembled, we can now attempt to reconstruct 
the abundance pattern of a `typical' VMP DLA, 
and to consider the clues it may provide on the 
nucleosynthesis by the earliest generation of stars. 
To achieve this goal, we are obviously required to 
select a reference element other that H, 
as we have no means to determine how much H 
was mixed with the nucleosynthetic products 
from the earliest generations of stars. 
Rather, the ratio of two \emph{metals} 
provides the best handle for determining 
the properties of the generation of 
stars from which they were synthesised. 

With our goal to probe early nucleosynthesis 
borne in mind, the most appropriate reference element 
is O, for the following reasons: 
(1) the dominant O yield comes from a single source 
    -- massive stars. Thus, the origin of O is well-understood; 
(2) O is the most abundant metal in the Universe; and 
(3) at the lowest metallicities, where we expect to 
    uncover the signature from early nucleosynthesis, 
    several O lines become unsaturated in DLAs. Thus, 
    it is relatively straightforward to measure [O/H].
Taking oxygen then as the reference element, 
we have constructed the typical abundance 
pattern for a VMP DLA by determining the 
mean $\langle \rm{X/O} \rangle$ ratio 
for each available element, X, 
and then referring this mean value to the adopted 
solar scale (i.e. we take the log of the mean, 
[$\langle$X/O$\rangle$]). 
These mean values are listed in 
Table~\ref{tab:typ_dla} along with the 
dispersion in the available measurements
($\sigma_{\rm [\langle X/O\rangle]}$). 
In the last column of Table~\ref{tab:typ_dla}, 
we indicate the total number of DLAs that 
were used in determining the mean X/O. 

\begin{figure}
  \centering
  \includegraphics[angle=0,width=80mm]{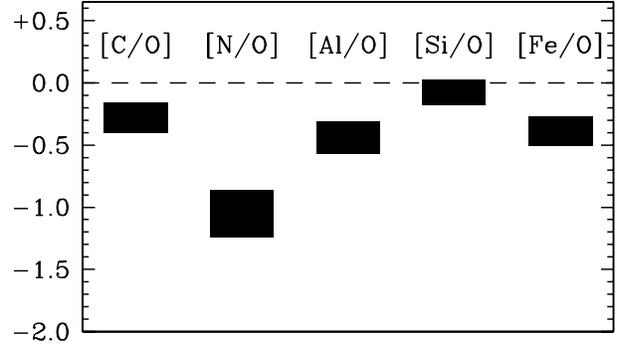}
  \caption{
  The abundance pattern of a typical VMP DLA is illustrated by the black boxes, 
  where the height in each box represents the dispersion in the population. 
  The dashed line corresponds to the solar abundance ratios. 
  }
  \label{fig:typ_dla}
\end{figure}

The corresponding abundance pattern is 
illustrated in Figure~\ref{fig:typ_dla}. 
In the following subsection, we investigate the 
most likely origin of the metals in VMP DLAs 
by directly comparing this typical abundance 
pattern to model yield calculations of 
both Population II and Population III stars. 
Before continuing, however, it is important to keep in mind 
that the  `typical'  [$\langle$N/O$\rangle$]
in Table~\ref{tab:typ_dla} may be biased high,
because it does not include a number of upper limits,
where the \NI\ lines are too weak to be measured.
As discussed above (Section~\ref{sec:co}), 
it is also possible that [$\langle$C/O$\rangle$] may be 
biased towards lower values than the true mean.

According to our sample definition 
(see Section~\ref{sec:lit_dlas}), 
this `typical' abundance pattern is 
based on DLAs where [Fe/H] $\le -2.0$. 
By restricting this sample to only 
contain DLAs with [Fe/H] $\le -2.5$, we found that 
the typical abundance for all elements considered 
here changes by no more than $0.02$ dex, except in 
the case of [$\langle$N/O$\rangle$], which is lower 
by $0.14$ dex. Thus, our choice of metallicity cut 
does not introduce a significant bias into the 
typical VMP DLA abundance pattern.

\subsection{Clues to early episodes of nucleosynthesis}

There is increasing evidence to suggest 
that the most metal-poor DLAs may 
retain the signature from the 
earliest episodes of star formation 
\citep{Ern06,Pet08,Coo11}. In this 
picture, the most metal-poor DLAs condensed 
directly out of material that was enriched by either:
(1) an external halo that distributed 
its products over large cosmological volumes 
via multi-SN events \citep{MadFerRee01}, or perhaps 
(2) just a few SNe from the halo in which 
the DLA now resides \citep{Bla11}.

Cosmological simulations of galaxy formation
support the possibility that such DLAs still retain 
the chemical signature of early enrichment \citep{Pon08,Tes09}; 
the most metal-poor DLAs 
arise in low mass halos that have undergone 
little to no \emph{in situ} star formation. 
It is possible, however, that VMP DLAs acquired 
some of their metals at later times from nearby 
sources that delivered metals into the IGM 
via galactic superwinds (see e.g. \citealt{OppDav08}), 
which may complicate the interpretation. Perhaps 
the most straightforward way to discriminate 
between these enrichment scenarios is to compare 
the model yields of both Population III and 
Population II stars with that of the 
typical VMP DLA described in 
Section~\ref{sec:typ_VMPDLA}. 

\begin{figure}
  \centering
  {\includegraphics[angle=0,width=80mm]{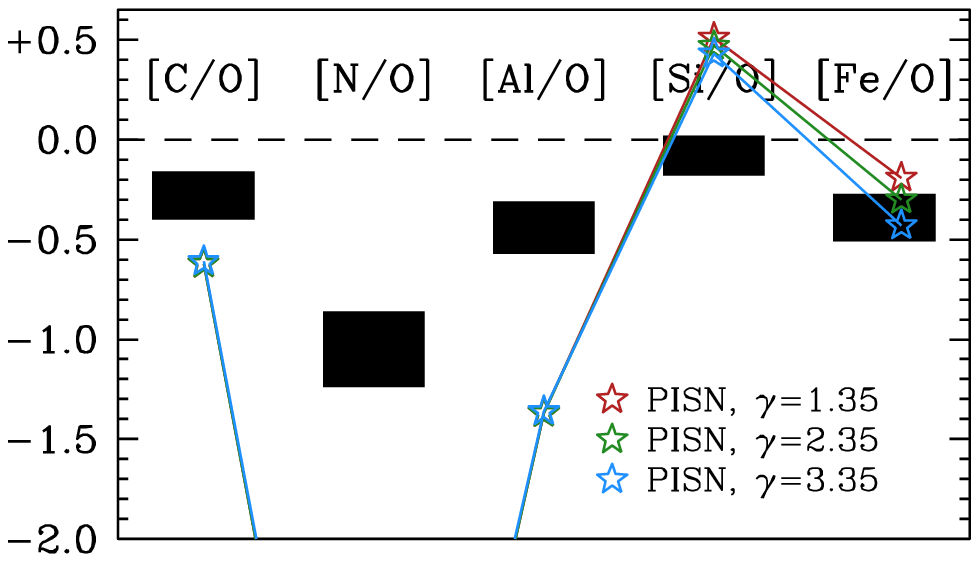}}\\
    {\vspace{0.4cm}}
  {\includegraphics[angle=0,width=80mm]{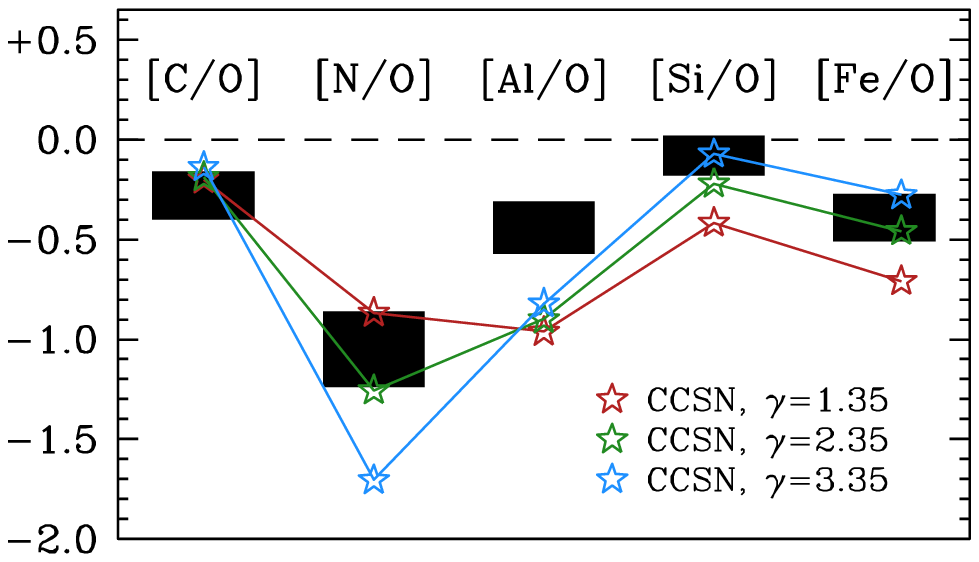}}\\
    {\vspace{0.4cm}}
  {\includegraphics[angle=0,width=80mm]{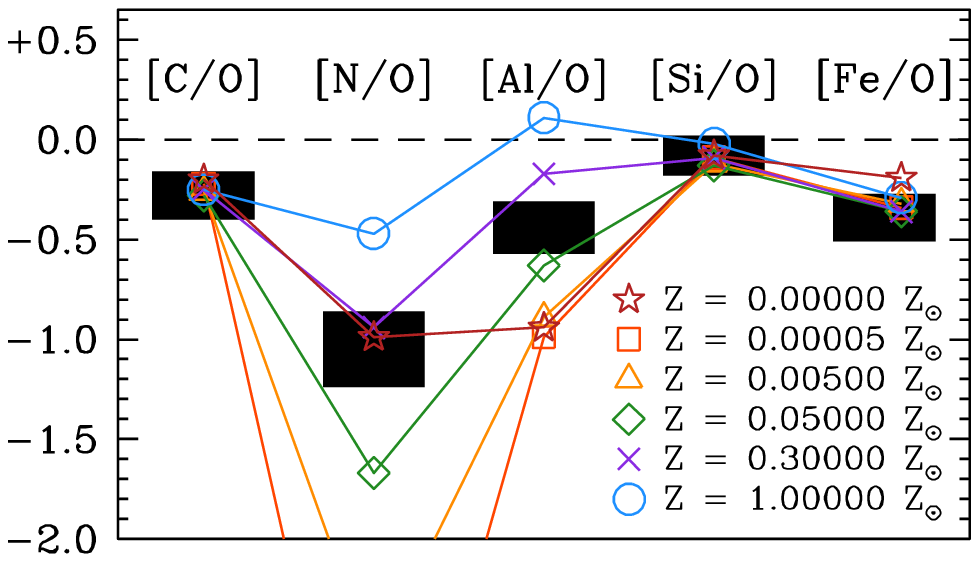}}
  \caption{
  The abundance pattern of the `typical' VMP DLA (black boxes; cf. Figure~\ref{fig:typ_dla}) 
  is compared to nucleosynthesis models of massive stars (symbols connected by lines). 
  \emph{Top panel:}
  The symbols represent the yields from pair-instability supernovae (PISN) of zero metallicity 
  stars \citep{HegWoo02} for three indices of a power law IMF 
  (red, green, blue corresponds to $\gamma=1.35,2.35,3.35$ respectively). 
  \emph{Middle panel:}
  Same as above, except for core-collapse supernovae (CCSN) models of zero metallicity stars \citep{HegWoo10}. 
  \emph{Bottom panel:}
  Comparing the typical DLA abundance pattern with the model explosive yields of massive Population II stars \citep{ChiLim04}. 
  A range of metallicities is considered, as indicated by the accompanying legend. 
  In all cases for this bottom panel, we plot the integrated yields 
  from the Salpeter IMF with a power law index of $\gamma=1.35$.
  }
  \label{fig:yield_models}
\end{figure}

We consider three sources that could be responsible 
for the metals in VMP DLAs:
(1) massive metal-free stars, with main sequence 
masses in the range $140-260$ M$_{\odot}$ that 
explode as pair-instability SNe (PISN; \citealt{HegWoo02}); 
(2) massive metal-free stars, with progenitor 
masses in the range $10-100$ M$_{\odot}$ that 
explode as core-collapse SNe (CCSN; \citealt{HegWoo10}); and 
(3) massive Population II (and I) stars, with progenitor 
masses in the range $13-35$ M$_{\odot}$, covering 
a range in metallicity \citep{ChiLim04}, also ending 
their lives as CCSN. To determine the dominant source 
of the metals in VMP DLAs, we integrate these model 
yields over a Salpeter-like power law IMF, 
d$N$/d$M \propto M^{-(1+\gamma)}$ (where $\gamma=1.35$ for a Salpeter IMF), 
and consider three values for the power law index in the 
case of zero metallicity ($\gamma=1.35,2.35,3.35$). 
For massive Population II stars, we consider only a 
Salpeter IMF, with $\gamma=1.35$.
The results of these calculations are shown in 
Figure~\ref{fig:yield_models}.

Let us first consider the yields from metal-free 
stars with masses in the range $140-260$ M$_{\odot}$. 
These stars explode as PISN, the physics of which is 
well-understood \citep{HegWoo02,UmeNom02}. 
The calculated yields from PISN are thus the least model 
dependent of the three cases considered here. 
Qualitatively, these models could 
have been ruled out on the basis of the near-solar 
[Si/O] that is typical of the VMP DLA population; 
PISN are expected to yield supersolar [Si, S, Ar, Ca/O]. 
Indeed, for the range of $\gamma$ considered here,
such SNe provide a poor fit to the typical VMP DLA population
(top panel of Figure~\ref{fig:yield_models}). 

We now turn to models of massive stars ($10-100$ M$_{\odot}$) 
that end their lives as CCSN.
Unfortunately, the explosion mechanism of CCSN is 
poorly understood, and several unknown physical 
effects need to be parameterized and 
suitably adjusted to find the best solution 
for a given set of data. In particular, one 
usually parameterizes the explosion energy, 
the degree of mixing between the stellar layers 
during the explosion, and the amount of material 
that falls back onto the central remnant. 

The most recent suite of published CCSN 
yields for metal-free stars are those by 
\citet{HegWoo10}. These computations 
provide a detailed account of the 
nucleosynthetic products over a large 
range of progenitor masses ($10-100$ M$_{\odot}$)
with a typical mass resolution of 0.1 M$_{\odot}$. 
As part of their study, \citet{HegWoo10} compiled 
a database of these model yields that are imported into their 
\textsc{starfit}\footnote{\textsc{starfit} is written with the \textsc{interactive data language} 
software and is available from:\\ http://homepages.spa.umn.edu/$\sim$alex/znuc/}
software. This software is designed to objectively 
sieve through the vast parameter space and select the 
explosion parameters that best fit the data.

To maintain consistency with the other 
yield models that are considered here, 
we compare the abundance pattern of the typical VMP DLA 
to the expected yields for the three power law indices 
of a Salpeter-like IMF ($\gamma=1.35,2.35,3.35$). 
We therefore froze the remaining parameters to the `standard' case, 
which corresponds to a constant explosion energy ($1.2\times10^{51}$ erg) 
for all masses in the range $10-100$ M$_{\odot}$ 
(see \citealt{HegWoo10} for further details). 
This standard case sets the piston location of the explosion 
to be at the base of the oxygen burning shell (where the 
entropy per baryon $\simeq4$), and applies a mixing 
boxcar filter with a width which is $10\%$ the He core size. 
The material that falls back onto the central remnant is 
not parameterized in this code, but is instead calculated 
by switching the piston off $100$\,s after the explosion 
and defining an inner boundary condition where material 
is accreted. 

All three fits are reproduced in the 
middle panel of Figure~\ref{fig:yield_models}. 
The `standard' case with the yields from metal-free
stars seems to produce a reasonable agreement with
the observed metal ratios in the `typical' VMP DLAs,
although Al is discrepant by $\sim 0.5$\, dex and,
interestingly, IMF slopes steeper than Salpeter seem to 
fit the data best.
Of course, 
by relaxing some of the default constraints it may be possible
to improve the fit further, but we have refrained from
doing so, given that there are still 
many uncertainties in accurately modelling the 
physics behind the explosion. Some of these 
uncertainties are only now beginning to be
addressed in some detail 
(see \citealt{Jog10b} and references therein).

Finally, we consider the set of model CCSN yields published 
by \citet{ChiLim04}, which allow us to test whether or 
not Population II stars can also account for the origin 
of the metals in VMP DLAs. Before comparing the models 
by Chieffi \&\ Limongi to the typical VMP DLA, it is worth noting the important 
differences between this code and the one described 
above by \citet{HegWoo10}. Aside from the obvious difference 
in metallicity, the models by \citet{ChiLim04} target the 
mass range $13-35$ M$_{\odot}$ with a relatively coarser 
mass resolution of 5 M$_{\odot}$. In addition, this code 
parameterizes the amount of material that falls back onto the 
central remnant. In their standard case, this prescription 
requires $0.1$ M$_{\odot}$ of $^{56}$Ni to be ejected from 
the star, and thus all material interior to this mass 
coordinate is `accreted' by the remnant. Finally, this code 
is yet to implement a `mixing parameter' to model the mixing 
that takes place between the stellar layers during the 
SN explosion. Adopting their standard case, which has an 
explosion energy of $1.2\times10^{51}$ erg, we integrate the 
model yields over a Salpeter-like IMF with $\gamma=1.35$. 
The results are shown in the bottom panel 
of Figure~\ref{fig:yield_models}. 

%

These calculations adopt an initial chemical 
composition that is simply scaled from the solar abundance pattern. 
According to \citet{ChiLim04}, the model yields for stars with 
an initial metallicity of Z $\leq 0.005$ Z$_{\odot}$ are not strongly 
dependent on the initial composition of the star. It is important to note, 
however, that by introducing an $\alpha$-enhancement to the 
initial metallicity (which is perhaps more realistic than simply scaling the 
solar abundance pattern), the yields for the odd atomic number elements 
are increased (see their Figure~1). Thus, our calculations
may underestimate the N/O and Al/O ratios.
Furthermore, at higher metallicities 
(Z $> 0.005$ Z$_{\odot}$), the yields  
\emph{do} depend on the initial composition of the stars. 

Inspecting the bottom panel of Figure~\ref{fig:yield_models}, 
it can been seen that, at face value, 
the Z/Z$_{\odot}$ $= 0.0, 0.05$ and $0.3$ 
models provide reasonable fits to the abundance pattern 
of a typical VMP DLA. We also note the broad 
agreement between the metal-free models by \citet{HegWoo10} 
and \citet{ChiLim04}. 
On the other hand, the Z/Z$_{\odot}$ $= 0.05$ and $0.3$ 
models also seem to provide reasonable fits to the abundance pattern. 
However, once the stars have reached metallicities greater
than 1/20 of solar, the metal ratios in the gas should be compared with
the predictions of full chemical evolution models which are
beyond the scope of this paper.

Given the current (largely) model-dependent 
nature of these calculations, we are unable 
to draw firm conclusions at this stage. 
Whilst the above models suggest that 
metal-free stars could have synthesised the metals that now 
reside in VMP DLAs, we cannot rule out the possibility 
that Population II stars are also responsible. 
We suspect that it will be necessary to measure the 
abundances of additional metals in order to better 
distinguish between Population II and Population III models,
since the ratios of the most abundant elements 
([C/O], [Si/O] and [Fe/O]) provide the weakest constraints 
on the nature of the objects that synthesised them. 
Some of the largest differences between these models 
are exhibited by the iron-peak elements, and in particular, 
Ti, Ni and Zn. The detection of these rarer elements in the 
most metal-poor DLAs will have to wait for the advent of the
next generation of extremely large optical telescopes. 

Finally, we note that the models used here to compare 
with a typical VMP DLA are still quite dependent on 
unknown physics; the single largest uncertainty in 
these models is the explosion mechanism. Additional physics 
also needs to be included, 
such as the mixing  induced by stellar rotation 
\citep{MeyEksMae06,Hir07,Mey10,Jog10a} and the Rayleigh-Taylor instability 
\citep{JogWooHeg09}. These effects will presumably be considered 
in the next generation of fine-grid nucleosynthesis models, 
when the limitations of computing power will hopefully be 
less of a concern. 

\subsection{Comparison with data of medium spectral resolution}
\label{sec:medres}

Finally,  we compare DLA abundance determinations obtained 
from high ($R\sim40000$) and medium ($R\sim5000$) 
spectral resolution data. Such a comparison is 
motivated by the realization that, even with 
efficient echelle spectrographs on 8--10\,m telescopes.
it is typically 
necessary to integrate on a single  
QSO for the equivalent of one night 
in order to obtain the S/N ratio required to 
measure elemental abundances from high 
resolution spectra. By settling for lower resolutions,
the exposure times are greatly reduced;
for example, most of the QSOs in the survey
by \citet{Pen10} were observed for about one hour
with the Echelle Spectrograph and Imager (ESI).
Clearly, it is of interest to test how similar 
the abundance measurements are for the 
most metal-poor DLAs, if we were to forgo the 
accuracy of high spectral resolution 
in order to secure a larger sample. 

\begin{table}
\centering
    \caption{\textsc{column densities estimated from high and medium spectral resolution spectra}}
    \begin{tabular}{@{}lccc}
    \hline
Ion
& $\log N_{\rm coo}^{\rm a}$
& $\log N_{\rm pen,m}^{\rm b}$
& $\log N_{\rm pen,c}^{\rm c}$\\
\hline
\multicolumn{4}{c}{\textbf{J0831$+$3358}}\\
\AlII &  $12.19\pm0.06$  &  $\ldots$        &  $>11.56$        \\
\SiII &  $13.75\pm0.04$  &  $\ldots$        &  $>13.03$        \\
\FeII &  $13.33\pm0.06$  &  $\ldots$        &  $>13.13$        \\
\multicolumn{4}{c}{}\\
\multicolumn{4}{c}{\textbf{J1001$+$0343}}\\
\CII  &  $13.58\pm0.02$  &  $13.63\pm0.06$  &  $13.76\pm0.13$  \\
\OI   &  $14.25\pm0.02$  &  $13.90\pm0.07$  &  $13.98\pm0.08$  \\
\SiII &  $12.86\pm0.01$  &  $12.70\pm0.05$  &  $12.82\pm0.12$  \\
\multicolumn{4}{c}{}\\
\multicolumn{4}{c}{\textbf{J1037$+$0139}}\\
\OI   &  $15.06\pm0.04$  &  $\ldots$        &  $>14.48$        \\
\AlII &  $12.32\pm0.03$  &  $\ldots$        &  $>12.09$        \\
\SiII &  $13.97\pm0.03$  &  $\ldots$        &  $>13.69$        \\
\FeII &  $13.53\pm0.02$  &  $\ldots$        &  $>13.33$        \\
\hline
    \end{tabular}
    \smallskip

\flushleft{$^{\rm a}${Our results.}\\ }
$^{\rm b}${Column densities measured by \citet{Pen10}.}\\
$^{\rm c}${Column density estimates by \citet{Pen10} after applying a saturation correction.}\\
\label{tab:pen_com}
\end{table}

In this context, there are two main concerns that potentially limit 
the accuracy of abundance measurements 
from medium (as opposed to high) 
spectral resolution data.
First, VMP DLAs typically have line widths 
$\simlt10$ km s$^{-1}$ \citep{Led06,Mur07,Pro08}, 
which are unresolved at $R = 5000$. One must 
therefore appeal to a curve-of-growth analysis
appropriate to 
a single absorbing cloud, which is
often an oversimplification
(see our Table~\ref{tab:cloud_models}, 
and also \citealt{Pro06}).
Second, the relevant absorption lines may be saturated, 
and the degree of saturation may be difficult 
to estimate correctly, even with a well-defined 
curve-of-growth. 

As it happens, three VMP DLAs from the present work
are in common with the lower resolution survey of
\citet{Pen10}, providing us with the means to compare
column densities from the two sets of spectra, as in
Table~\ref{tab:pen_com}. For two of the DLAs,
J0831$+$3358 and J1037$+$0139, 
\citet{Pen10} reported lower limits on the column densities 
of the available metal ions.
As can be seen from Table~\ref{tab:pen_com},
while these lower limits are always consistent
with the values measured from our echelle 
spectra, they fall short of the true column density 
by widely differing amounts, from as little as $-0.13$\,dex
to as much as $-0.72$\,dex. This wide range 
significantly reduces
the usefulness of the lower limits.

Turning now to the DLA at $z_{\rm abs}=3.07841$ towards J1001$+$0343, 
we recall that our UVES spectra show that the
absorption arises in a single component 
with Doppler parameter  $b=7.0\pm0.1$ km s$^{-1}$ 
(see Figure~\ref{fig:J1001p0343}). 
This value is not too dissimilar from $b=7.5$\,km~s$^{-1}$
estimated by \citet{Pen10} 
from a curve-of-growth analysis. 
Indeed, the two analyses give consistent estimates of 
the \SiII\ column density (after \citet{Pen10} apply a 
saturation correction). 
The \CII\ and \OI\ lines, however, tell a different story. 
For this DLA, \citet{Pen10} need not have applied a 
saturation correction to the \CII\,$\lambda1334$ line, 
since it is not strongly saturated 
(see Figure~\ref{fig:J1001p0343}). 
Indeed, prior to applying such a correction, their 
column density estimate was in broad agreement with that 
derived here. Conversely, \OI\,$\lambda1302$ is 
closer to saturation and, even with the correction
applied by  \citet{Pen10}, these authors' estimate 
falls short of the value deduced here by nearly a factor of two.
The combined effect is an overestimate of  [C/O]
by $0.45$\,dex. 

Based on this example, there appear to be non-negligible 
uncertainties in the derivation of element ratios 
from spectra at $R \sim 5000$.
While these uncertainties did not affect the principle 
goal of the study by \citet{Pen10} -- to uncover the most metal-poor 
DLAs -- it would appear that high resolution observations 
are indeed necessary to measure element abundances in 
VMP DLAs with an accuracy better than a factor of $\sim2$.

\section{Summary and Conclusions}
\label{sec:conc}
 
We have conducted a survey for very metal-poor 
DLAs to shed light on the earliest episodes of 
nucleosynthesis in our Universe. 
Our sample includes seven new 
DLAs observed with high resolution spectrographs 
($R \simgt 30000$); when combined with the 
five metal-poor DLAs previously reported from 
this programme \citep{Pet08,Coo11} and an additional 
ten DLAs from the literature, it constitutes the 
largest survey to date for 
DLAs with a metallicity [Fe/H] $<-2.0$. 
From the analysis of these data, 
we draw the following conclusions.\\

\noindent ~~(i) Having now doubled the sample of DLAs 
where the C/O ratio is measured from unsaturated 
absorption lines, we confirm that DLAs exhibit 
near-solar values of C/O at the lowest metallicities 
probed. Furthermore, we find good agreement in the 
C/O ratio observed in our sample of DLAs and in recent 
compilations of the most metal-poor Galactic halo stars. 
We argue that such good agreement points to a universal 
origin for the C/O `excess' in this regime.

\smallskip

\noindent ~~(ii) For the first time, we investigate the 
[O/Fe] ratio in very metal-poor DLAs. For 
20 DLAs with [Fe/H] $<-2.0$, we find a small dispersion
around a mean value  
[$\langle$O/Fe$\rangle$]\,$=+0.39\pm0.12$. 
We have also presented tentative evidence for a rise 
in the [O/Fe] ratio when [Fe/H] $\simlt-3.0$.

\smallskip

\noindent ~~(iii) In view of the long-standing debate 
as to the behaviour of the [O/Fe] ratio in metal-poor 
Galactic halo stars, we have compared the stellar 
trend to that observed in our sample of DLAs. 
We find good agreement between stars and DLAs when 
the stellar oxygen abundance is measured from the 
[\OI]\,$\lambda6300$ line (after correcting for 3D 
effects). Based on the available DLA samples, we conclude 
that [O/Fe] is essentially flat in the metallicity 
interval $-3.0 \simlt$\,[Fe/H]\,$\simlt -1.0$, with the possibility of
an increase at yet lower metallicities.

\smallskip

\noindent  ~~(iv) We have constructed the abundance pattern 
of a typical very metal-poor DLA for the five most commonly observed 
metals, using O as a reference. We find that Si/O is just 
below solar ([$\langle$Si/O$\rangle$] $= -0.08$), 
whilst [$\langle$C/O$\rangle$] $= -0.28$ and 
[$\langle$Fe/O$\rangle$] $= -0.39$.
The largest deviations from a solar scaled abundance pattern 
are exhibited by N and Al, with 
[$\langle$N,Al/O$\rangle$] $= -1.05, -0.44$. 

\smallskip
  
\noindent ~~(v) One of the main aims of this work was to 
investigate the origin of the metals in the most metal-poor 
DLAs. To achieve this goal, we compared the  
abundance pattern of a `typical' VMP  DLA with those
expected from 
model calculations using the yields of Population II and Population III 
stars. For the few elements considered here, 
we find a reasonable agreement between the abundance pattern 
of the typical VMP DLA and the `standard model' of a population of 
metal-free stars (i.e. a top-heavy initial mass function where 
all stars explode as core-collapse supernovae with an 
energy of $1.2\times10^{51}$ erg). However, given that we only 
have access to a handful of metals, we cannot unambiguously 
rule out (an additional contribution from) more metal-rich 
Population II stars. On the other hand, we are able to firmly 
conclude that the typical very metal-poor DLA was not solely 
enriched by pair-instability supernovae from very massive
metal-free stars.

Our ongoing programme to measure the abundances in the most 
metal-poor DLAs complements local studies of Galactic metal-poor 
halo stars. The good agreement we have found 
between these two populations suggests a universal origin for 
their metals. The results presented here
emphasize the importance of 
measuring elemental abundances in the most metal-poor DLAs; 
these systems present us with a unique window of opportunity 
to probe the nucleosynthesis by some of the earliest structures 
in the Universe. 

\section*{Acknowledgements}
We are grateful to the relevant time assignment committees for
their continuing support of this demanding observational
programme, and to the staff astronomers at the VLT and Keck  
Observatories for their competent assistance with the observations.
We also thank an anonymous referee who provided valuable 
comments that improved the presentation of this work. 
Tom Barlow and Michael Murphy generously shared their
echelle data reduction software.
Valuable advice and help with various 
aspects of the work described in this paper was provided by Bob Carswell, 
Paul Hewett, and Regina Jorgenson.
We thank the Hawaiian
people for the opportunity to observe from Mauna Kea;
without their hospitality, this work would not have been possible.
RC is jointly funded by the Cambridge Overseas 
Trust and the Cambridge Commonwealth/Australia Trust 
with an Allen Cambridge Australia Trust Scholarship.
CCS's research is partly supported by grants
AST-0606912 and AST-0908805 from the US National Science Foundation.


\begin{appendix}

\section{VMP DLA column densities}
\label {app:1}

To facilitate comparison with future data sets,
Table~\ref{tab:mp_dlas} lists ion column densities
for the DLAs considered in this study. The corresponding
element abundances are collected in Table~\ref{tab:mp_dlas_abund}.

\begin{table*}
\centering
\begin{minipage}[c]{0.99\textwidth}
    \caption{\textsc{C, N, O, Al, Si, and Fe column densities in VMP DLAs}}
    \begin{tabular}{@{}lrcccccccr}
    \hline
    \hline
   \multicolumn{1}{c}{QSO}
& \multicolumn{1}{c}{$z_{\rm abs}$} 
& \multicolumn{1}{c}{$\log N$\/(H\,{\sc i})}
& \multicolumn{1}{c}{$\log N$\/(C\,{\sc ii})}
& \multicolumn{1}{c}{$\log N$\/(N\,{\sc i})}
& \multicolumn{1}{c}{$\log N$\/(O\,{\sc i})}
& \multicolumn{1}{c}{$\log N$\/(Al\,{\sc ii})}
& \multicolumn{1}{c}{$\log N$\/(Si\,{\sc ii})}
& \multicolumn{1}{c}{$\log N$\/(Fe\,{\sc ii})}
& \multicolumn{1}{c}{Ref.$^a$}\\
    \multicolumn{1}{c}{}
& \multicolumn{1}{c}{}
& \multicolumn{1}{c}{(cm$^{-2}$)}
& \multicolumn{1}{c}{(cm$^{-2}$)}
& \multicolumn{1}{c}{(cm$^{-2}$)}
& \multicolumn{1}{c}{(cm$^{-2}$)}
& \multicolumn{1}{c}{(cm$^{-2}$)}
& \multicolumn{1}{c}{(cm$^{-2}$)}
& \multicolumn{1}{c}{(cm$^{-2}$)}
& \multicolumn{1}{c}{}\\    
  \hline
\multicolumn{10}{l}{\textbf{Our Metal-Poor DLA Sample}}\\
J0035$-$0918    & 2.34010  & $20.55\pm0.10$ & $15.47\pm0.15$ & $13.51\pm0.06$  & $14.96\pm0.08$  & $11.73\pm0.05$  & $13.41\pm0.04$  & $12.98\pm0.07$  &  2     \\
J0311$-$1722    & 3.73400  & $20.30\pm0.06$ & $14.02\pm0.08$ & $\le13.07$      & $14.70\pm0.08$  & \ldots          & $13.31\pm0.07$  & $\le13.76$      &  1     \\
J0831$+$3358    & 2.30364  & $20.25\pm0.15$ & \ldots         & $\le12.78$      & $14.93\pm0.05$  & $12.19\pm0.06$  & $13.75\pm0.04$  & $13.33\pm0.06$  &  1,4   \\
Q0913$+$072     & 2.61843  & $20.34\pm0.04$ & $13.98\pm0.05$ & $12.29\pm0.12$  & $14.63\pm0.01$  & $11.78\pm0.03$  & $13.30\pm0.01$  & $12.99\pm0.01$  &  3     \\
J1001$+$0343    & 3.07841  & $20.21\pm0.05$ & $13.58\pm0.02$ & $\le12.50$      & $14.25\pm0.02$  & \ldots          & $12.86\pm0.01$  & $12.50\pm0.14$  &  1     \\
J1016$+$4040    & 2.81633  & $19.90\pm0.11$ & $13.66\pm0.04$ & $\le12.76$      & $14.13\pm0.03$  & \ldots          & $12.90\pm0.05$  & \ldots          &  3     \\
J1037$+$0139    & 2.70487  & $20.50\pm0.08$ & \ldots         & $13.27\pm0.04$  & $15.06\pm0.04$  & $12.32\pm0.03$  & $13.97\pm0.03$  & $13.53\pm0.02$  &  1     \\
J1340$+$1106    & 2.50792  & $20.09\pm0.05$ & \ldots         & $12.80\pm0.04$  & $15.02\pm0.03$  & $12.27\pm0.02$  & $13.75\pm0.02$  & $13.49\pm0.02$  &  1     \\
J1340$+$1106    & 2.79583  & $21.00\pm0.06$ & \ldots         & $14.04\pm0.02$  & $16.04\pm0.04$  & $13.24\pm0.03$  & $14.68\pm0.02$  & $14.32\pm0.01$  &  1     \\
J1419$+$0829    & 3.04973  & $20.40\pm0.03$ & \ldots         & $13.28\pm0.02$  & $15.17\pm0.02$  & \ldots          & $13.83\pm0.01$  & $13.54\pm0.03$  &  1     \\
J1558$+$4053    & 2.55332  & $20.30\pm0.04$ & $14.22\pm0.06$ & $12.66\pm0.07$  & $14.54\pm0.04$  & $11.92\pm0.06$  & $13.32\pm0.02$  & $13.07\pm0.06$  &  3     \\
Q2206$-$199     & 2.07624  & $20.43\pm0.04$ & $14.41\pm0.03$ & $12.79\pm0.05$  & $15.05\pm0.03$  & $12.18\pm0.01$  & $13.65\pm0.01$  & $13.33\pm0.01$  &  3     \\
    &          &       &          &                 &          &          &             &        &        \\
\multicolumn{10}{l}{\textbf{Literature DLAs}}\\
Q0000$-$2620    & 3.39012  & $21.41\pm0.08$ & \ldots         & $14.70\pm0.02$  & $16.42\pm0.10$  & \ldots          & $15.06\pm0.02$  & $14.87\pm0.03$  &  5     \\
Q0112$-$306     & 2.41844  & $20.50\pm0.08$ & \ldots         & $13.16\pm0.04$  & $14.95\pm0.08$  & \ldots          & $13.62\pm0.02$  & $13.33\pm0.05$  &  6     \\
J0140$-$0839    & 3.69660  & $20.75\pm0.15$ & $14.13\pm0.08$ & $\le12.38$      & $14.69\pm0.01$  & $11.82\pm0.04$  & $13.51\pm0.09$  & $12.77\pm0.19^{\rm b}$&  7     \\
J0307$-$4945    & 4.46658  & $20.67\pm0.09$ & \ldots         & $13.57\pm0.12$  & $15.91\pm0.17$  & $13.36\pm0.06$  & $14.68\pm0.07$  & $14.21\pm0.17$  &  8     \\
Q1108$-$077     & 3.60767  & $20.37\pm0.07$ & \ldots         & $\le12.84$      & $15.37\pm0.03$  & \ldots          & $14.34\pm0.02$  & $13.88\pm0.02$  &  6     \\
J1337$+$3153    & 3.16768  & $20.41\pm0.15$ & $13.98\pm0.06$ & $\le12.80$      & $14.43\pm0.09$  & $12.00\pm0.05$  & $13.24\pm0.05$  & $13.14\pm0.26$  &  9     \\
J1558$-$0031    & 2.70262  & $20.67\pm0.05$ & \ldots         & $14.46\m$       & $15.86\m$       & \ldots          & $14.24\m$       & $14.11\m$       &  10    \\
Q1946$+$7658    & 2.84430  & $20.27\pm0.06$ & \ldots         & $12.59\pm0.04$  & $14.82\pm0.01$  & \ldots          & $13.60\pm0.01$  & $13.24\pm0.01$  &  11    \\
Q2059$-$360     & 3.08293  & $20.98\pm0.08$ & \ldots         & $13.95\pm0.02$  & $16.09\pm0.04$  & \ldots          & $14.86\pm0.05$  & $14.48\pm0.02$  &  6     \\ 
J2155$+$1358    & 4.21244  & $19.61\pm0.10$ & $13.95\pm0.06$ & \ldots          & $14.50\pm0.05$  & $11.92\pm0.17$  & $13.25\pm0.04$  & $12.93\pm0.23$  &  12    \\
    \hline
    \end{tabular}
    \smallskip

$^{\rm a}${References---1: This work;
2:  \citet{Coo11};
3:  \citet{Pet08};
4:  \citet{Pen10};
5:  \citet{Mol00}; 
6:  \citet{PetLedSri08};
7:  \citet{Ell10};
8:  \citet{Des01};
9: \citet{Sri10};
10: \citet{OMe06};
11: \citet{Pro02};
12: \citet{Des03}.
}\\
$^{\rm b}${\citet{Ell10} quote a $3\sigma$ upper limit to the \FeII\ column density of 
$\log N$\/(Fe\,{\sc ii})/cm$^{-2} < 12.73$. We have since rereduced these data 
(as described in Section~\ref{sec:obs}), and detected the \FeII\,$\lambda1608$ 
line at the $4\sigma$ level. The column density for \FeII\ quoted here is derived 
using the optically thin limit approximation.}
\\
$\m${An error estimate for this measurement was not provided by the authors.}
\\
\label{tab:mp_dlas}
\end{minipage}
\end{table*}

\section{Oxygen and Iron in DLAs and Stars}
\label {app:2}

In Tables~\ref{tab:app_ofe_dlas} and
\ref{tab:app_ofe_stars}, we list the values of  
[O/Fe] and [Fe/H] plotted in Figure~\ref{fig:ofe}
for DLAs and Galactic stars respectively, together
with the corresponding errors.
The stellar measurements are all from  
the [\OI]\,$\lambda6300$ line 
(taken from \citealt{Nis02,Cay04,Gar06}), 
with appropriate 3D corrections 
based on the work by 
\citet{Nis02} and \citet{ColAspTra07};
see text in Section~\ref{sec:ofe} for further details.

\begin{table*}
\centering
\begin{minipage}[c]{0.99\textwidth}
    \caption{\textsc{[Fe/H] and [O/Fe]  in VMP DLAs}}
    \begin{tabular}{@{}lccccr}
    \hline
    \hline
   \multicolumn{1}{c}{QSO Name}
& \multicolumn{1}{c}{$z_{\rm abs}$} 
& \multicolumn{1}{c}{$\log N$\/(H\,{\sc i})}
& \multicolumn{1}{c}{[Fe/H]}
& \multicolumn{1}{c}{[O/Fe]}
& \multicolumn{1}{c}{Ref.$^a$}\\
    \multicolumn{1}{c}{}
& \multicolumn{1}{c}{}
& \multicolumn{1}{c}{(cm$^{-2}$)}
& \multicolumn{1}{c}{}
& \multicolumn{1}{c}{}
& \multicolumn{1}{c}{}\\    
  \hline
Q0000$-$2620    & 3.39012  & $21.41\pm0.08$ & $-2.01\pm0.09$ & $+0.33\pm0.10$  &  2    \\
J0035$-$0918    & 2.34010  & $20.55\pm0.10$ & $-3.04\pm0.12$ & $+0.76\pm0.11$  &  3    \\
Q0112$-$306     & 2.41844  & $20.50\pm0.08$ & $-2.64\pm0.09$ & $+0.40\pm0.09$  &  4    \\
J0140$-$0839    & 3.69660  & $20.75\pm0.15$ & $-3.45\pm0.24$ & $+0.70\pm0.19$  &  5    \\
J0307$-$4945    & 4.46658  & $20.67\pm0.09$ & $-1.93\pm0.19$ & $+0.48\pm0.24$  &  6    \\
J0311$-$1722    & 3.73400  & $20.30\pm0.06$ & $\le-2.03$     & $\ge-0.26$      &  1    \\
J0831$+$3358    & 2.30364  & $20.25\pm0.15$ & $-2.39\pm0.16$ & $+0.38\pm0.08$  &  1,7  \\
Q0913$+$072     & 2.61843  & $20.34\pm0.04$ & $-2.82\pm0.04$ & $+0.42\pm0.02$  &  8    \\
J1001$+$0343    & 3.07841  & $20.21\pm0.05$ & $-3.18\pm0.15$ & $+0.53\pm0.14$  &  1    \\
J1037$+$0139    & 2.70487  & $20.50\pm0.08$ & $-2.44\pm0.08$ & $+0.31\pm0.04$  &  1    \\
Q1108$-$077     & 3.60767  & $20.37\pm0.07$ & $-1.96\pm0.07$ & $+0.27\pm0.04$  &  4    \\
J1337$+$3153    & 3.16768  & $20.41\pm0.15$ & $-2.74\pm0.30$ & $+0.07\pm0.28$  &  9    \\
J1340$+$1106    & 2.50792  & $20.09\pm0.05$ & $-2.07\pm0.05$ & $+0.31\pm0.04$  &  1    \\
J1340$+$1106    & 2.79583  & $21.00\pm0.06$ & $-2.15\pm0.06$ & $+0.50\pm0.04$  &  1    \\
J1419$+$0829    & 3.04973  & $20.40\pm0.03$ & $-2.33\pm0.04$ & $+0.41\pm0.04$  &  1    \\
J1558$-$0031    & 2.70262  & $20.67\pm0.05$ & $-2.03^{\rm b}$ & $+0.53^{\rm b}$  &  10   \\
J1558$+$4053    & 2.55332  & $20.30\pm0.04$ & $-2.70\pm0.07$ & $+0.25\pm0.07$  &  8    \\
Q1946$+$7658    & 2.84430  & $20.27\pm0.06$ & $-2.50\pm0.06$ & $+0.36\pm0.02$  &  11   \\
Q2059$-$360     & 3.08293  & $20.98\pm0.08$ & $-1.97\pm0.08$ & $+0.39\pm0.04$  &  4    \\ 
J2155$+$1358    & 4.21244  & $19.61\pm0.10$ & $-2.15\pm0.25$ & $+0.35\pm0.24$  &  12   \\
Q2206$-$199     & 2.07624  & $20.43\pm0.04$ & $-2.57\pm0.04$ & $+0.50\pm0.03$  &  8    \\
    \hline
    \end{tabular}
    \smallskip

$^{\rm a}${References---1: This work;
2:  \citet{Mol00}; 
3:  \citet{Coo11};
4:  \citet{PetLedSri08};
5:  \citet{Ell10};
6:  \citet{Des01};
7:  \citet{Pen10};
8:  \citet{Pet08};
9: \citet{Sri10};
10: \citet{OMe06};
11: \citet{Pro02};
12: \citet{Des03}.
}\\
$^{\rm b}${An error estimate for this measurement was not provided by the authors.}
\\
\label{tab:app_ofe_dlas}
\end{minipage}
\end{table*}

\begin{table*}
\centering
\begin{minipage}[c]{0.99\textwidth}
    \caption{\textsc{[Fe/H] and [O/Fe]  in metal-poor stars}}
    \begin{tabular}{@{}lccccr}
    \hline
    \hline
   \multicolumn{1}{c}{Star Name}
& \multicolumn{1}{c}{$T_{\rm eff}$\,$^{\rm a}$} 
& \multicolumn{1}{c}{$\log g$\,$^{\rm b}$}
& \multicolumn{1}{c}{[Fe/H]}
& \multicolumn{1}{c}{[O/Fe]}
& \multicolumn{1}{c}{Ref.$^c$}\\
    \multicolumn{1}{c}{}
& \multicolumn{1}{c}{(K)}
& \multicolumn{1}{c}{(cgs)}
& \multicolumn{1}{c}{}
& \multicolumn{1}{c}{}
& \multicolumn{1}{c}{}\\    
  \hline
HD\,$2796$      &  4950    &  1.50          & $-2.37\pm0.10$ & $+0.23\pm0.13$  &  1  \\
HD\,$3567$      &  6000    &  4.07          & $-1.11\pm0.06$ & $+0.29\pm0.06$  &  2  \\
HD\,$4306$      &  4990    &  3.04          & $-2.24\pm0.10$ & $+0.47\pm0.11$  &  3  \\
HD\,$26169$     &  4972    &  2.49          & $-2.19\pm0.10$ & $+0.34\pm0.09$  &  3  \\
HD\,$27928$     &  5044    &  2.67          & $-2.05\pm0.10$ & $+0.22\pm0.11$  &  3  \\
HD\,$45282$     &  5352    &  3.15          & $-1.46\pm0.10$ & $+0.33\pm0.07$  &  3  \\
HD\,$97320$     &  5976    &  4.16          & $-1.16\pm0.06$ & $+0.20\pm0.12$  &  2  \\
HD\,$108317$    &  5300    &  2.76          & $-2.16\pm0.10$ & $+0.49\pm0.12$  &  3  \\
HD\,$111980$    &  5694    &  3.99          & $-1.03\pm0.06$ & $+0.17\pm0.07$  &  2  \\
HD\,$122563$    &  4600    &  1.10          & $-2.71\pm0.10$ & $+0.31\pm0.13$  &  1  \\
HD\,$126587$    &  4712    &  1.66          & $-2.76\pm0.10$ & $+0.13\pm0.14$  &  3  \\
HD\,$126681$    &  5524    &  4.48          & $-1.12\pm0.06$ & $+0.33\pm0.08$  &  2  \\
HD\,$128279$    &  5336    &  2.95          & $-2.10\pm0.10$ & $+0.17\pm0.23$  &  3  \\
HD\,$132475$    &  5818    &  3.95          & $-1.39\pm0.06$ & $+0.33\pm0.09$  &  2  \\
HD\,$140283$    &  5690    &  3.69          & $-2.32\pm0.06$ & $+0.50\pm0.19$  &  2  \\
HD\,$160617$    &  5931    &  3.77          & $-1.72\pm0.06$ & $+0.22\pm0.15$  &  2  \\
HD\,$166913$    &  6039    &  4.11          & $-1.50\pm0.06$ & $+0.22\pm0.24$  &  2  \\
HD\,$186478$    &  4700    &  1.30          & $-2.49\pm0.10$ & $+0.47\pm0.10$  &  1  \\
HD\,$189558$    &  5613    &  3.91          & $-1.07\pm0.06$ & $+0.31\pm0.05$  &  2  \\
HD\,$205650$    &  5733    &  4.39          & $-1.12\pm0.06$ & $+0.30\pm0.09$  &  2  \\
HD\,$213657$    &  6114    &  3.85          & $-1.86\pm0.06$ & $+0.30\pm0.24$  &  2  \\
HD\,$218857$    &  5015    &  2.78          & $-1.72\pm0.10$ & $+0.18\pm0.09$  &  3  \\
HD\,$274939$    &  5090    &  2.79          & $-1.43\pm0.10$ & $+0.35\pm0.05$  &  3  \\
HD\,$298986$    &  6071    &  4.21          & $-1.30\pm0.06$ & $+0.23\pm0.17$  &  2  \\
BD\,$-18\decdeg5550$  &  4750    &  1.40    & $-2.94\pm0.10$ & $+0.08\pm0.24$  &  1  \\
BD\,$-01\decdeg2582$  &  5072    &  2.92    & $-2.03\pm0.10$ & $+0.31\pm0.11$  &  3  \\
BD\,$+17\decdeg3248$  &  5250    &  1.40    & $-1.99\pm0.10$ & $+0.46\pm0.11$  &  1  \\
BD\,$+23\decdeg3130$  &  5170    &  3.00    & $-2.29\pm0.06$ & $+0.38\pm0.15$  &  2  \\
CS\,$22186-035$ &  4900    &  1.50          & $-2.88\pm0.10$ & $+0.26\pm0.26$  &  1  \\
CS\,$22873-055$ &  4550    &  0.70          & $-2.87\pm0.10$ & $+0.19\pm0.13$  &  1  \\
CS\,$22891-209$ &  4700    &  1.00          & $-3.16\pm0.10$ & $+0.41\pm0.16$  &  1  \\
CS\,$22892-052$ &  4850    &  1.60          & $-2.91\pm0.10$ & $+0.14\pm0.26$  &  1  \\
CS\,$22896-154$ &  5250    &  2.70          & $-2.58\pm0.10$ & $+0.64\pm0.23$  &  1  \\
CS\,$22948-066$ &  5100    &  1.80          & $-3.01\pm0.10$ & $+0.54\pm0.23$  &  1  \\
CS\,$22949-037$ &  4900    &  1.50          & $-3.81\pm0.10$ & $+1.54\pm0.13$  &  1  \\
CS\,$22953-003$ &  5100    &  2.30          & $-2.73\pm0.10$ & $+0.44\pm0.23$  &  1  \\
CS\,$22966-057$ &  5300    &  2.20          & $-2.52\pm0.10$ & $+0.70\pm0.20$  &  1  \\
CS\,$22968-014$ &  4850    &  1.70          & $-3.42\pm0.10$ & $+0.51\pm0.27$  &  1  \\
CS\,$29491-053$ &  4700    &  1.30          & $-2.92\pm0.10$ & $+0.43\pm0.17$  &  1  \\
CS\,$29495-041$ &  4800    &  1.50          & $-2.71\pm0.10$ & $+0.37\pm0.13$  &  1  \\
CS\,$29516-024$ &  4650    &  1.20          & $-2.94\pm0.10$ & $+0.28\pm0.20$  &  1  \\
CS\,$29518-051$ &  5200    &  2.60          & $-2.58\pm0.10$ & $+0.63\pm0.23$  &  1  \\
CS\,$30325-094$ &  4950    &  2.00          & $-3.17\pm0.10$ & $+0.36\pm0.33$  &  1  \\
CS\,$31082-001$ &  4825    &  1.50          & $-2.79\pm0.10$ & $+0.28\pm0.15$  &  1  \\
    \hline
    \end{tabular}
    \smallskip

$^{\rm a}${Stellar effective temperature.}
\\
$^{\rm b}${Surface gravity.}
\\
$^{\rm c}${References---1: \citet{Cay04};
2: \citet{Nis02};
3: \citet{Gar06}.
}\\
\label{tab:app_ofe_stars}
\end{minipage}
\end{table*}

\end{appendix}

\label{lastpage}


\begin{thebibliography}{99}

\bibitem[\protect\citeauthoryear{Akerman et al.}{2004}]{Ake04}
Akerman C.~J., Carigi L., Nissen P.~E., Pettini M., Asplund M., 2004, A\&A, 414, 931

\bibitem[\protect\citeauthoryear{Akerman et al.}{2005}]{Ake05}
Akerman C.~J., Ellison S.~L., Pettini M., Steidel C.~C., 2005, A\&A, 440, 499

\bibitem[\protect\citeauthoryear{Aoki et al.}{2007}]{Aok07}
Aoki W., Beers T.~C., Christlieb N., Norris J.~E., Ryan S.~G., Tsangarides S., 2007, ApJ, 655, 492

\bibitem[\protect\citeauthoryear{Asplund}{2005}]{Asp05}
Asplund M., 2005, ARA\&A, 43, 481

\bibitem[\protect\citeauthoryear{Asplund et al.}{2009}]{Asp09}
Asplund M., Grevesse N., Sauval A.~J., Scott P., 2009, ARA\&A, 47, 481

\bibitem[\protect\citeauthoryear{Becker et al.}{2006}]{Bec06} 
Becker G.~D., Sargent W.~L.~W., Rauch M., Simcoe R.~A., 2006, ApJ, 640, 69

\bibitem[\protect\citeauthoryear{Becker et al.}{2011}]{Bec11}
Becker G.~D., Sargent W.~L.~W., Rauch M., Calverley A.~P., 2011, arXiv, arXiv:1101.4399

\bibitem[\protect\citeauthoryear{Beers \& Christlieb}{2005}]{BeeChr05}
Beers T.~C., Christlieb N., 2005, ARA\&A, 43, 531

\bibitem[\protect\citeauthoryear{Bensby \& Feltzing}{2006}]{BenFel06}
Bensby T., Feltzing S., 2006, MNRAS, 367, 1181

\bibitem[\protect\citeauthoryear{Bland-Hawthorn et al.}{2011}]{Bla11}
Bland-Hawthorn J., et al., 2011, ApJ, in press

\bibitem[\protect\citeauthoryear{Bromm \& Larson}{2004}]{BroLar04}
Bromm V., Larson R.~B., 2004, ARA\&A, 42, 79

\bibitem[\protect\citeauthoryear{Bromm \& Loeb}{2003}]{BroLoe03}
Bromm V., Loeb A., 2003, Nature, 425, 812

\bibitem[\protect\citeauthoryear{Carollo et al.}{2011}]{Car11}
Carollo D., Beers T.~C., Bovy J., Sivarani T., Norris J.~E., Freeman K.~C., Aoki W., Lee Y.~S., 2011, arXiv, arXiv:1103.3067

\bibitem[\protect\citeauthoryear{Cayrel et al.}{2004}]{Cay04}
Cayrel R., et al., 2004, A\&A, 416, 1117

\bibitem[\protect\citeauthoryear{Chiappini et al.}{2006}]{Chi06}
Chiappini C., Hirschi R., Meynet G., Ekstr{\"o}m S., Maeder A., Matteucci F., 2006, A\&A, 449, L27

\bibitem[\protect\citeauthoryear{Chieffi \& Limongi}{2002}]{ChiLim02}
Chieffi A., Limongi M., 2002, ApJ, 577, 281

\bibitem[\protect\citeauthoryear{Chieffi \& Limongi}{2004}]{ChiLim04}
Chieffi A., Limongi M., 2004, ApJ, 608, 405

\bibitem[\protect\citeauthoryear{Christlieb et al.}{2002}]{Chr02}
Christlieb N., et al., 2002, Nature, 419, 904 

\bibitem[\protect\citeauthoryear{Collet, Asplund, \& Trampedach}{2007}]{ColAspTra07}
Collet R., Asplund M., Trampedach R., 2007, A\&A, 469, 687

\bibitem[\protect\citeauthoryear{Cooke et al.}{2011}]{Coo11}
Cooke R., Pettini M., Steidel C.~C., Rudie G.~C., Jorgenson R.~A., 2011, MNRAS, 412, 1047

\bibitem[\protect\citeauthoryear{Dekker et al.}{2000}]{Dek00}
Dekker H., D'Odorico S., Kaufer A., Delabre B., Kotzlowski H., 2000, SPIE, 4008, 534

\bibitem[\protect\citeauthoryear{Dessauges-Zavadsky et al.}{2001}]{Des01}
Dessauges-Zavadsky M., D'Odorico S., McMahon R.~G., Molaro P., Ledoux C., P{\'e}roux C., Storrie-Lombardi L.~J., 2001, A\&A, 370, 426

\bibitem[\protect\citeauthoryear{Dessauges-Zavadsky et al.}{2003}]{Des03}
Dessauges-Zavadsky M., P{\'e}roux C., Kim T.-S., D'Odorico S., McMahon R.~G., 2003, MNRAS, 345, 447

\bibitem[\protect\citeauthoryear{Drawin}{1969}]{Dra69}
Drawin, H. W. 1969, Z. Phys., 225, 483

\bibitem[\protect\citeauthoryear{Ellison et al.}{2010}]{Ell10}
Ellison S.~L., Prochaska J.~X., Hennawi J., Lopez S., Usher C., Wolfe A.~M., Russell D.~M., Benn C.~R., 2010, MNRAS, 406, 1435

\bibitem[\protect\citeauthoryear{Erni et al.}{2006}]{Ern06}
Erni, P., Richter, P., Ledoux, C., \& Petitjean, P.\ 2006, A\&A, 451, 19 

\bibitem[\protect\citeauthoryear{Fabbian et al.}{2009a}]{Fab09a}
Fabbian D., Nissen P.~E., Asplund M., Pettini M., Akerman C., 2009a, A\&A, 500, 1143

\bibitem[\protect\citeauthoryear{Fabbian et al.}{2009b}]{Fab09b}
Fabbian D., Asplund M., Barklem P.~S., Carlsson M., Kiselman D., 2009b, A\&A, 500, 1221

\bibitem[\protect\citeauthoryear{Ferland et al.}{1998}]{Fer98}
Ferland G.~J., Korista K.~T., Verner D.~A., Ferguson J.~W., Kingdon J.~B., Verner E.~M., 1998, PASP, 110, 761

\bibitem[\protect\citeauthoryear{Field \& Steigman}{1971}]{FieSte71}
Field G.~B., Steigman G., 1971, ApJ, 166, 59

\bibitem[\protect\citeauthoryear{Frebel et al.}{2005}]{Fre05}
Frebel A., et al., 2005, Nature, 434, 871

\bibitem[\protect\citeauthoryear{Frebel, Johnson, \& Bromm}{2007}]{FreJohBro07}
Frebel A., Johnson J.~L., Bromm V., 2007, MNRAS, 380, L40

\bibitem[\protect\citeauthoryear{Frebel}{2010}]{Fre10}
Frebel A., 2010, Astron. Nachr., 331, 474

\bibitem[\protect\citeauthoryear{Fulbright \& Johnson}{2003}]{FulJoh03}
Fulbright J.~P., Johnson J.~A., 2003, ApJ, 595, 1154

\bibitem[\protect\citeauthoryear{Garc{\'{\i}}a P{\'e}rez et al.}{2006}]{Gar06}
Garc{\'{\i}}a P{\'e}rez A.~E., Asplund M., Primas F., Nissen P.~E., Gustafsson B., 2006, A\&A, 451, 621

\bibitem[\protect\citeauthoryear{Greggio}{2010}]{Gre10}
Greggio L., 2010, MNRAS, 406, 22

\bibitem[\protect\citeauthoryear{Haardt \& Madau}{2001}]{HarMad01}
Haardt F., Madau P., 2001, in Neumann D.~M., Tran J.~T.~V., eds, Clusters of Galaxies and the High Redshift Universe Observed in X-ray, preprint (astro-ph/0106018)

\bibitem[\protect\citeauthoryear{Heger \& Woosley}{2002}]{HegWoo02}
Heger A., Woosley S.~E., 2002, ApJ, 567, 532

\bibitem[\protect\citeauthoryear{Heger \& Woosley}{2010}]{HegWoo10}
Heger A., Woosley S.~E., 2010, ApJ, 724, 341

\bibitem[\protect\citeauthoryear{Hirschi}{2007}]{Hir07}
Hirschi R., 2007, A\&A, 461, 571

\bibitem[\protect\citeauthoryear{Jenkins \& Tripp}{2006}]{JenTri06}
Jenkins, E.~B., \& Tripp, T.~M. 2006, ApJ, 637, 548

\bibitem[\protect\citeauthoryear{Joggerst, Almgren, \& Woosley}{2010b}]{Jog10b}
Joggerst C.~C., Almgren A., Woosley S.~E., 2010b, ApJ, 723, 353

\bibitem[\protect\citeauthoryear{Joggerst et al.}{2010a}]{Jog10a}
Joggerst C.~C., Almgren A., Bell J., Heger A., Whalen D., Woosley S.~E., 2010a, ApJ, 709, 11 

\bibitem[\protect\citeauthoryear{Joggerst, Woosley, \& Heger}{2009}]{JogWooHeg09}
Joggerst C.~C., Woosley S.~E., Heger A., 2009, ApJ, 693, 1780

\bibitem[\protect\citeauthoryear{Jorgenson et al.}{2009}]{Jor09}
Jorgenson R.~A., Wolfe A.~M., Prochaska J.~X., Carswell R.~F., 2009, ApJ, 704, 247

\bibitem[\protect\citeauthoryear{Karlsson, Bromm, \& Bland-Hawthorn}{2011}]{KarBroBla11}
Karlsson T., Bromm V., Bland-Hawthorn J., 2011, arXiv, arXiv:1101.4024

\bibitem[\protect\citeauthoryear{Kiselman}{1993}]{Kis93}
Kiselman D., 1993, A\&A, 275, 269

\bibitem[\protect\citeauthoryear{Ledoux et al.}{2006}]{Led06}
Ledoux, C., Petitjean, P., Fynbo, J.~P.~U., M{\o}ller, P., \& Srianand, R.\ 2006, A\&A, 457, 71

\bibitem[\protect\citeauthoryear{Ledoux, Petitjean, \& Srianand}{2003}]{LedPetSri03}
Ledoux C., Petitjean P., Srianand R., 2003, MNRAS, 346, 209

\bibitem[\protect\citeauthoryear{Lodders, Plame, \& Gail}{2009}]{LodPlaGai09}
Lodders K., Plame H., Gail H.-P., 2009, Tr{\"u}mper~J.E., ed., Landolt-B\"{o}rnstein, New Series,
Abundances of the Elements in the Solar System. Springer-Verlag, Berlin, p. 44

\bibitem[\protect\citeauthoryear{Madau, Ferrara, \& Rees}{2001}]{MadFerRee01}
Madau P., Ferrara A., Rees M.~J., 2001, ApJ, 555, 92

\bibitem[\protect\citeauthoryear{Mannucci, Della Valle, \& Panagia}{2006}]{ManDelPan06}
Mannucci F., Della Valle M., Panagia N., 2006, MNRAS, 370, 773

\bibitem[\protect\citeauthoryear{Mashonkina et al.}{2011}]{Mas11}
Mashonkina L., Gehren T., Shi J.-R., Korn A.~J., Grupp F., 2011, A\&A, 528, A87

\bibitem[\protect\citeauthoryear{McWilliam}{1997}]{McW97}
McWilliam A., 1997, ARA\&A, 35, 503

\bibitem[\protect\citeauthoryear{Meynet et al.}{2010}]{Mey10}
Meynet G., Hirschi R., Ekstrom S., Maeder A., Georgy C., Eggenberger P., Chiappini C., 2010, A\&A, 521, A30

\bibitem[\protect\citeauthoryear{Meynet, Ekstr{\"o}m, \& Maeder}{2006}]{MeyEksMae06}
Meynet G., Ekstr{\"o}m S., Maeder A., 2006, A\&A, 447, 623

\bibitem[\protect\citeauthoryear{Molaro et al.}{2000}]{Mol00}
Molaro P., Bonifacio P., Centuri{\'o}n M., D'Odorico S., Vladilo G., Santin P., Di Marcantonio P., 2000, ApJ, 541, 54

\bibitem[\protect\citeauthoryear{Morton}{2003}]{Mor03}
Morton, D.~C. 2003, ApJS, 149, 205

\bibitem[\protect\citeauthoryear{Murphy et al.}{2007}]{Mur07}
Murphy, M.~T., Curran, S.~J., Webb, J.~K., M{\'e}nager, H., \& Zych, B.~J.\ 2007, MNRAS, 376, 673

\bibitem[\protect\citeauthoryear{Nissen et al.}{2007}]{Nis07}
Nissen P.~E., Akerman C., Asplund M., Fabbian D., Kerber F., Kaufl H.~U., Pettini M., 2007, A\&A, 469, 319

\bibitem[\protect\citeauthoryear{Nissen et al.}{2002}]{Nis02}
Nissen P.~E., Primas F., Asplund M., Lambert D.~L., 2002, A\&A, 390, 235

\bibitem[\protect\citeauthoryear{Norris et al.}{2007}]{Nor07}
Norris J.~E., Christlieb N., Korn A.~J., Eriksson K., Bessell M.~S., Beers T.~C., Wisotzki L., Reimers D., 2007, ApJ, 670, 774

\bibitem[\protect\citeauthoryear{Noterdaeme et al.}{2008}]{Not08}
Noterdaeme P., Ledoux C., Petitjean P., Srianand R., 2008, A\&A, 481, 327

\bibitem[\protect\citeauthoryear{Noterdaeme et al.}{2009}]{Not09}
Noterdaeme P., Petitjean P., Ledoux C., Srianand R., 2009, A\&A, 505, 1087

\bibitem[\protect\citeauthoryear{O'Meara et al.}{2005}]{OMe05}
O'Meara J.~M., Burles S., Prochaska J.~X., Prochter G., Bernstein R., 2005, Williams~P., Shu~C.-G., Menard~B., eds, Proc. IAU Symp. 199, 
Chemical history at $z \ge 2.0$: first results from the Magellan+Keck survey of Lyman limit systems. Cambridge University Press, Cambridge, p.463

\bibitem[\protect\citeauthoryear{O'Meara et al.}{2006}]{OMe06}
O'Meara J.~M., Burles S., Prochaska J.~X., Prochter G.~E., Bernstein R.~A., Burgess K.~M., 2006, ApJ, 649, L61

\bibitem[\protect\citeauthoryear{Oppenheimer \& Dav{\'e}}{2008}]{OppDav08}
Oppenheimer B.~D., Dav{\'e} R., 2008, MNRAS, 387, 577


\bibitem[\protect\citeauthoryear{Penprase et al.}{2008}]{Pen08}
Penprase B.~E., Sargent W.~L.~W., Martinez I.~T., Prochaska J.~X., Beeler D.~J., 2008, AIP Conf. Proc., 990, 499

\bibitem[\protect\citeauthoryear{Penprase et al.}{2010}]{Pen10}
Penprase B.~E., Prochaska J.~X., Sargent W.~L.~W., Toro-Martinez I., Beeler D.~J., 2010, ApJ, 721, 1

\bibitem[\protect\citeauthoryear{P{\'e}roux et al.}{2007}]{Per07}
P{\'e}roux C., Dessauges-Zavadsky M., D'Odorico S., Kim T.-S., McMahon R.~G., 2007, MNRAS, 382, 177

\bibitem[\protect\citeauthoryear{P{\'e}roux et al.}{2005}]{Per05}
P{\'e}roux C., Dessauges-Zavadsky M., D'Odorico S., Sun Kim T., McMahon R.~G., 2005, MNRAS, 363, 479

\bibitem[\protect\citeauthoryear{P{\'e}roux et al.}{2003a}]{Per03a}
P{\'e}roux C., McMahon R.~G., Storrie-Lombardi L.~J., Irwin M.~J., 2003a, MNRAS, 346, 1103

\bibitem[\protect\citeauthoryear{P{\'e}roux et al.}{2003b}]{Per03b}
P{\'e}roux C., Dessauges-Zavadsky M., D'Odorico S., Kim T.-S., McMahon R.~G., 2003b, MNRAS, 345, 480

\bibitem[\protect\citeauthoryear{P{\'e}roux et al.}{2001}]{Per01}
P{\'e}roux C., Storrie-Lombardi L.~J., McMahon R.~G., Irwin M., Hook I.~M., 2001, AJ, 121, 1799

\bibitem[\protect\citeauthoryear{Petitjean, Ledoux, \& Srianand}{2008}]{PetLedSri08}
Petitjean P., Ledoux C., Srianand R., 2008, A\&A, 480, 349

\bibitem[\protect\citeauthoryear{Pettini et al.}{1997}]{Pet97}
Pettini M., King D.~L., Smith L.~J., Hunstead R.~W., 1997, ApJ, 478, 536

\bibitem[\protect\citeauthoryear{Pettini et al.}{2008}]{Pet08}
Pettini M., Zych B.~J., Steidel C.~C., Chaffee F.~H., 2008, MNRAS, 385, 2011

\bibitem[\protect\citeauthoryear{Pontzen et al.}{2008}]{Pon08}
Pontzen A., et al., 2008, MNRAS, 390, 1349

\bibitem[\protect\citeauthoryear{Prochaska}{2006}]{Pro06}
Prochaska J.~X., 2006, ApJ, 650, 272

\bibitem[\protect\citeauthoryear{Prochaska et al.}{2008}]{Pro08}
Prochaska, J.~X., Chen, H.-W., Wolfe, A.~M., Dessauges-Zavadsky, M., \& Bloom, J.~S.\ 2008, ApJ, 672, 59

\bibitem[\protect\citeauthoryear{Prochaska et al.}{2002}]{Pro02}
Prochaska J.~X., Henry R.~B.~C., O'Meara J.~M., Tytler D., Wolfe A.~M., Kirkman D., Lubin D., Suzuki N., 2002, PASP, 114, 933

\bibitem[\protect\citeauthoryear{Prochaska et al.}{2003}]{Pro03}
Prochaska J.~X., Gawiser E., Wolfe A.~M., Cooke J., Gelino D., 2003, ApJS, 147, 227

\bibitem[\protect\citeauthoryear{Prochaska et al.}{2007}]{Pro07}
Prochaska J.~X., Wolfe A.~M., Howk J.~C., Gawiser E., Burles S.~M., Cooke J., 2007, ApJS, 171, 29

\bibitem[\protect\citeauthoryear{Prochaska \& Wolfe}{2002}]{ProWol02}
Prochaska J.~X., Wolfe A.~M., 2002, ApJ, 566, 68

\bibitem[\protect\citeauthoryear{Prochaska \& Wolfe}{2009}]{ProWol09}
Prochaska J.~X., Wolfe A.~M., 2009, ApJ, 696, 1543

\bibitem[\protect\citeauthoryear{Ryan et al.}{2005}]{Rya05}
Ryan S.~G., Aoki W., Norris J.~E., Beers T.~C., 2005, ApJ, 635, 349

\bibitem[\protect\citeauthoryear{Simcoe, Sargent, \& Rauch}{2004}]{SimSarRau04}
Simcoe R.~A., Sargent W.~L.~W., Rauch M., 2004, ApJ, 606, 92

\bibitem[\protect\citeauthoryear{Spite et al.}{2005}]{Spi05}
Spite M., et al., 2005, A\&A, 430, 655

\bibitem[\protect\citeauthoryear{Srianand et al.}{2010}]{Sri10}
Srianand R., Gupta N., Petitjean P., Noterdaeme P., Ledoux C., 2010, MNRAS, 405, 1888

\bibitem[\protect\citeauthoryear{Suda et al.}{2008}]{Sud08}
Suda T., et al., 2008, PASJ, 60, 1159

\bibitem[\protect\citeauthoryear{Tescari et al.}{2009}]{Tes09}
Tescari E., Viel M., Tornatore L., Borgani S., 2009, MNRAS, 397, 411

\bibitem[\protect\citeauthoryear{Tomkin et al.}{1992}]{Tom92}
Tomkin J., Lemke M., Lambert D.~L., Sneden C., 1992, AJ, 104, 1568

\bibitem[\protect\citeauthoryear{Tsujimoto \& Bekki}{2011}]{TsuBek11}
Tsujimoto T., Bekki K., 2011, arXiv, arXiv:1103.0033

\bibitem[\protect\citeauthoryear{Umeda \& Nomoto}{2002}]{UmeNom02}
Umeda H., Nomoto K., 2002, ApJ, 565, 385

\bibitem[\protect\citeauthoryear{Umeda \& Nomoto}{2003}]{UmeNom03}
Umeda H., Nomoto K., 2003, Nature, 422, 871

\bibitem[\protect\citeauthoryear{Vladilo et al.}{2001}]{Vla01} 
Vladilo, G., Centuri{\'o}n, M., Bonifacio, P., \& Howk, J.~C.\ 2001, ApJ, 557, 1007 

\bibitem[\protect\citeauthoryear{Vladilo}{2002}]{Vla02}
Vladilo G., 2002, A\&A, 391, 407

\bibitem[\protect\citeauthoryear{Vladilo}{2004}]{Vla04}
Vladilo G., 2004, A\&A, 421, 479

\bibitem[\protect\citeauthoryear{Vogt et al.}{1994}]{Vog94}
Vogt S.~S., et al., 1994, SPIE, 2198, 362

\bibitem[\protect\citeauthoryear{Wheeler, Sneden, \& Truran}{1989}]{WheSneTru89}
Wheeler J.~C., Sneden C., Truran J.~W., Jr., 1989, ARA\&A, 27, 279

\bibitem[\protect\citeauthoryear{Wolfe, Gawiser, \& Prochaska}{2005}]{WolGawPro05}
Wolfe A.~M., Gawiser E., Prochaska J.~X., 2005, ARA\&A, 43, 861

\bibitem[\protect\citeauthoryear{Wolfe et al.}{1986}]{Wol86}
Wolfe A.~M., Turnshek D.~A., Smith H.~E., Cohen R.~D., 1986, ApJS, 61, 249

\end{thebibliography}
\end{document}